\documentclass[aps,prx,twocolumn,longbibliography]{revtex4-1}
\usepackage{latexsym,epsfig,amssymb,amsfonts,amsmath,color,graphicx,bbm}
\usepackage[utf8]{inputenc}
\usepackage{hyperref}
\hypersetup{
	colorlinks   = true,
	citecolor    = black
}

\newcommand{\ket}[1]{\left | \, #1 \right \rangle}

\newcommand{\bra}[1]{\left \langle #1 \, \right |}

\newcommand{\eqr}[1]{Eq.~(\ref{#1})}
\newcommand{\fir}[1]{Fig.~\ref{#1}}

\newcommand{\secr}[1]{Sec.~\ref{#1}}

\allowdisplaybreaks

\begin{document}

\title{Enhancement of super-exchange pairing in the periodically-driven Hubbard model}

\author{J. R. Coulthard$^{1}$, S. R. Clark$^{2,3}$, S. Al-Assam$^{1}$, A. Cavalleri$^{3,1}$ and D. Jaksch$^{1,4}$}
\affiliation{$^1$Clarendon Laboratory, University of Oxford, Parks Road, Oxford OX1 3PU, United Kingdom}
\affiliation{$^2$Department of Physics, University of Bath, Claverton Down, Bath BA2 7AY, United Kingdom}
\affiliation{$^3$Max Planck Institute for the Structure and Dynamics of Matter, University of Hamburg CFEL, Hamburg, Germany}
\affiliation{$^4$Centre for Quantum Technologies, National University of Singapore, 3 Science Drive 2, Singapore 117543}

\date{\today}

\begin{abstract}
Recent experiments performed on cuprates and alkali-doped fullerides have demonstated that key signatures of superconductivity can be induced above the equilibrium critical temperature by optical modulation. These observations in disparate physical systems may indicate a general underlying mechanism. Multiple theories have been proposed, but these either consider specific features, such as competing instabilities, or focus on conventional BCS-type superconductivity. Here we show that periodic driving can enhance electron pairing in strongly-correlated systems. Focusing on the strongly-repulsive limit of the doped Hubbard model, we investigate in-gap, spatially inhomogeneous, on-site modulations. We demonstrate that such modulations substantially reduce electronic hopping, while simultaneously sustaining super-exchange interactions and pair hopping via driving-induced virtual charge excitations. We calculate real-time dynamics for the one-dimensional case, starting from zero and finite temperature initial states, and show that enhanced singlet--pair correlations emerge quickly and robustly in the out-of-equilibrium many-body state. Our results reveal a fundamental pairing mechanism that might underpin optically induced superconductivity in some strongly correlated quantum materials.

%We show that periodic driving can enhance electron pairing in strongly-correlated systems. Focusing on the strong-coupling limit of the doped Hubbard model we investigate in-gap, spatially inhomogeneous, on-site modulations and demonstrate that they substantially reduce electronic hopping without suppressing super-exchange interactions and pair hopping. We calculate real-time dynamics for the one-dimensional case, starting from zero and finite temperature initial states, and show that enhanced singlet--pair correlations emerge quickly and robustly in the out-of-equilibrium many-body state. Our results reveal a fundamental pairing mechanism that might underpin optically induced superconductivity in some strongly correlated quantum materials.
\end{abstract}


\maketitle
\section{Introduction}
Controlling the structural and electronic properties of a solid by resonantly driving a single low-energy degree of freedom is emerging as a transformative tool in materials science \cite{Rini2007}. Such excitations often play a decisive role in stabilising various broken-symmetry states, and driving them opens up the possibility to switch between phases. This not only includes the melting of equilibrium long-ranged order, like charge-density-waves \cite{Fausti2011,Schmitt2008,Miyano1997,Cavalleri2001,Perfetti2008}, magnetic order \cite{Beaurepaire1996,Stanciu2007,Ehrke2011}, and orbital order \cite{Ehrke2011,Forst2011a}, but even more remarkably, \emph{inducing} order, such as superconductivity, out of equilibrium \cite{Mitrano2016, Denny2015}.

To date, light-induced superconductivity has been observed in several cuprates \cite{Hu2014,Kaiser2014a,Mankowsky2014} and an alkali-doped fullerene \cite{Mitrano2016} all with quite distinct physics. This raises the question of how ubiquitous such effects are, and what mechanism(s) might underpin their appearance. So far, theoretical exploration has concentrated on a minimal Fr{\"o}hlich-type model of phonon-mediated superconductivity \cite{Sentef2015,Knap2015} subjected to a driving induced quench of the electronic hopping amplitude. This was envisaged as occurring from a modified electronic structure due to non-linear phonon coupling \cite{Sentef2015,Subedi2014}, or from polaronic suppression due to phonon squeezing \cite{Knap2015}. In either case, this results in an increase in the density of states at the Fermi level, giving a corresponding increase in the superconducting coupling constant. Despite the slow collective dynamics and elevated electron-phonon scattering, fast enhancements of the superconducting order parameter were predicted.

\begin{figure}
\setlength{\unitlength}{\linewidth}
% \begin{picture}(0.99, 0.25)
% \put(0,0){\includegraphics[width=0.96\linewidth]{./UndrivenHubbardNoPairing.pdf}}
% \put(0.0, 0.2){(a)}
% \put(0.1, 0.165){$t$}
% \put(0.35 ,0.165){$J$}
% \put(0.48 ,0.14){$U \gg t$}
% \put(0.75 ,0.165){$\alpha J$}
% \end{picture}

% \begin{picture}(0.99, 0.45)
% \put(0,0){\includegraphics[width=0.96\linewidth]{./DrivenHubbardWithPairing5.pdf}}
% \put(0.0,0.38){(b)}

% \put(0.06, 0.34){$\tilde{t} \ll t$}
% %\put(0.30 ,0.34){$J$}
% \put(0.345 ,0.34){$J$}
% \put(0.48 ,0.32){$U \gg t$}
% \put(0.76 ,0.345){$ \alpha J$}
% \put(0.52 ,0.13){$\Omega$}
% \end{picture}
\includegraphics[width=0.96\linewidth]{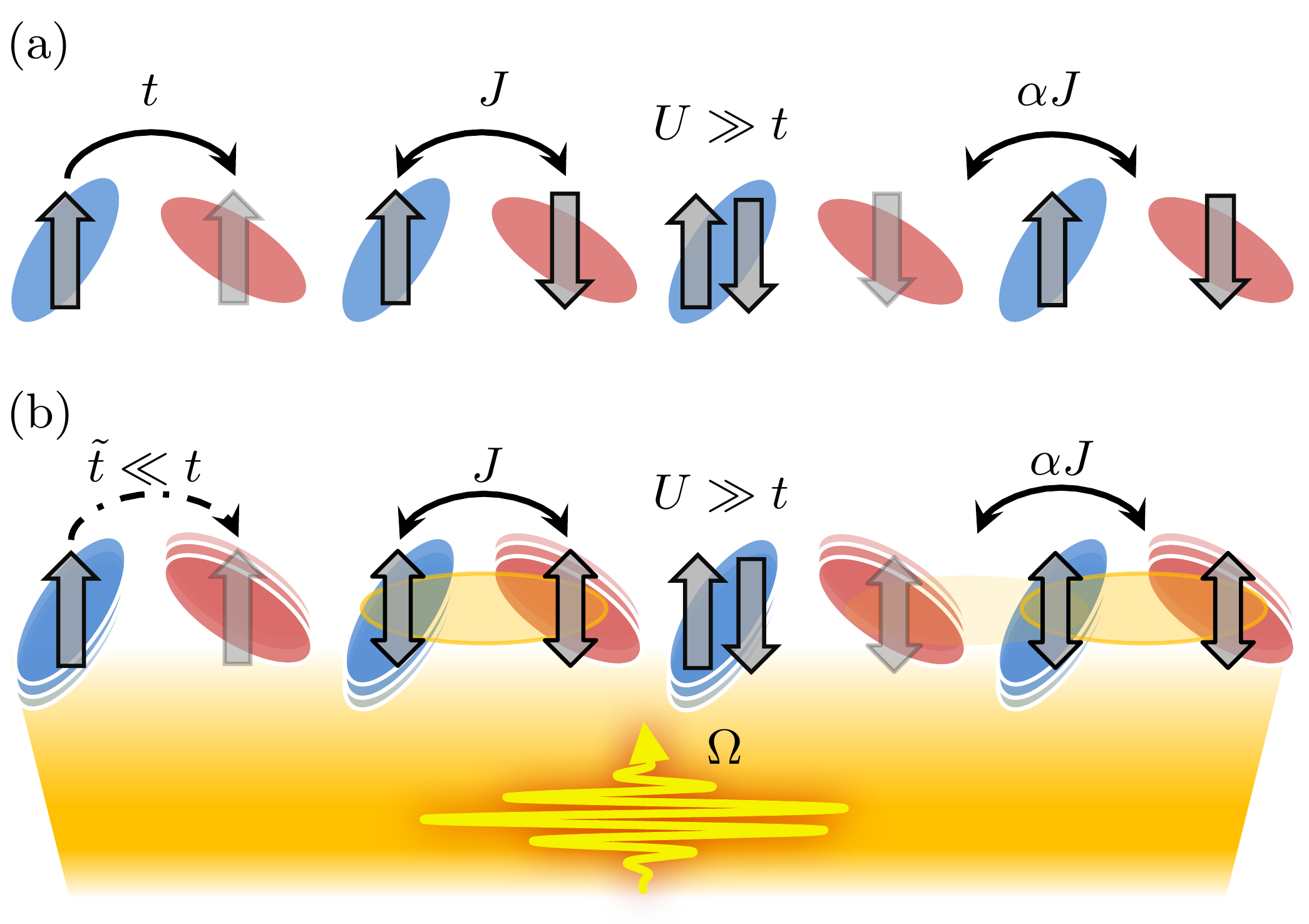}
\caption{(a) A bipartite chain with electron hopping $t$, exchange interaction $J$ and onsite correlation $U \gg t$. Singlet pairs move through the lattice with effective hopping rate $\alpha J$. (b) Exciting on-site vibrational modes with $t<\Omega<U$ greatly suppresses hopping $\tilde{t} \ll t$, but leaves super-exchange interaction $J$ and pair hopping $\alpha J$ approximately unchanged, resulting in an enhancement of nearest-neighbour singlet pairing.\label{fig1}}
\end{figure}

In this work we propose a qualitatively different mechanism for driving-enhanced superconductivity in strongly correlated lattice systems.
We show that the modulation of site energies in a bipartite lattice with a frequency $\Omega$ inside the charge-transfer gap $U$ {\em slows down} electron hopping $t \rightarrow \tilde{t} < t$ without reducing the super-exchange interactions $J$ or pair hopping $\alpha J$. This is because the absorption and re-emission of quanta from the driving field creates virtual charge-density excitations, leading to additional exchange interactions. 
Driving-induced contributions therefore break the usual relation $J \propto t^2/U$, and compensate for the suppressed hopping contribution. As a result, the normally subordinate $J$ can induce a strong pairing effect and enhance long-range pair correlations, even in one spatial dimension. 
%This driving-induced contribution compensates for the suppressed contribution from the hopping and yields an overall $J$ and pair hopping $\alpha J$ that is approximately constant. The driving thus breaks the usual relation $J \propto t^2/U$, allowing the normally subordinate $J$ to induce a strong pairing effect, and enhance long-range pair correlations even in one spatial dimension. 
We find that on the moderate timescales assessed, the resulting non-equilibrium states are not significantly heated by the driving field. Instead, driving can {\em substantially reduce} the effective temperature of an initial thermal state on experimentally relevant timescales, akin to many-body adiabatic cooling.
%We show that the modulation of site energies in a bipartite lattice with frequency $\Omega$ inside the charge-transfer gap $U$ {\em slows down} electron hopping $t$ without reducing super-exchange interactions $J$ or pair hopping $\alpha J$ (see \fir{fig1}). It does this by reducing the single-particle tunneling $t \rightarrow \tilde{t}$, thus reducing the usual kinetic superexchange $J_{\rm kin} = 4\tilde{t}^2/U$, while simultaneously providing an \emph{additional} driving-induced superexchange contribution $J_{\rm light}$. This balances the suppression and maintains the overall strength of the superexchange and pair-hopping. The driving thus breaks the usual relation $J \propto t^2/U$, allowing the normally subordinate $J$ to induce a strong pairing effect, and enhance long-range pair correlations even in one spatial dimension. 
%We find that on the moderate timescales assessed, the resulting non-equilibrium states are not significantly heated by the driving field. Instead, driving can {\em substantially reduce} the effective temperature of an initial thermal state on experimentally relevant timescales, akin to many-body adiabatic cooling.

Specifically, we investigate the Hubbard model with a bare hopping rate $t$ and a large on-site interaction $U\gg t$, subjected to periodic driving of frequency $\Omega$ that modulates the on-site energy in time $\tau$ with a spatially alternating pattern. We numerically confirm that this leads to dynamical enhancement of pairing for a one-dimensional system with realistic finite-frequency driving $t < \Omega < U$, ramped up on adiabatic and non-adiabatic time-scales, and at zero and finite initial temperatures. We also analytically study the system in Floquet theory \cite{Shirley1965,Dunlap1986, Buchleitner2002,Bukov2015,Mentink2014,Creffield2002,Eckardt2005} and derive an effective static model to provide qualitative insights into the non-equilibrium dynamics at low temperatures. Note that, in contrast to many studies using Floquet theory, $\Omega$ is not the largest frequency in our setup.

While motivated by experiments, our aim here is to explore a fundamental principle from a minimal, periodically driven, strongly-correlated model, as opposed to material-specific ab initio calculations. Nonetheless, organic superconductors (such as charge transfer salts) under THz driving are likely candidate systems in which the mechanism proposed here may be observed. In these materials, THz pulses can resonantly excite an infrared-active local molecular vibration often located in the in-gap regime \cite{Kaiser2014b,Singla2015}. The ensuing ``sloshing'' motion of the molecule is sufficiently large in amplitude and heavy that its leading order effect is to couple to the electronic states via a time periodic modulation of the on-site energy. These massive vibrations can therefore be considered classical oscillators for which back-action from the electronic system can be safely ignored. While the laser pulse itself excites coherently and uniformly across the system, the material is assumed to possess a two-molecule unit cell \footnote{For the main effects described here, a two-site periodicity is not essential, and may be realised with higher spatial periodicities.}, as in \fir{fig1}, so that the modulation induced on the two interlocking sub-lattices $a$ and $b$ differs in amplitude and/or phase. Such a unit cell might be composed of different molecules or identical molecules with differing orientations due to the stacking morphology \cite{Mori1998}. Similar physics can be cleanly realized in optical lattices filled with ultracold fermionic atoms by ``shaking" the lattice \cite{Eckardt2009,Jotzu2014,Bloch2008, Lewenstein2012, Struck2012}, with very recent work ongoing in this direction \cite{Desbuquois2017}, or by exploiting Raman transitions between internal atomic states \cite{Bilitewski2016}. Our results provide a mechanism by which super-exchange physics may be better exposed in these systems.

This paper is organized as follows. In Sec.~\ref{sec:DHM} we introduce the driven Hubbard model, describe the Floquet basis and work out the quasi-energies for a small system via exact numerical diagonalization. We then use time-dependent density matrix renormalisation group (DMRG) methods in Sec.~\ref{sec:DFP} to study the real-time dynamics of the driven Hubbard model in one spatial dimension. In Sec.~\ref{sec:tJM} we derive an effective static model whose ground and thermal states are used to approximate the non-equilibrium states of the driven Hubbard model. Finally, we conclude in Sec.~\ref{sec:Conc}.

\section{The Driven Hubbard Model}\label{sec:DHM}
The focus of this work is the driven Hubbard model Hamiltonian  (taking $\hbar = 1$)
\begin{equation}
\hat{H}(\tau) = \hat{H}_{\rm hub} + \hat{H}_{\rm drive}(\tau),
\end{equation}
where $\hat{H}_{\rm hub} = \hat{H}_{\rm hop} + \hat{H}_{\rm int} -\mu \sum_{j} \hat{n}_{j}$ and contributions given by
\begin{eqnarray}
\hat{H}_{\rm hop} &=& -t \sum_{\langle ij \rangle \sigma} ( \hat{c}^{\dagger}_{i, \sigma} \hat{c}_{j, \sigma} + \textrm{H.c.}), \\
\hat{H}_{\rm int} &=& U \sum_{j} \hat{n}_{j, \uparrow} \hat{n}_{j, \downarrow}, \\
\hat{H}_{\rm drive}(\tau) &=& \frac{V_a}{2}\sin(\Omega\tau - \Delta\phi) \sum_{j \in a} \hat{n}_j \nonumber \\
&&\qquad + \frac{V_b}{2}\sin(\Omega\tau + \Delta\phi) \sum_{j \in b} \hat{n}_j. \label{driving}
\end{eqnarray}
Here $\hat{c}_{i,\sigma}$ with $\sigma=\uparrow,\downarrow$ is the fermionic annihilation operator for a spin-$\sigma$ electron on site $j$, $\hat{n}_{j,\sigma} = \hat{c}^\dagger_{j,\sigma}\hat{c}_{j,\sigma}$, $\hat{n}_j = \hat{n}_{j, \uparrow} + \hat{n}_{j, \downarrow}$, and $\langle ij\rangle$ denotes nearest-neighbour sites on a bipartite lattice composed of $a$ and $b$ sub-lattices. We denote the lattice filling by $\bar n = \sum_j \langle \hat{n}_j \rangle/L$ where $L$ is the number of lattice sites. The hopping amplitude is $t$, $U$ the on-site Coulomb repulsion, and $\mu$ the chemical potential. The driving $\hat{H}_{\rm drive}(\tau)$ describes a time $\tau$ periodic single particle Hamiltonian with driving frequency $\Omega$, corresponding sub-lattice driving amplitudes $V_{a(b)}$, and a phase difference of $2 \Delta \phi$. For simplicity we assume $V_a = V_b = V$ for the driving amplitude, which may be a function of time $V(\tau)$, and constant phase $\Delta \phi = \pi/2$ throughout the paper. However, the qualitative features of our results are expected more generally (see Appendix~\ref{sec:app_floquet}). We next use a Floquet analysis to start investigating the dynamics induced by $\hat H(\tau)$.

\subsection{Floquet analysis} \label{sec:floquet}
Floquet theory \cite{Shirley1965, Bukov2015} is based on the time analog of Bloch's theorem and is applicable here since $\hat{H}(\tau+T) = \hat{H}(\tau)$ with $T = 2\pi/\Omega$. It gives that a complete set of solutions to the time-dependent Schr{\"o}dinger equation $[\hat{H}(\tau) - {\rm i}\partial_\tau]\ket{\Psi(\tau)} = 0$ can then be written as $\ket{\psi_\eta(\tau)} = \exp(-{\rm i}\epsilon_\eta \tau)\ket{\phi_\eta(\tau)}$. The $T$-periodic Floquet states $\ket{\phi_\eta(\tau+T)} = \ket{\phi_\eta(\tau)}$ are solutions to the eigenvalue equation
\begin{equation}
[\hat{H}(\tau) - {\rm i}\partial_\tau]\ket{\phi_\eta(\tau)} = \epsilon_\eta\ket{\phi_\eta(\tau)}\,, \label{evalue}
\end{equation}
with associated real quasi-energies $\epsilon_\eta$ that are defined up to integer multiples of $\Omega$. Periodicity means that the quasi-energy spectrum possesses a zone-like structure where physically distinct eigenstates lie within a quasi-energy range $E- \frac{1}{2}\Omega < \epsilon_\eta \leq E + \frac{1}{2}\Omega$, where the choice of $E$ is arbitrary but often taken as $E=0$.

The Hermitian Floquet Hamiltonian $\hat{H}_{\rm F} = \hat{H}(\tau) - {\rm i}\partial_\tau$ acts on an extended Hilbert space $\mathcal{H}\otimes\mathcal{T}$ which augments the original Hilbert space $\mathcal H$ by the space $\mathcal{T}$ of square-integrable $T$-periodic functions in time. This extended Hilbert space, whose elements are denoted as $|\chi\rangle\!\rangle$, is endowed with a suitable scalar product by time-averaging over a period $T$ as
\begin{equation}
\langle\!\langle \chi | \xi \rangle\!\rangle = \frac{1}{T}\int_0^T \langle  \chi(\tau) | \xi(\tau)  \rangle \, {\rm d}\tau, \label{ext_scalar_prod}
\end{equation}
where $\ket{\chi(\tau)}$ and $\ket{\xi(\tau)}$ are any $T$-periodic states in $\mathcal{H}$, and $\langle  \chi(\tau) | \xi(\tau)  \rangle$ is the conventional scalar product for $\mathcal{H}$.

We take the Fock basis $\ket{\{n_{j,\sigma}\}}$ of the lattice system, where $n_{j,\sigma}$ spin-$\sigma$ electrons occupy site $j$, and construct an orthonormal Floquet-Fock basis of $\mathcal{H}\otimes\mathcal{T}$ as \cite{Eckardt2005}
\begin{equation} \label{floquet_basis}
|\{n_{j,\sigma}\},m\rangle\!\rangle = \ket{\{n_{j,\sigma}\}}e^{{\rm i}m\Omega\tau + {\rm i}\frac{V}{2 \Omega} \sin(\Omega \tau) (\sum_{j \in a}n_j - \sum_{j \in b}n_j)} \,.
\end{equation}
These basis states include phases for the $m$-th Fourier component and those associated with transforming into the frame rotating with $\hat{H}_{\rm drive}(\tau)$.

The matrix elements of the Floquet Hamiltonian $\hat{H}_{\rm F}$ in this basis are
\begin{widetext}
\begin{eqnarray}
\langle\!\langle \{n'_{j,\sigma}\},m' |\hat{H}_{\rm F}|\{n_{j,\sigma}\},m \rangle\!\rangle &=& \zeta_{m'-m} ~\langle \{n'_{j,\sigma}\} |\hat{H}_{\rm hop}| \{n_{j,\sigma}\}\rangle  + \, \delta_{m,m'}\left[\langle \{n'_{j,\sigma}\}|\hat{H}_{\rm int}|\{n_{j,\sigma}\}\rangle + m\Omega\right]. \label{floquet_ham}
\end{eqnarray}
\end{widetext}
The couplings $\zeta_{m'-m}$ are given by (also see Appendix~\ref{sec:app_floquet})
\begin{equation}
\zeta_{m'-m}  = s^{m'-m}\mathcal{J}_{m'-m}(\nu), \label{driven_couplings}
\end{equation}
where $\mathcal{J}_n$ is the $n$-th order Bessel function of the first kind and $\nu = V /\Omega$. They depend on the difference in the Fourier components $m'-m$, and also on the driving parameter $\nu$ as well as the change in sub-lattice $a$ occupation $s = \sum_{j \in a}(n'_j -n_j)=\pm 1$ for the Fock states being connected.

This matrix representation of $\hat{H}_{\rm F}$ has a natural block structure with respect to the Fourier index $m$ labelling the Floquet sector replicas of the system. The diagonal blocks are a matrix representation of $\mathcal{J}_{0}(\nu)\hat{H}_{\rm hop} + \hat{H}_{\rm int}$ in the Fock basis, i.e.~$\hat{H}_{\rm hub}$ with a renormalised hopping amplitude $\tilde{t} \equiv t \mathcal{J}_{0}(\nu)$, and shifted in energy by $m\Omega$. Correspondingly, the off-diagonal blocks coupling different $m$ sectors are a matrix representation of $\zeta_{m'-m} \hat{H}_{\rm hop}$. In the remainder of this section we numerically investigate the Floquet Hamiltonian for a small system.

\subsection{Small system}
Using a small, numerically exactly diagonalisable one-dimensional lattice we calculate the quasi-energy spectrum $\epsilon_\eta$ as a function of $\nu$ from \eqr{evalue}. We concentrate on the quasi-energy states in the $m=0$ Floquet sector that emerge from the low-energy sector of $\hat{H}_{\rm hub}$. In \fir{fig2}(a) the results for high-frequency driving $\Omega \gg U,t$ where the Floquet sectors are energetically well separated and decouple in a perturbative sense. The width of the spectrum initially shrinks with increasing $\nu$ indicating a reduction of the driven hopping amplitude $\tilde{t} < t$, consistent with the $\zeta_0 = \mathcal{J}_0(\nu)$ dependence. When the driving strength reaches $\nu = \nu_0 \approx 2.4$ where $\mathcal{J}_0(\nu_0) = 0$, the electron hopping and the super-exchange are both fully suppressed. The dynamics of the system are therefore frozen, as signified by the collapse of the spectrum to a $\epsilon_\eta = 0$ degeneracy at $\nu_0$. With further increases of $\nu$ the hopping amplitude $\tilde{t}$ becomes negative, and the quasi-energy spectrum correspondingly broadens. This so-called “dynamical localisation" and band-flipping are well known single-particle effect that have been demonstrated experimentally, e.g. in optical lattices \cite{Dunlap1986, Lignier2007, Bloch2008}.

A qualitatively different result occurs for in-gap driving $t < \Omega < U$, as shown in \fir{fig2}(b). The quasi-energy spectrum is again seen to reduce in bandwidth with increasing $\nu$ initially, indicating that $\tilde{t}$ is still being suppressed. 
However, in contrast to high-frequency limit the spectrum retains a finite width proportional to $J = 4t^2/U$, and is shifted down by approximately $J$, even when $\nu \approx \nu_0$. Rather than being frozen out, the dynamics in this driving regime are now being governed by the normally subordinate super-exchange energy $J$ that appears to be unsuppressed. This observation motivates the further numerical studies presented in the next section. There we gather evidence that the driven system has an increased susceptibility to pair formation and long-range correlations for driving $\nu \lesssim \nu_0$. This then leads us to derive an effective static $t$--$J$ model that describes the singlet-pair dynamics in the driven state with good accuracy even for strong driving $\nu > 1$.

%However, in contrast to high-frequency limit the spectrum retains a finite width proportional to $J = 4 t^2/U$ (related to the driving-induced contribution to the pair-hopping, $J_{\rm light}$), and is shifted down by approximately $J$, even when $\nu \approx \nu_0$. 
%Rather than being frozen out, the dynamics in this driving regime are now being governed by the normally subordinate super-exchange energy $\tilde{J} \approx J$ that appears to be unsuppressed. This observation motivates the further numerical studies presented in the next section. There we gather evidence that the driven system has an increased susceptibility to pair formation and long-range correlations for driving $\nu \lesssim \nu_0$. This then leads us to derive an effective static $t$--$J$ model that describes the singlet-pair dynamics in the driven state with good accuracy even for strong driving $\nu > 1$.

\begin{figure}[t]
% \setlength{\unitlength}{\linewidth}
% \begin{picture}(0.98, 0.7)
% \put(0,0){\includegraphics[width=0.90\linewidth]{./quasi_en_hf_lf}}
% \put(0.81, 0.62){(a)}
% \put(0.81, 0.34){(b)}
% \put(0.73, 0.28){$\approx 4J$}
% \put(0.705, 0.27){$\boldsymbol{\downarrow}$}
% \put(0.705, 0.19){$\boldsymbol{\uparrow}$}
% \end{picture}
\includegraphics[width=0.90\linewidth]{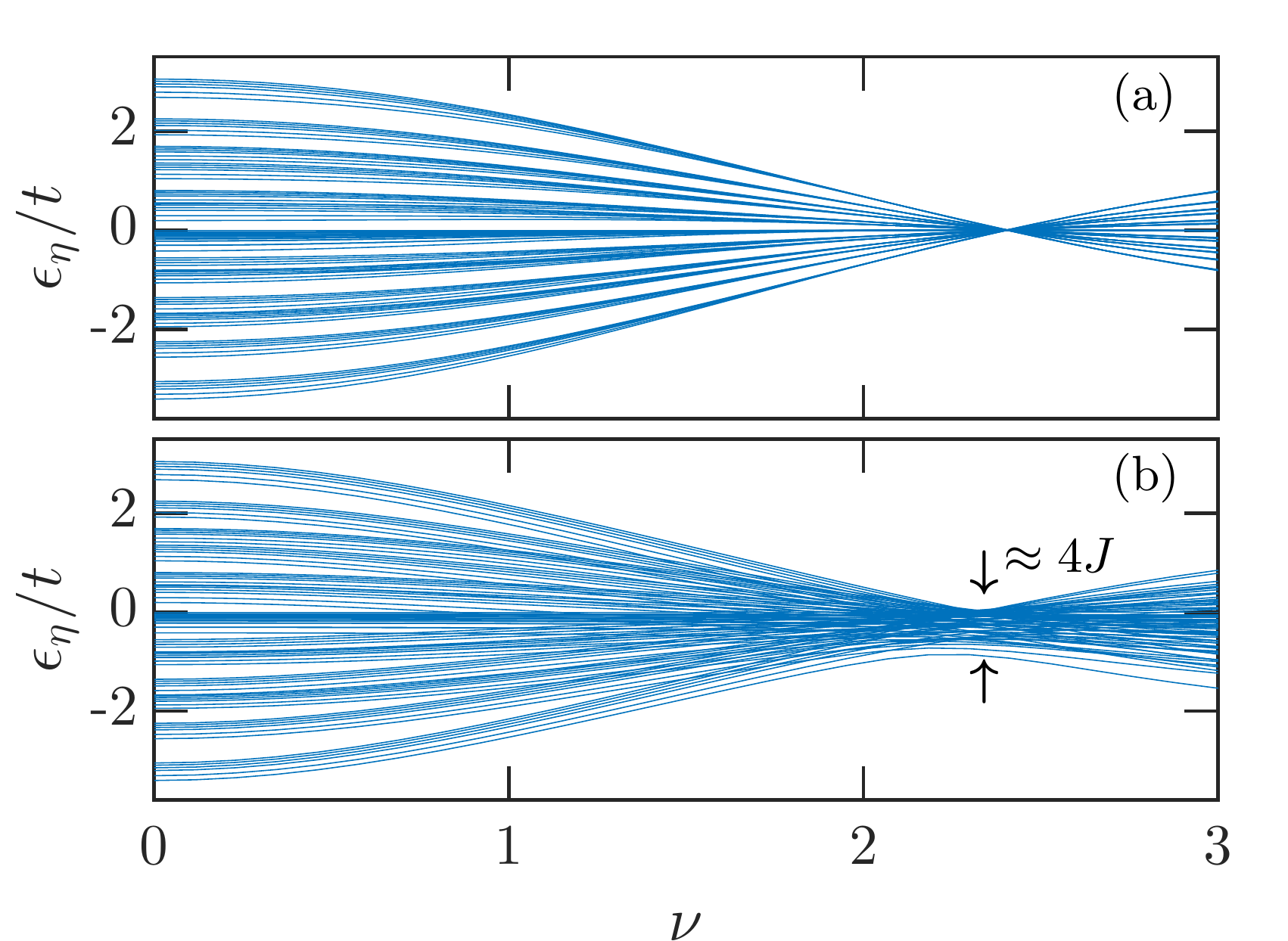}
%\vspace{-1.5em}
\caption{The singly-occupied subspace of the Floquet quasi-energy spectrum $\epsilon_\eta$, computed exactly for $L=6$ sites, with $2$ up electrons and $2$ down electrons. In (a), where $\Omega = 100t$, $U = 20t$,  the quasi-energy spectrum collapses to a single point at $\nu_0 \approx 2.4$, signalling complete suppression of $\tilde{t}$ relative to $U$. In (b), where $\Omega = 6t$, and $U = 20t$, the spectrum does not collapse to a point, but instead strongly resembles the spectrum of a $t$--$J$ Hamiltonian with $J > \tilde{t}$.
\label{fig2}}
\end{figure}

\section{Driving enhanced fermion pairing}\label{sec:DFP}
Here we consider a one dimensional driven Hubbard model, and study directly its real-time dynamics when slowly ramping up the driving amplitude, first at zero temperature for both infinite and finite systems, and then at finite temperature for a finite system. Our numerics are based on highly accurate time-dependent DMRG methods \cite{Schollwock2011,Verstraete2008,Vidal2003,Vidal2007} as implemented in the Tensor Network Theory (TNT) Library \cite{Alassam2016} and are described in more detail in Appendix~\ref{sec:app_num}.

\subsection{Zero temperature} \label{sec:puredriving}

\begin{figure}[t]
\begin{center}
% \begin{picture}(0.49, 0.40)
% \put(0,0){\includegraphics[width=0.49\linewidth]{./PureDrivenDensitySF.png}}
% \put(0.11,0.28){\scriptsize (a)}
% \end{picture}
% \begin{picture}(0.49, 0.40)
% \put(0,0){\includegraphics[width=0.49\linewidth]{./PureDrivenSpinSF.png}}
% \put(0.11,0.28){\scriptsize (b)}
% \end{picture}
% \begin{picture}(0.49, 0.40)
% \put(0,0){\includegraphics[width=0.49\linewidth]{./PureDrivenSingletSF.png}}
% \put(0.11,0.28){\scriptsize (c)}
% %\put(0,0){\includegraphics[width=0.49\linewidth]{./Figures/SingletSF0Plot}}
% \end{picture}
% \begin{picture}(0.49, 0.40)
% \put(0,0){\includegraphics[width=0.49\linewidth]{./PureRSSingletSlice}} %on log scale
% \put(0.12,0.325){\scriptsize (d)}
% \end{picture}
\includegraphics[width=0.98\linewidth]{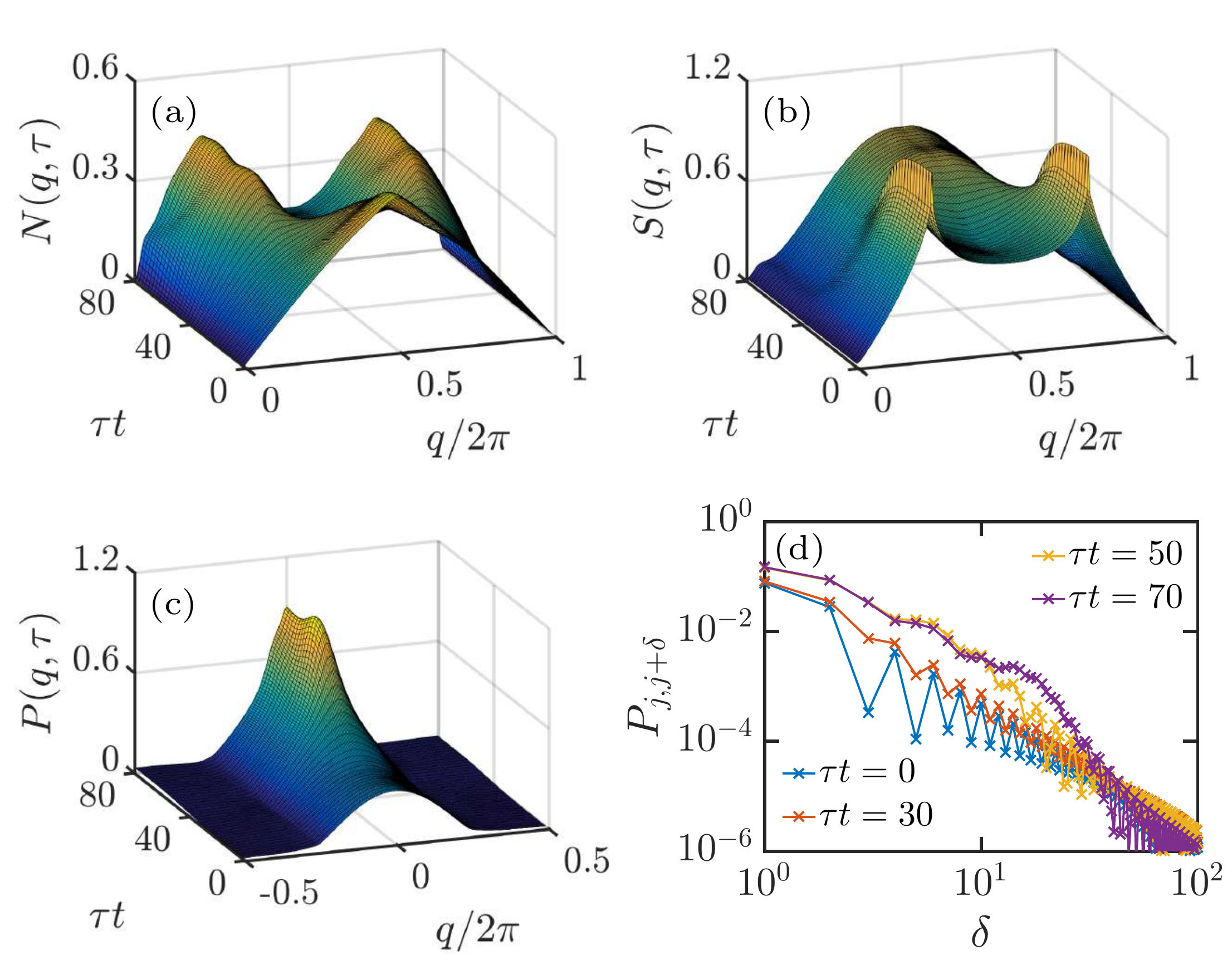}

\caption{An infinite Hubbard chain with $U = 20t$, $\mu = -0.2t$ and driving frequency $\Omega = 6t$. The driving amplitude is ramped according to $\nu(\tau) = (\nu(\infty)/2) \left[\tanh\left((\tau - \tau_0)/\tau_{\rm r}\right) + 1\right] $ with final driving strength $\nu(\infty) = 2.2$, ramp time $\tau_{\rm r} = 15/t$, and $\tau_0 = 25/t$. The plots show (a) the density-density structure factor $N(q,\tau)$, (b) the spin structure factor $S(q,\tau)$ and (c) the singlet pairing structure factor $P(q,\tau)$ as a function of quasi-momentum $q$ and time $\tau$. In (a)--(c) we have averaged over the frequency $\Omega$ oscillations, e.g. that are visible in the line-outs shown in \fir{fig4}(a). Residual low frequency oscillations in these quantities are due to the finite ramping time $\tau_{\rm r}$. In (d) the real-space pairing correlation function $P_{j,j + \delta}$ at various time slices is shown.} \label{fig3}
\end{center}
\end{figure}

We calculate the real-time dynamics for the translationally-invariant infinite system starting from the ground state, using a chemical potential $\mu = -0.2t$, resulting in an approximately quarter-filled system. The driving amplitude is ramped up with in-gap driving frequency $t < \Omega < U$. We characterise the driving induced non-equilibrium state $\ket{\psi(\tau)}$ by studying its density-density correlations $$N_{ij}(\tau) = \langle \hat{n}_i\hat{n}_j\rangle  -\langle \hat{n}_i\rangle \langle \hat{n}_j\rangle\,,$$ the spin-spin correlations $$S_{ij}(\tau) =  \langle \hat{S}^z_i\hat{S}^z_j\rangle\,,$$ with $\hat{S}^z_i = (\hat{n}_{i,\uparrow} - \hat{n}_{i,\downarrow})/2\,,$ and the nearest-neighbour singlet-paring correlations $$P_{ij}(\tau) =  \langle \hat{b}^\dagger_{i,i+1}\hat{b}_{j,j+1}\rangle\,.$$ Here the operators $\hat{b}^\dagger_{ij}$  ($\hat{b}_{ij}$), given by
\begin{equation}
\hat{b}^\dagger_{ij} = \frac{1}{\sqrt{2}}(\hat{c}^\dagger_{i, \uparrow}\hat{c}^\dagger_{j, \downarrow}-\hat{c}^\dagger_{i, \downarrow}\hat{c}^\dagger_{j, \uparrow})\,,
\end{equation}
create (annihilate) a singlet electron pair at sites $i$ and $j$, and
$\langle \cdot \rangle = \bra{\psi(\tau)} \cdot \ket{\psi(\tau)}$. We mostly concern ourselves with the corresponding structure factors,
\begin{equation}
%X(q) = \frac{1}{L} \sum_{jk} X_{jk} {\rm e}^{{\rm i} q(j-k)},
X(q) = \sum_{k} X_{j,j+k} {\rm e}^{{\rm i} q k},
\end{equation}
for the translationally-invariant infinite system, where $X$ is any of the quantities $N$, $S$ or $P$, and $q$ is the dimensionless quasi-momentum.

We plot the density structure factor $N(q, \tau)$  in \fir{fig3}(a). The initial kink at $q = 4k_{\rm F}$, where $k_{\rm F} = \bar n \pi/2$ is the non-interacting Fermi wave-vector, is suppressed in favour of peaks at $q = \pm 2k_{\rm F}$ as the driving increases. This signifies a doubling of the wavelength of Friedel oscillations in the system's density-density correlations and is indicative of the formation of bound pairs. We interpret the slope of $N(q, \tau)$ as $q \rightarrow 0$ as a dynamical version of the Luttinger parameter $K_\rho(\tau)$ \cite{Giamarchi2003}. In equilibrium $N(q) \approx K_\rho q/\pi$ \cite{Ejima2005,Moreno2011}, and in \fir{fig3}(a) it is apparent that the gradient around $q=0$ increases at later times corresponding to an increase in $K_\rho(\tau)$. Eventually $K_{\rho(\tau)}$ exceeds $1$ signifying the formation of an attractive Luttinger liquid.

The spin structure factor, shown in \fir{fig3}(b), begins with sharp peaks at $2k_{\rm F}$, indicating that the Hubbard ground state has a tendency towards antiferromagnetic order. At quarter-filling, this order is incommensurate with the lattice period. In the presence of periodic driving the initial peaks are suppressed and instead a peak at $q=\pi$ forms, consistent with the formation of islands of \emph{commensurate} antiferromagnetic order. The broadness of this emerging peak shows that the underlying order is not quasi-long-ranged yet. Indeed, its form is similar to $S(q) = \bar n (1-\cos q)$ expected for a gas of free nearest-neighbour singlet pairs \cite{Chen1993}.

We obtain the most direct evidence of pairing via the singlet structure factor, and in particular its uniform $P(q=0,\tau)$ component which contains contributions from both long-range and short-range correlations. In \fir{fig3}(c) a broad peak about $q=0$ is seen initially, which under driving eventually increases in magnitude by a factor of more than three, and sharpens. This is consistent with $K_\rho(\tau) > 1$ and suggests that the driving has formed a quasi-condensate of singlet pairs in momentum space. To isolate the long-range contribution to $P(q=0,\tau)$ we examine the real-space singlet correlations in \fir{fig3}(d). These correlations confirm a suppression of $2 k_{\rm F}$ modulations at short times, followed by a significant enhancement of the pairing correlations. They are found to spread through the system at a rate of approximately $t/4 \sim J$, consistent with a pair-hopping $\alpha J$ as shown in \fir{fig1} \cite{Lauchli2008}. We observe that the enhancement reaches a range of approximately $20-30$ sites on the timescales considered. %We therefore expect that a finite system of a few $10$'s of sites will adequately capture the physics of the driven system. Indeed, in Appendix~\ref{sec:app_num} we compare and find a negligible difference between the infinite and finite systems. 

\begin{figure}[t!]
\vspace{1.4em}
\setlength{\unitlength}{\linewidth}
\begin{center}
% \begin{picture}(0.49, 0.40)
% \put(0,0){\includegraphics[width=0.49\linewidth]{./SingletSF0}}
% \put(0.425,0.33){\scriptsize (a)}
% \end{picture}
% \begin{picture}(0.49, 0.40)
% \put(0,0){\includegraphics[width=0.49\linewidth]{./LuttingerParam}}
% \put(0.425,0.33){\scriptsize (b)}
% \end{picture}
% \begin{picture}(0.49, 0.40)
% \put(0,0){\includegraphics[width=0.49\linewidth]{./RampSpeed}}
% \put(0.425,0.11){\scriptsize (c)}
% \end{picture}
% \begin{picture}(0.49, 0.40)
% \put(0.0,0){\includegraphics[width=0.49\linewidth]{./VaryFrequencyLine}}
% \put(0.32, 0.19){$\downarrow$}
% \put(0.24, 0.29){$\uparrow$}
% \put(0.43,0.11){\scriptsize (d)}
% \end{picture}
\includegraphics[width=0.98\linewidth]{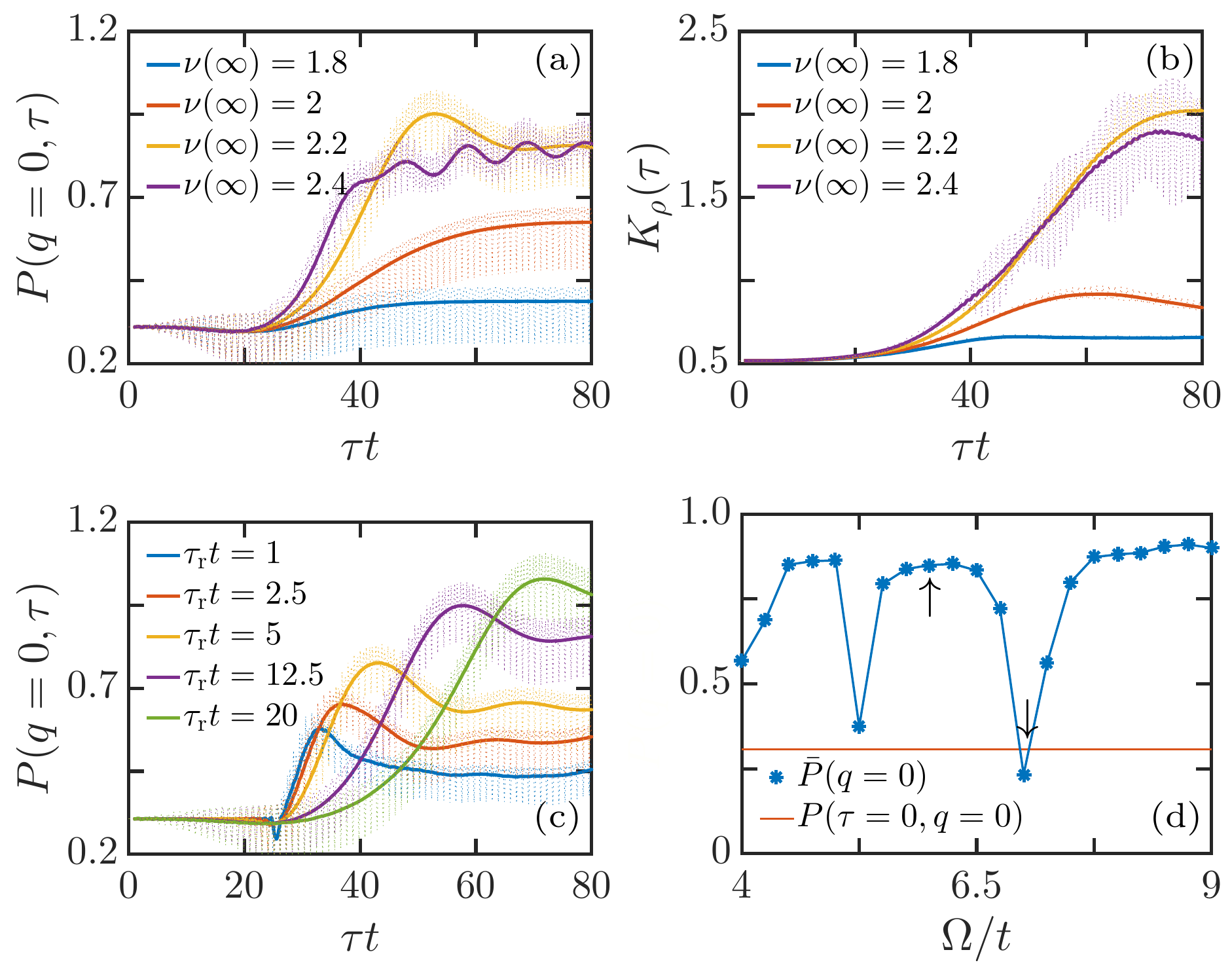}
\caption{ (a) The height of the singlet structure factor peak $P(q=0, \tau)$ at various final driving strengths. The instantaneous value is indicated by a dotted line, while the moving time average is denoted by a solid line. (b) The dynamical Luttinger parameter $K_{\rho}(\tau)$ extracted from the slope of $N(q,\tau)$ at $q \rightarrow 0$. (c) The peak of the singlet structure factor $P(q=0,\tau)$ for various ramp times. (d) The final height of the singlet structure factor averaged from $\tau = 60/t$ to $80/t$, $\bar{P}(q=0)$ as a function of the driving frequency $\Omega$. The solid blue line is drawn to guide the eye. Arrows mark frequencies $\Omega = 6t$ and $\Omega = 7t$. All parameters not explicitly given in the plots are the same as in \fir{fig3}.}\label{fig4}
\end{center}
\end{figure}

In \fir{fig4} we demonstrate the robustness of the pairing dynamics for different driving parameters. As shown in \fir{fig4}(a) the magnitude of $P(q=0,\tau)$ oscillates with frequency $\Omega$ about a mean value which is greatly enhanced with increasing driving strength $\nu$ as long as $\nu < \nu_0$. The dynamical Luttinger parameter displayed in \fir{fig4}(b) increases with driving and exceeds unity for sufficiently strong driving. This suggests the onset of attractive interactions within the system. However, for the largest final driving strengths $\nu ( \infty ) \approx \nu_0$, where the tunneling is suppressed most strongly, $P(q=0,\tau)$ reaches a maximum height at intermediate times and then reduces. This behaviour is indicative of the pair state being unstable to decay towards a resonating valence bond (RVB) type state as we will discuss in detail in the next section.

The dependence of $P(q=0,\tau)$ on the ramping time $\tau_{\rm r}$ is shown in \fir{fig4}(c). Roughly speaking the ramp is expected to cross over to adiabatic when the time-derivative of its amplitude $\dot \nu \approx \nu(\infty) / \tau_r \approx J$, i.e. when it is sufficiently smaller than the dominant energy scale of the final Hamiltonian. This is not satisfied for the fastest ramps shown in \fir{fig4}(c), yet within their ramping profile a moderate enhancement of the pairing is still induced and is sustained for longer times. For the slowest ramps considered the increases in the enhancement begin to saturate, suggesting they are close to adiabatic for this system.

In \fir{fig4}(d) we show the final pairing enhancement as a function of driving frequency. A number of resonance dips are seen where no singlet-pairing enhancement is observed. The origin of these is traced back to level crossings in the Floquet quasi-energy spectrum as the ramp is traversed, which is discussed in more detail in Appendix~\ref{sec:floquet_crossings}. In short, at these frequencies $m>0$ Floquet replicas of the upper Hubbard band, composed of states containing high-energy doubly occupied configurations, cross the $m=0$ lower Hubbard band, composed of states with predominantly singly-occupied configurations. As a result, the driving induces resonant transitions between these states and $m\Omega$ of energy is absorbed, creating double-occupancies and destroying nearest-neighbour pairs. However, away from these resonances the substantial enhancement reported for $\Omega = 6t$ is observed over a wide range of frequencies $\Omega < U$. 

Given the relatively slow speed of the spread of pairing correlations shown in \fir{fig3}(a), we expect that a finite system of a few $10$'s of sites will adequately capture the physics of the driven system, particularly for finite temperature systems where the range of correlations is suppressed by thermal fluctuations. Indeed, in Appendix~\ref{sec:app_num} we directly compare infinite and finite time-dependent DMRG calculations and confirm that there is a negligible difference between the structure factors of the infinite and finite systems.

\subsection{Finite temperature}
% Write about comparison to the infinite system here
The relevance of our observations at zero temperature to real materials hinges on whether this effect survives at finite temperatures. To answer this, we use the finite temperature extension to time-dependent DMRG \cite{Verstraete2004, Zwolak2004}. We then compute the coherent evolution with periodic driving via $\hat{H}(\tau)$ for an initial thermal state of $\hat{H}_{\rm hub}$ at inverse temperature $\beta_0$. For concreteness, we restrict our considerations here to a finite system of $L$ sites. %As discussed in \secr{sec:puredriving}, we expect that the physics of the system on the timescales we consider will be adequately described be a finite-sized system, particularly at finite-temperature where the range of correlations is suppressed by thermal fluctuations. Hence, for simplicity and numerical tractability, we here restrict our considerations to a finite system of length $L=24$.

\begin{figure}[t!]
\setlength{\unitlength}{\linewidth}
\begin{center}
% \begin{picture}(0.49, 0.40)
% \put(0,0){\includegraphics[width=0.49\linewidth]{./ThermalSingletSF.png}}
% \put(0.11,0.28){\scriptsize (a)}
% \end{picture}
% \begin{picture}(0.49, 0.40)
% \put(0,0){\includegraphics[width=0.49\linewidth]{./ThermalSingletSFCentre.pdf}}
% \put(0.42,0.11){\scriptsize (b)}
% \end{picture}
\includegraphics[width=0.98\linewidth]{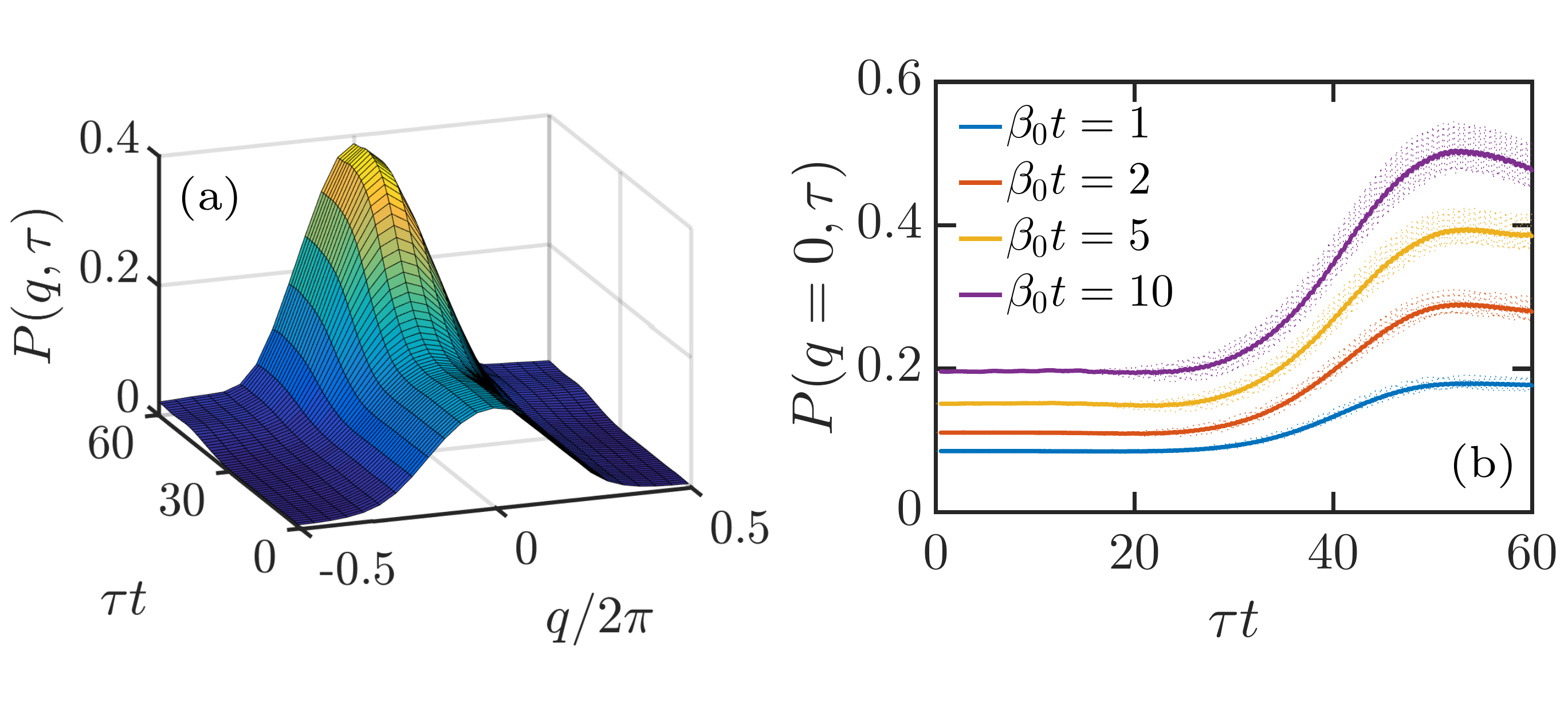}
\caption{(a) The singlet structure factor $P(q, \tau)$ with an initial temperature $\beta_0 = 5/t$. As in \fir{fig3}, we have averaged over the frequency $\Omega$ oscillations. (b) The height of the peak $P(q=0, \tau)$, for several initial temperatures $\beta_0$. The dotted line shows the instantaneous value, while the solid line shows the moving time average. These results were computed for $L=24$ sites. All other parameters not explicitly given in the plots are the same as in \fir{fig3}.} \label{fig5}
\end{center}
\end{figure}

Figure \ref{fig5}(a) shows that enhanced pairing persists at finite temperature, albeit with a peak of reduced height and broadened width compared to the zero temperature case. In \fir{fig5}(b), we show the singlet structure factor $P(q=0,\tau)$ as a function of time for various initial values of $\beta_0$. Perhaps counterintuitively, the driven state consistently exhibits enhanced singlet pairing correlations even when the initial temperature $1/\beta_{0} \sim t \gg J$ far exceeds the pair binding energy. In the next section we will introduce an effective time-independent model to qualitatively capture all the physics underlying the numerical results discussed so far. In the context of this effective model the thermal enhancement can be viewed as a many-body version of adiabatic cooling.

The reason for this is already apparent from \fir{fig2}(b). As the driving amplitude $\nu$ approaches $\nu_0$, the bandwidth of the quasi-energy spectrum in the $m=0$ Floquet sector is squashed. This is described by changes in the effective model's parameters with $\nu$, and from \fir{fig4}(c) we saw that for sufficiently slow ramping, $\dot \nu \ll J$, this change will be adiabatic. Consequently, the driving substantially reduces the energy gaps between the many-body eigenstates of this model while keeping their thermal populations unchanged. Therefore the driven state remains approximately thermal, but at a significantly lower temperature. A similar effect is used in cold-atom systems, where an adiabatic increase of the lattice depth results in a lowering of the temperature \cite{Blakie2005}. Indeed it has been shown that even for instantaneous quenches, one can obtain cooling in a wide variety of physical systems \cite{Cardy2009}.  

\section{Effective $t$--$J$ model}\label{sec:tJM}
The numerical results presented in the previous section indicate that super-exchange interaction $J$ and pair hopping $\alpha J$, as in \fir{fig1}, play a significant role in the driven dynamics when $t < \Omega < U$. Moreover, from our calculations we find (see \fir{fig:entropy}(b)) that the negligible double occupation in the initial state remains small in the driven state, once $\Omega$ does not coincide with any resonance. This suggests that an effective $t$--$J$ model may represent an adequate foundation for a Floquet analysis in this driving regime \cite{Bukov2016a}. The $t$--$J$ model arises from $\hat{H}_{\rm hub}$ by perturbatively projecting double occupancies out via the standard approach \cite{Essler2005} to yield
\begin{eqnarray} \label{eqn:tjjmodel}
\hat{H}_{\rm tJ \alpha} &&= \mathbb{P}_0 \left[\hat{H}_{\rm hop} + \hat{H}_{\rm ex} + \hat{H}_{\rm pair}\right]\mathbb{P}_0,
\end{eqnarray}
with super-exchange and pair hopping contributions
\begin{eqnarray}
\hat{H}_{\rm ex} &=& -J \sum_{\langle ij \rangle}\hat{b}^\dagger_{ij}\hat{b}_{ij}, \\
\hat{H}_{\rm pair} &=& -\alpha J \sum_{\langle ijk \rangle}^{i \neq k} \left( \hat{b}^\dagger_{ij}\hat{b}_{jk} + {\rm H.c.} \right).
\end{eqnarray}
Here, the operator $\mathbb{P}_0 = \prod_{j=1}^L (1-\hat{n}_{j,\uparrow}\hat{n}_{j,\downarrow})$ projects onto the subspace of Fock states without any double occupancies. The bracket $\langle ijk \rangle$ denotes sums over nearest neighbour sites, not double-counting bonds. The two-site super-exchange $\hat{H}_{\rm ex}$ term binds nearest-neighbour singlet pairs together with an energy $J$. Correspondingly, the three-site pair hopping term $\hat{H}_{\rm pair}$ describes the motion of these pairs without breaking their bond.

In equilibrium the parameters of the $t$--$J$ model Hamiltonian $\hat{H}_{\rm tJ \alpha}$ relate to the original Hubbard model through $J = 4t^2/U$ and $\alpha = 1/2$. The validity of the $t$--$J$ model relies on the strongly interacting limit $t\ll U$, and thus mandates $J \ll t$. We now proceed with a Floquet analysis of the driven $t$--$J$ Hamiltonian defined as
\begin{equation}
\hat{H}'(\tau) = \hat{H}_{\rm tJ \alpha} + \mathbb{P}_0\hat{H}_{\rm drive}(\tau)\mathbb{P}_0,
\end{equation}
identical to that described in Sec.~\ref{sec:floquet}. Removing $U$ as an explicit energy scale allows us to examine the behaviour with $\Omega$ as the largest energy scale of the resulting model, while continually assuming in-gap driving $\Omega<U$. This approach will provide us with a simple, intuitive picture of the physics for in-gap driving. A more detailed description, handling the interplay between finite $U$ and $\Omega$ in perturbation theory, is contained in the Appendix~\ref{sec:app_pert}. There, starting from the driven Hubbard model, we explicitly work out the contributions to the exchange interaction from the electron hopping and from driving-induced virtual charge excitations. Those calculations corroborate the picture arising from the $t$--$J$ model for sufficiently weak driving strength $
\nu$ and $\Omega < U$.

\subsection{Floquet analysis}
By calculating a matrix representation of the Floquet Hamiltonian for the driven $t$--$J$ model $\hat{H}'(\tau)$, using the Floquet basis introduced in \ref{sec:floquet}, we immediately find that the diagonal blocks are given by the matrix representation of $\mathbb{P}_0 [\mathcal{J}_0(\nu)\hat{H}_{\rm hop} + \hat{H}_{\rm ex} + \hat{H}_{\rm pair}]\mathbb{P}_0$. Thus, the hopping is still suppressed, while the driving dependent phase factors in the Floquet basis cancel out for $\hat{H}_{\rm ex}$ and $\hat{H}_{\rm pair}$, as they did for $\hat{H}_{\rm int}$ in the driven Hubbard model, leaving these terms unaffected. Assuming $\Omega \gg t$, so the Floquet sectors decouple, we find that the effective Hamiltonian describing the $m=0$ sector is a $t$--$J$ model with hopping $\tilde t = \mathcal{J}_0(\nu) t$, with $J$ and $\alpha$ unchanged. The ratio ${J}/{\tilde t}$ therefore becomes strongly driving dependent
\begin{equation}\label{simpleresult}
J/\tilde t = \frac{4 t}{U \mathcal{J}_0(\nu)},
%= \tilde{J}_{\rm kin}/\tilde{t} + \tilde{J}_{\rm light}/\tilde{t} \,,
\end{equation}
% where
% \begin{equation}\label{simpleresult2}
% \tilde{J}_{\rm light}/\tilde{t} = 2 \frac{4t}{U \mathcal{J}_{0}(\nu)} \sum_{m=1}^{\infty} \mathcal{J}_{m}(\nu)^2 .
% \end{equation}
and breaks the usual relation $J=4t^2/U \ll t$. This allows the realisation of the regime $J/\tilde t > 1$ that is normally inaccessible. Using this result, a slow increase in the driving strength $\dot \nu \ll J$ corresponds to adiabatically moving through the phase diagram of the $t$--$J$ model into this new parameter regime. For this reason, we now examine the equilibrium properties of this effective model and compare them to the driven steady states obtained in \secr{sec:DFP}.

\subsection{Phase diagram of $t$--$J$ model} \label{sec:phasediagram}

\begin{figure}[t!]
\setlength{\unitlength}{\linewidth}
\begin{center}
% \begin{picture}(0.49, 0.40)
% \put(0,0){\includegraphics[width=0.50\linewidth]{./SingletSFExampleFit}}
% \put(0.42,0.33){\scriptsize (a)}
% \end{picture}
% \begin{picture}(0.49, 0.40)
% \put(0,0){\includegraphics[width=0.49\linewidth]{./SpinSFExampleFit}}
% \put(0.425,0.33){\scriptsize (b)}
% \end{picture}
% \begin{picture}(0.49, 0.40)
% \put(0,0){\includegraphics[width=0.49\linewidth]{./EffectivetJFit}}
% \put(0.43,0.12){\scriptsize (c)}
% \end{picture}
% \begin{picture}(0.49, 0.40)
% \put(0,0){\includegraphics[width=0.485\linewidth]{./EffectiveTemperature}}
% \put(0.43,0.12){\scriptsize (d)}
% \end{picture}
\includegraphics[width=0.98\linewidth]{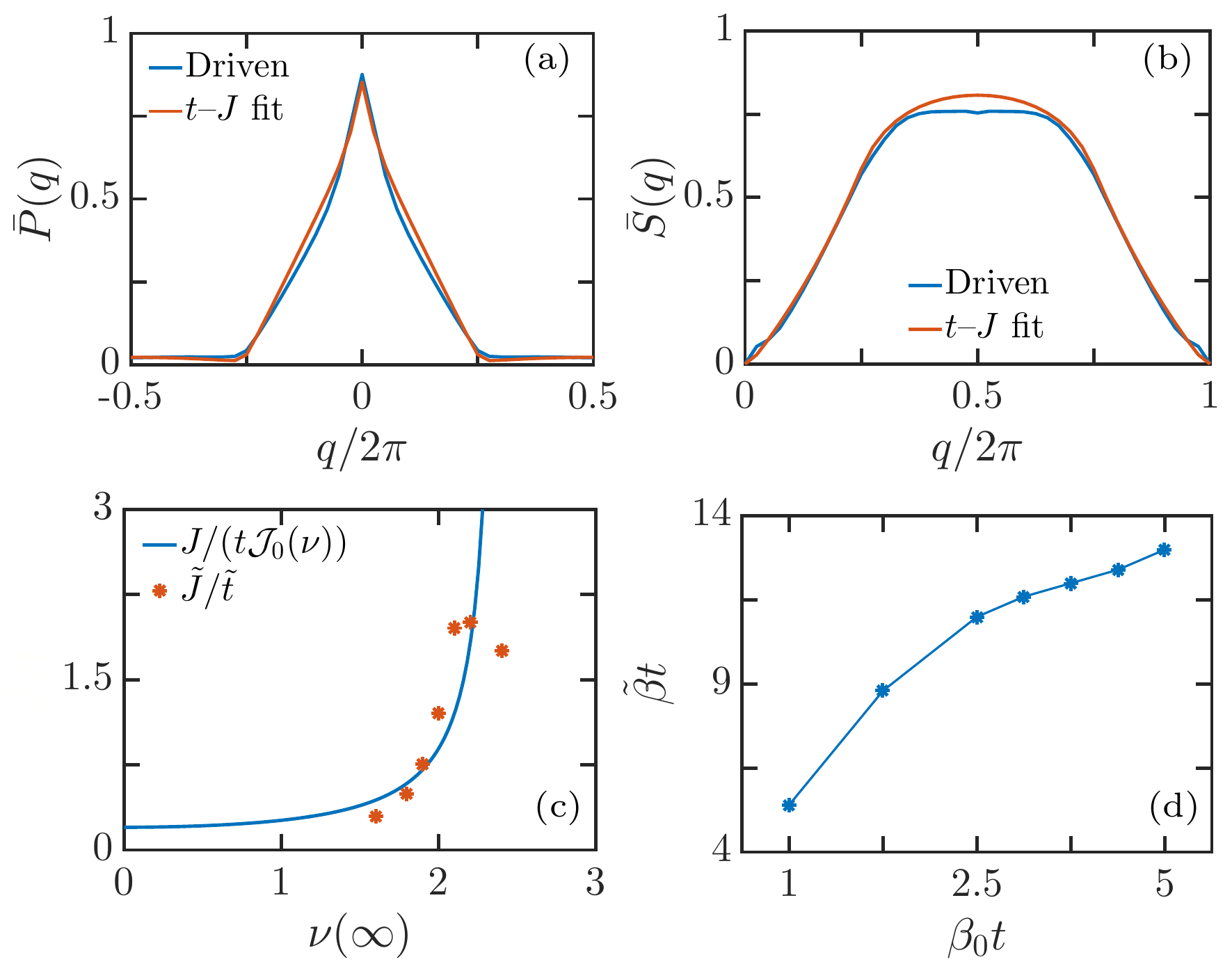}
\caption{The time average between $\tau t = 60-80$ of the driven state's structure factors for (a) singlet pairing $\bar{P}(q)$ and (b) spin $\bar{S}(q)$ from \fir{fig3}(b)--(c) are shown alongside the same quantities for a $t$--$J$ ground state with $\alpha = 1/2$ and an effective $J_{\rm eff}/t_{\rm eff}$ that gives the closest match to $\bar{P}(q)$. For $\nu(\infty) = 2.2$, the effective $J_{\rm eff}/t_{\rm eff} = 2.0$ Repeating this fit for the sequence of driving strengths $\nu$ shown in \fir{fig4}(a) gives $J_{\rm eff}/t_{\rm eff}$ plotted in (c), with the solid curve reporting the prediction of \eqr{simpleresult}. In (d) $\bar{P}(q)$ of the driven initial thermal state at inverse temperature $\beta_0$ was compared to $t$--$J$ model thermal states, with $J$ fixed to its equilibrium value and the ratio $J/t$ fixed by the results in \fir{fig6}(c), to extract an effective inverse temperature $\beta_{\rm eff}$. A solid line is drawn to guide the eye. The computations for (a)-(c) were carried for $L=40$ sites, and for (d) were carried out with $L=24$ sites.} \label{fig6}
\end{center}
\end{figure}

In one-dimension the ground state of the $t$--$J$ model, for any $\alpha$ and below half-filling $\bar{n} < 1$, is metallic with $K_\rho = 0.5$ as $J/t \rightarrow 0$. As $ J/ t$ increases, $K_\rho$ also increases monotonically. When the hopping $t \ll J$ then $\hat{H}_{\rm ex} + \hat{H}_{\rm pair}$ together resemble a Hamiltonian for hard-core bosons \footnote{We note that while the $\hat{b}_{ij}$ and $\hat{b}_{ij}^{\dagger}$ operators commute on disjoint pairs of sites they do not obey bosonic commutation relations if the pairs overlap.}. This similarity suggests that for sufficiently large $ J/ t$, super-exchange mediates the formation of nearest-neighbour singlet pairs with binding energy $- J$, while the pair hopping gives them a bandwidth proportional to $\alpha J$. This is consistent with the form of the quasi-energy spectrum near $\nu \approx \nu_0$ shown in \fir{fig2}(b).

The pairing effect is captured by simple energetic arguments. Solving the two electron problem the binding energy of nearest-neighbour singlet pairs is given by \cite{Ammon1995}
\begin{equation}
E_{\rm pair} = - J - 2 \alpha  J - \frac{4 t^2}{J+2  \alpha  J}.
\end{equation}
Thus, in the dilute limit, comparing $E_{\rm pair}$ to the energy for two free electrons $E_{\rm free} = -4 t$, suggests that pair formation will occur for $ J/t \geq 2/(1+2 \alpha)$. These nearest-neighbour singlet pairs subsequently quasi-condense to form a superconductor, characterised by dominant quasi-long-range singlet pair correlations \cite{Spalek1987}
\begin{equation}
\langle \hat{b}^{\dagger}_{j+x, j+1+x} \hat{b}_{j,j+1} \rangle \sim x^{-(1+1/K_\rho)},
\end{equation}
with $K_\rho > 1$ signifying an attractive Luttinger liquid.

For yet larger $J/ t$, the formation of phase-separated electron-rich and hole-rich regions signalled by a diverging $K_\rho$ might be expected \cite{Moreno2011}. The exact solution of a Heisenberg spin-chain gives the antiferromagnetic bond energy $E_{\rm bonds} = -2 J\ln(2)$. For $ \alpha = 1/2$ we see that $E_{\rm pair} < E_{\rm bonds}$ for all values of $ J/ t$, meaning that singlet pairs will \emph{never} freeze into larger antiferromagnetic clusters. The pair hopping term destabilises antiferromagnetic clusters, even when $ J/ t \gg 1$, since it homogenises holes throughout the system akin to hole repulsion. However, for $\bar n \geq 0.5$ superconductivity is not expected to persist up to $ J/ t \rightarrow \infty$. Instead an RVB ``singlet gas'' appears with short-ranged pair correlations. This picture of the equilibrium properties of the 1D $t$--$J$ model is borne out by comprehensive exact diagonalization \cite{Ogata1992}, DMRG \cite{Moreno2011} and quantum Monte Carlo calculations \cite{Ammon1995}.

The numerical results displayed in \fir{fig3}, \fir{fig4} are broadly consistent with these equilibrium properties of the $t$--$J$ model spanning a wide range of $ J/ t$ values. In particular we may now attribute the increased peak height $P(q=0,\tau)$ and emergence of quasi-longer-ranged pair correlations to the driving elevating $ J/ t$ sufficiently slowly to near-adiabatically transition the system from its initial metallic phase into the superconducting phase. This comparison is made more quantitatively in \fir{fig6}(a)--(b) where the singlet pairing and spin structure factors from \fir{fig3}(b)--(c) are closely matched to those of a $t$--$J$ model superconducting ground state with $J/t \approx 2$. For the purposes of comparison with the $t$--$J$ model we, for simplicity, computed the driven Hubbard states and $t$--$J$ ground states with finite size time-dependent DMRG algorithm with $L$ sites.%For strong driving strengths $\nu \rightarrow \nu_0$, where $J/\tilde t \gg 1$, the transition to an RVB singlet gas is evidenced by less pronounced increases in $P(q=0,\tau)$ and short-ranged correlations.

By repeating the fitting of the zero temperature driven state singlet correlations to a $t$--$J$ model ground state, we extract an effective super-exchange to hopping ratio $J_{\rm eff}/t_{\rm eff}$ as a function of driving strength $\nu$.  The results in \fir{fig6}(c) show decent agreement with \eqr{simpleresult}, and indicate that the superconducting phase can be reached with $\nu(\infty) \approx 2$. We also fit the singlet structure factors of the driven state obtained from a finite initial temperature to thermal states of the $t$--$J$ model to obtain an effective temperature $\beta_{\rm eff}$. The results, shown in \fir{fig6}(d) show that the driving substantially decreases the effective temperature, confirming our interpretation as adiabatically cooling the system.

\subsection{Floquet heating}

\begin{figure}
\setlength{\unitlength}{\linewidth}
\begin{center}
% \begin{picture}(0.49, 0.40)
% \put(0,0){\includegraphics[width=0.49\linewidth]{./2SiteEntropy}}
% \put(0.41,0.32){\scriptsize (a)}
% \end{picture}
% \begin{picture}(0.49, 0.40)
% \put(0,0){\includegraphics[width=0.49\linewidth]{./Doublons}}
% \put(0.41, 0.32){\scriptsize{(b)}}
% \end{picture}
% \vspace{-0.6cm}
\includegraphics[width=0.98\linewidth]{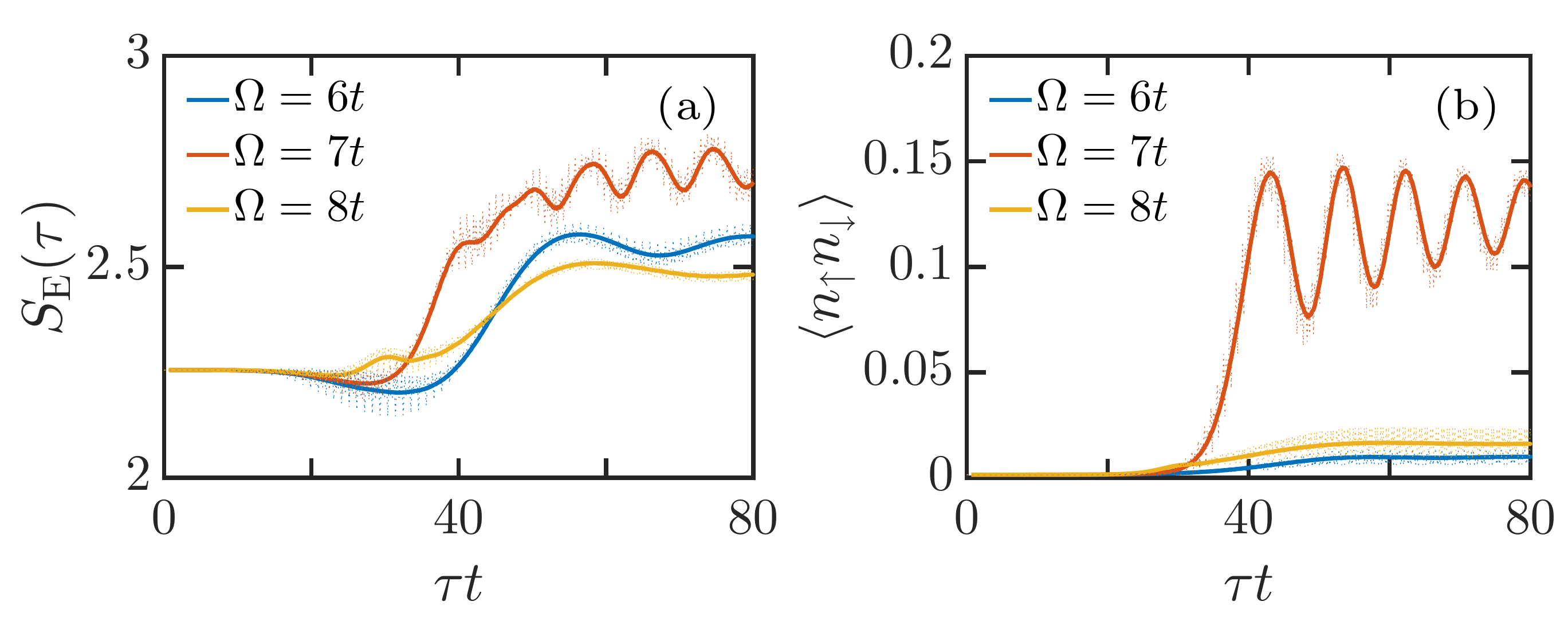}
\end{center}
\caption{(a) The entanglement entropy $S_{\rm E}$ between a two-site block and the rest of the zero temperature, infinite system as a function of time is shown. (b) The average double-occupancy per site $\langle n_{\uparrow} n_{\downarrow} \rangle$ is plotted. The dotted line shows the instantaneous value, while the solid line shows the moving time average. The system parameters not explicitly stated are the same as those in \fir{fig3}. Frequencies $\Omega = 6t$ and $\Omega = 7t$ shown here correspond to the arrows drawn in \fir{fig4}(d).} \label{fig:entropy}
\end{figure}

In general, periodic driving of a generic many-body system is expected to cause so-called Floquet heating, even far away from resonances. Finite $\Omega$ corrections will result in the system being described by an effective Hamiltonian possessing non-negligible ``unphysical'' terms that are both spatially non-local and multi-body. Eigenstates of such a Hamiltonian will be highly delocalised in the eigenbasis of more physical short-ranged few-body Hamiltonians, like the Hubbard and $t$--$J$ models. Thus the eigenstate thermalisation hypothesis (ETH) \cite{Deutsch1991,Srednicki1994,Rigol2008} suggests that in the asymptotic long-time limit, independent on its initial state, any finite frequency drive results in all physical observables of the system becoming indistinguishable from those of a featureless infinite temperature state \cite{Lazarides2014,DAlessio2014,Ponte2015}. However, the ETH does not predict the rate of Floquet heating \cite{Maricq1982}.

Despite operating in a finite frequency driving regime, on the accessible simulation times the numerical results presented here do not display significant heating effects. The time-averaged correlations emerging from the driven Hubbard model, while possessing some small quantitative discrepancies e.g. in \fir{fig6}(a)--(b), are nonetheless well captured by an effective $t$--$J$ model rather than a more pathological Hamiltonian. This suggests that the Floquet heating rate is small and that our findings are instead consistent with the notion of prethermalisation in driven systems \cite{Berges2004,Eckstein2009,Mathey2010,Bukov2016b}.  

To examine the Floquet heating rate further, we compute the entanglement entropy between a two-site block and the rest of the the infinite size, zero temperature chain, as shown in \fir{fig:entropy}(a). As expected, it increases when the driving is ramped up, showing that the driven state is more entangled compared to the initial Hubbard ground state. However, it quickly reaches a quasi-steady state, and does not show an appreciable further increase on the timescales considered, suggesting that Floquet heating is negligible here. The slow growth of entanglement in the system is precisely what allows our simulations to accurately track the dynamics for many 10's of hopping times. Even when the system is driven through the $\Omega = 7t$ resonance (shown in \fir{fig4}(d)), the entropy shows a negligible increase after the initial ramp. Although this resonant driving is rather energetic, creating and destroying doublons, it induces reversible dynamics \cite{Mendoza2017}, and thus does not correspond to heating. Nonetheless, the creation of relatively large numbers of doublons highlights the breakdown of the effective {$t$--$J$} model description in the resonant case, in contrast with the $\Omega = 6t$ drive, which avoids such resonances and maintains a small double-occupancy.

The appearance of very slow Floquet heating rates for generic many-body systems was examined in several recent studies \cite{Abanin2015a,Abanin2015b,Kuwahara2016,Mori2016}. They show that an effective static Hamiltonian $\hat{H}_*$ will describe the system up to times $\tau < \tau_*$ where $\tau_* \propto \exp(C\Omega/\varepsilon)$. Here $C$ is a numerical constant of order unity and $\varepsilon$ is an energy scale bounding the terms in the driven Hamiltonian \cite{Abanin2015b}. Our results suggest that for in-gap driving $\varepsilon \sim t$ giving a broad time window in which the system is described by an $\hat{H}_*$. This effective Hamiltonian may not be exactly a $t$--$J$ model, but it nonetheless appears to share its essential physical features, such as supporting a pairing phase. In such a broad time window, it is likely that dissipation and other couplings to the environment, which we do not model here, will govern the long time behaviour. Floquet heating is therefore unlikely to be of dominant practical importance in experiments.

\section{Conclusions}\label{sec:Conc}
We have shown that periodic in-gap driving of a strongly-correlated electronic system can slow electrons down while maintaining the normally subordinate super-exchange interaction, making $J$ the dominant energy scale. This effect manifests itself in a many-body one-dimensional setting as a distinct switching of pair correlations. We showed that the driven state is similar to what would be expected if $J/ t$ was enhanced to values which are considered unphysical in thermal equilibrium. Furthermore, these effects were found to be robust to finite ramp times of the driving and finite initial temperatures. Our future experimental and theory work is likely to focus on higher dimensional systems where dynamically enhanced pairing is expected to emerge at smaller values of $J/t$ \cite{Sorella2002}. The inclusion of competing instabilities, such as charge-density-waves, and the interplay of driving with dissipation \cite{Langemeyer2014,Hossein2014}, could then provide a fuller picture of this route to engineering light-induced superconducting states in quantum materials.

\section*{Acknowledgements}
This research is funded by the European Research Council under the European Union's Seventh Framework Programme (FP7/2007--2013)/ERC Grant Agreement no.~319286 Q-MAC. S.A. and D.J. acknowledge support from the EPSRC Tensor Network Theory grant (EP/K038311/1).

\appendix
%\section*{Appendices}

\section{Floquet Hamiltonian coupling} \label{sec:app_floquet}

In the main text we concentrated on the special case of driving with $V_a = V_b = V$ and constant phase $\Delta \phi = \pi/2$. Here we go back to the general driving $\hat{H}_{\rm drive}$ given in \eqr{driving} and define the Floquet-Fock basis as
\begin{eqnarray}
\ket{\{n_{j,\sigma}\},m} &=& \ket{\{n_{j,\sigma}\}}\exp\left[{\rm i}m\Omega\tau\right] \label{full_floquet_basis} \\
&& \times\exp\left[- {\rm i}\Delta_{a}(\tau)\sum_{j \in a}n_j - {\rm i}\Delta_{b}(\tau)\sum_{j \in b}n_j\right], \nonumber
\end{eqnarray}
where $m$ is the Fourier component and
\begin{eqnarray}
\Delta_{a}(\tau) &=& -\frac{V_{a}}{2\Omega}\cos(\Omega\tau - \Delta\phi), \nonumber \\
\Delta_{b}(\tau) &=& -\frac{V_{b}}{2\Omega}\cos(\Omega\tau + \Delta\phi).
\end{eqnarray}
The Floquet Hamiltonian is defined by the eigenvalue problem \eqr{evalue} which in the basis \eqr{full_floquet_basis} becomes
\begin{equation}
[\hat{H}(\tau) - {\rm i}\partial_\tau]\ket{\{n_{j,\sigma}\},m} = [\hat{H}_{\rm hop} + \hat{H}_{\rm int} + m\Omega\mathbbm{1}]\ket{\{n_{j,\sigma}\},m}. \nonumber
\end{equation}
We then use the extended scalar product \eqr{ext_scalar_prod} to evaluate the righthand side as $\langle\!\langle \{n'_{j,\sigma}\},m' |\hat{H}_{\rm hop} + \hat{H}_{\rm int} + m\Omega\mathbbm{1}|\{n_{j,\sigma}\},m \rangle\!\rangle$. The last two terms are diagonal in the Floquet-Fock basis so all the non-trivial physics is contained in the hopping matrix elements. These are given by
\begin{widetext}
\begin{eqnarray}
\langle\!\langle \{n'_{j,\sigma}\},m' |\hat{H}_{\rm hop}| \{n_{j,\sigma}\},m \rangle\!\rangle &=& \frac{1}{T}\int_0^T \exp\left[{\rm i}(m-m')\Omega\tau + {\rm i}s \Delta_{a}(\tau) - {\rm i}s\Delta_{b}(\tau)\right]\, {\rm d}\tau ~\langle \{n'_{j,\sigma}\} |\hat{H}_{\rm hop}| \{n_{j,\sigma}\} \rangle, \\
&=& \zeta_{m'-m}~\langle \{n'_{j,\sigma}\} |\hat{H}_{\rm hop}| \{n_{j,\sigma}\} \rangle.
\end{eqnarray}
Here we have used that $\langle \{n'_{j,\sigma}\} |\hat{H}_{\rm hop}| \{n_{j,\sigma}\} \rangle$ is non-zero only when an electron (of either spin) moves from a site in sub-lattice $a$ to a site in sub-lattice $b$, or the reverse. Thus we can denote the change in the occupation of sub-lattice $a$ as $\sum_{j \in a}(n'_j -n_j) = s$, and we know that the change in occupation of sub-lattice $b$ is $\sum_{j \in b}(n'_j -n_j) = -s$, with $s = \pm 1$. The Floquet coupling coefficient $\zeta_{m'-m}$ depends not only on the Fourier components but also on the parameters of the driving $V_a, V_b, \Delta\phi,\Omega$ and the Fock states being connected via $s$. To evaluate $\zeta_{m'-m}$ we first expand $\Delta_{a}(\tau) - \Delta_{b}(\tau)$ as
\begin{eqnarray}
\Delta_{a}(\tau) - \Delta_{b}(\tau) &=& \frac{(V_a + V_b)}{2\Omega}\sin(\Omega\tau)\sin(\Delta\phi) - \frac{(V_a - V_b)}{2\Omega}\cos(\Omega\tau)\cos(\Delta\phi).
\end{eqnarray}
Next we use a Jacobi-Anger expansion to breakup the exponentials of trigonometric functions and perform the time-averaging integration to obtain
\begin{eqnarray}
\zeta_{m'-m}
&=& \sum_{n=-\infty}^\infty (-{\rm i})^{n} \mathcal{J}_n\left(s\frac{(V_a - V_b)}{2\Omega}\cos(\Delta\phi)\right) \sum_{n'=-\infty}^\infty \mathcal{J}_{n'}\left(s\frac{(V_a + V_b)}{2\Omega}\sin(\Delta\phi)\right)\delta_{n+n',m'-m}. \label{full_driven_couplings}
\end{eqnarray}
\end{widetext}
This coupling appears in the Floquet Hamiltonian in \eqr{floquet_ham} for the most general driving parameters. For the special case used in the main text, $V_a = V_b=V$ and $\Delta\phi = \pi/2$, this reduces $\zeta_{m'-m}$ in \eqr{full_driven_couplings} to that given in \eqr{driven_couplings}. However, note that \eqr{full_driven_couplings} displays suppression of the hopping for a much wider range of driving parameters than this special case. As an example, in \fir{fig:zeta0} we take $\nu_a = V_a/\Omega = 2 \nu_b$ and plot $\zeta_0$, the hopping suppression factor in the limit $U \rightarrow \infty$, $\Omega \rightarrow \infty$, $U/\Omega < 1$. This shows that the hopping can still be suppressed to zero even in the case where $\nu_a \neq \nu_b$ and $\Delta\phi \neq \pi/2$.

\begin{figure}
\setlength{\unitlength}{\linewidth}
\begin{center}
\includegraphics[width=0.9\linewidth]{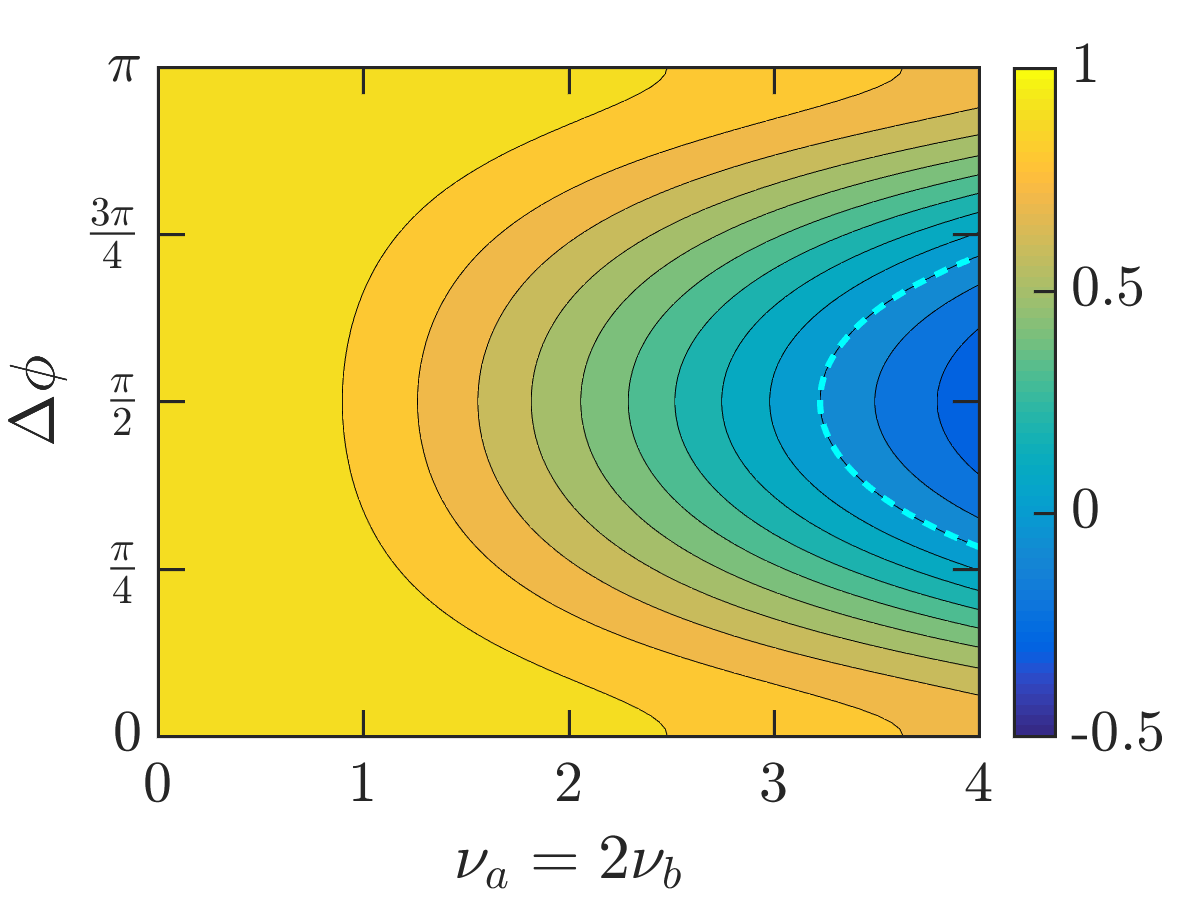}
\caption{The coefficient $\zeta_0$ with unequal $a$ and $b$ driving strengths as a function of $\nu_a$ and $\Delta\phi$. The dotted line marks the contour $\zeta_0 = 0$, where single particle hopping is suppressed to $0$.} \label{fig:zeta0}
\end{center}
\end{figure}

\section{Crossings in the Floquet Spectrum} \label{sec:floquet_crossings}

As seen in \fir{fig4}(d), there are several resonance frequencies at which pairing (and indeed all corrrelations) are destroyed rather than enhanced by the driving. This can be understood by looking at the Floquet quasi-energy spectrum. Since the spectrum is periodic high energy states in the upper Hubbard band are folded down into the 
first Brillouin zone $-\Omega/2 < U - m\Omega < \Omega/2$, as shown in \fir{fig:floquetcrossings} for a small system. For $\Omega = 6t$ folded upper Hubbard band states do not intersect the lower Hubbard band portion of the spectrum, and so a sufficiently slow ramp will adiabatically follow the Floquet state connected to the undriven ground state. However, for the $\Omega = 7t$ case, there is a level-crossing between these bands. Similar crossings are found for the other resonance frequencies in \fir{fig4}(d). y driving through these crossings, strong excitation of states in the upper Hubbard band is expected, destroying nearest-neighbour singlet correlations. Indeed for the larger system this is what is shown in \fir{fig:entropy}(b) where driving at $\Omega = 7t$ is found to result in 10 times larger increase in the average density of double occupancies than driving at $\Omega = 6t$. The smallness of the latter case supports our interpretation of the driven system with an effective $t$--$J$ model.

\begin{figure}[t!]
\setlength{\unitlength}{\linewidth}
\begin{center}
% \begin{picture}(0.9, 0.73)
% \put(0,0){\includegraphics[width=0.9\linewidth]{./QuasienergyCrossings}}
% \put(0.81, 0.42){(a)}
% \put(0.81, 0.34){(b)}
% \end{picture}
\includegraphics[width=0.98\linewidth]{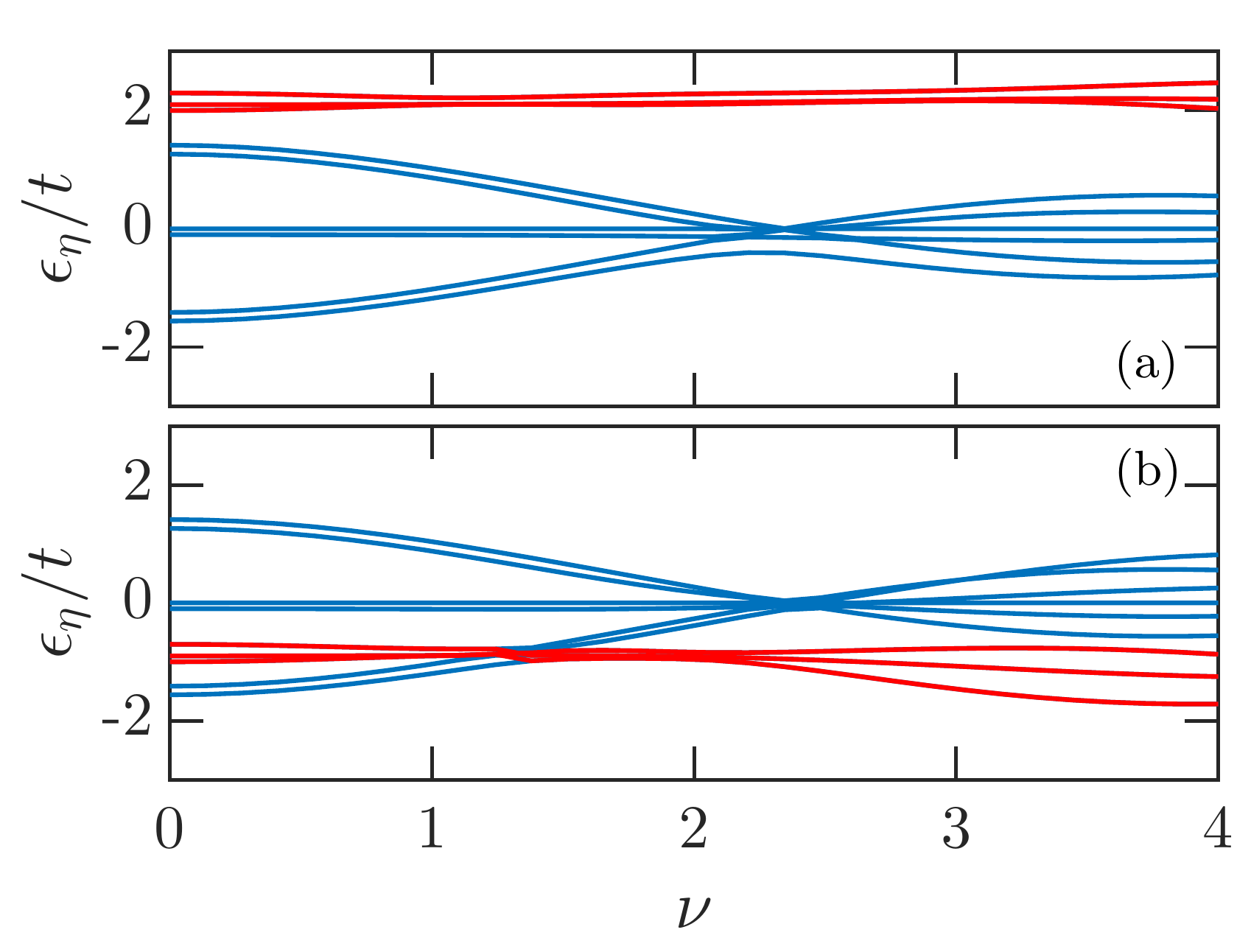}

\caption{The Floquet spectrum $\epsilon_{\eta}$ for an $L=6$ site system with $U=20t$, $\Delta \phi = \pi$. The doubly-occupied levels are drawn in red, the singly-occupied are drawn in blue. In (a), $\Omega = 6t$, and there are no crossings between singly and doubly-occuped states. In (b), $\Omega = 7t$ and, we observe a crossing between the upper and lower Hubbard bands. The two frequencies shown here correspond to the arrows drawn in \fir{fig4}(d).} \label{fig:floquetcrossings}
\end{center}
\end{figure}

\section{Details on numerical calculations} \label{sec:app_num}

For the zero temperature calculations, we computed the ground state of the Hubbard system using the infinite time-evolving block decimation (iTEBD) to evolve an initial translationally-invariant infinite matrix product state (iMPS) in imaginary time until a convergence in the singular values is reached \cite{Vidal2007}. Using a fourth-order Suzuki-Trotter decomposition, and successively refining the timestep to systematically eliminate the Trotter error and bring the final state into canonical form, we obtain an iMPS with a bond dimension of $\chi = 200$. We subsequently evolve this iMPS using iTEBD, and compute the expectation values given in \secr{sec:puredriving}. 

To accurately capture the driven state our iTEBD calculations were performed with iMPS bond dimensions up to $\chi = 400$ and a driving frequency dependent timestep $\Delta \tau = 2\pi/(50 \Omega)$. To resolve the fast oscillations in the correlation functions, they were sampled every $10$ timesteps. The correlations were robust to moderate changes in $\chi$ and $\Delta \tau$, and the cumulative truncation error during the time-evolution remained small.

In \fir{fig:infcomparison}, we show the spin and singlet structure factors for the driven system for finite size ($L=40$) calculation, with the filling $n = 1/2$, fixed using a U(1) symmetric MPS. We compare this to the results obtained for the  infinite system using iTEBD results computed without symmetries, and a chemical potential $\mu = -0.2t$ result. We find essentially no difference between the results for the finite and infinite systems, thus demonstrating that the results seen for the finite system are not a finite size effect. Consequently, for simplicity and computational tractability, we use a finite system for the purposes of finite temperature calculations and for comparisons to the $t$--$J$ model.

\begin{figure}
\setlength{\unitlength}{\linewidth}
\begin{center}
% \begin{picture}(0.49, 0.40)
% \put(0,0){\includegraphics[width=0.49\linewidth]{./InfSingletComparison}}
% \put(0.41,0.32){\scriptsize (a)}
% \end{picture}
% \begin{picture}(0.49, 0.40)
% \put(0,0){\includegraphics[width=0.49\linewidth]{./InfSpinComparison}}
% \put(0.41, 0.32){\scriptsize{(b)}}
% \end{picture}
\includegraphics[width=0.98\linewidth]{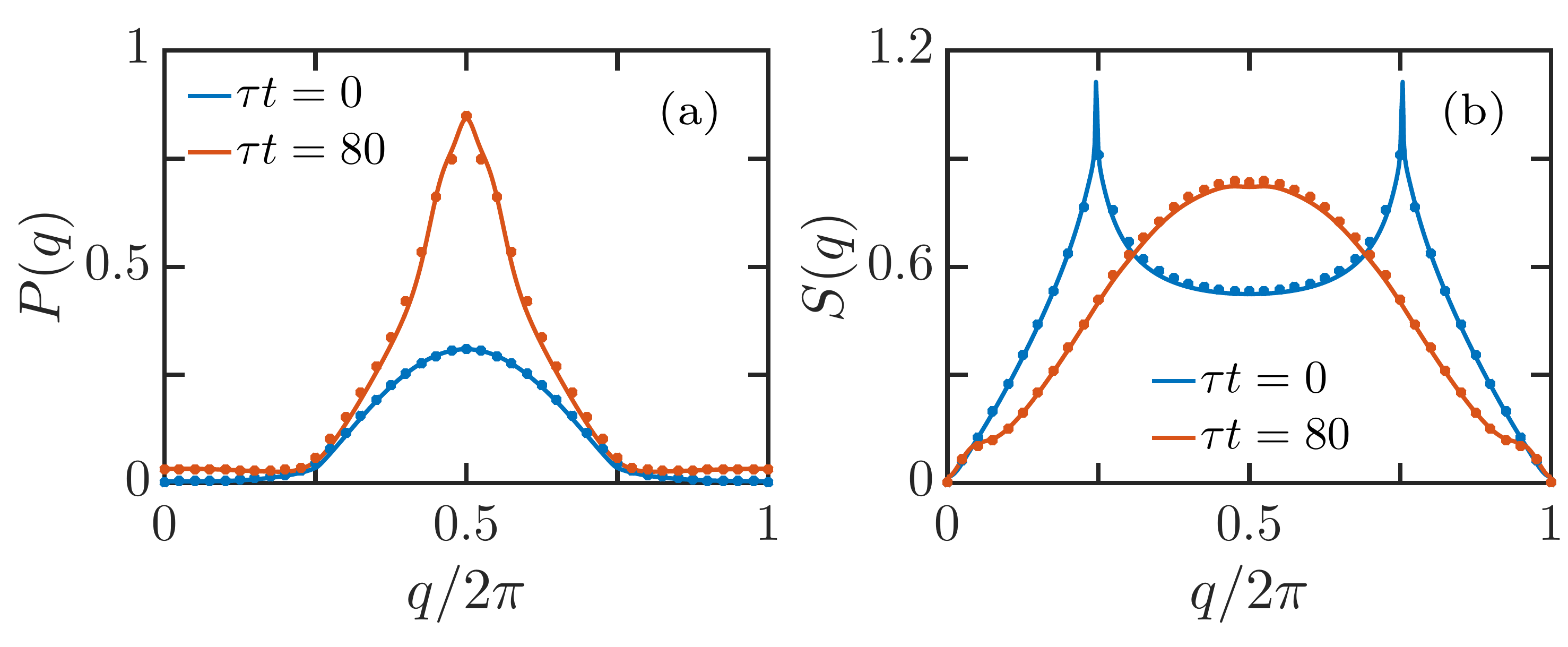}
\vspace{-0.6cm}
\end{center}
\caption{A comparison of the (a) singlet and (b) spin structure factors at two different times, computed for the same parameters as in \fir{fig3}. The points show results for the finite-size, $L=40$ results, while the line shows the infinite-size results.} \label{fig:infcomparison}
\end{figure}

To quantify the amount of Floquet heating in our system, we compute the entanglement entropy between a two-site block and the remainder of the infinite system, defined as 
\begin{equation}
S_{\rm E}(\tau) = -{\rm Tr} \left[\rho_{\rm C}(\tau) \log_{2}(\rho_{\rm C}(\tau)) \right],
\end{equation}
where $\rho_{\rm C}$ is the reduced density matrix for the two-site block. The behaviour of $S_{\rm E}(\tau)$ is shown in \fir{fig:entropy}(a) for various driving frequencies. After ramping up the driving, the entanglement entropy grows in time, quickly reaching a quasi-steady state. Note that the maximum possible value of $S_{\rm E}$ (for an infinite temperature state) is $\log_{2}(16) = 4$. The largest growth is seen for $\Omega = 7t$, which was already identified in \fir{fig4}(d) as a Floquet resonance. The additional entanglement makes the resonantly driven case more difficult to capture with an MPS ansatz due to a rapid growth of truncation error after the ramp \cite{Daley2005}. The growth in $S_{\rm E}(\tau)$ for frequencies away from resonances are broadly similar to each other and smaller. As we see no growth in $S_{\rm E}(\tau)$ after the ramp, we find no evidence of Floquet heating on the timescales considered here.  

For the finite temperature calculations, we obtained thermal states of a finite sized Hubbard model via imaginary-time TEBD. An initial matrix product operator (MPO) representation of the identity matrix (i.e. an infinite temperature state) was evolved in imaginary time with a timestep $\Delta \beta = 0.01/t$ to obtain an MPO representation of the desired thermal mixed state \cite{Zwolak2004}, tuning the chemical potentail to obtain desired filling $\bar{n} = 1/2$. We then performed real-time evolution with an MPO dimension up to $\chi = 400$ to describe the driven state.

\section{Different fillings}

\begin{figure}[t]
\setlength{\unitlength}{\linewidth}
\begin{center}
% \begin{picture}(0.49, 0.40)
% \put(0,0){\includegraphics[width=0.49\linewidth]{./SparseDrivenSpinSF.png}}
% \put(0.11,0.28){\scriptsize (a)}
% \end{picture}
% \begin{picture}(0.49, 0.40)
% \put(0,0){\includegraphics[width=0.49\linewidth]{./SparseDrivenSingletSF.png}}
% \put(0.11,0.28){\scriptsize (b)}
% \end{picture}
% \begin{picture}(0.49, 0.40)
% \put(0,0){\includegraphics[width=0.49\linewidth]{./DenseDrivenSpinSF.png}}
% \put(0.11,0.28){\scriptsize (c)}
% \end{picture}
% \begin{picture}(0.49, 0.40)
% \put(0,0){\includegraphics[width=0.49\linewidth]{./DenseDrivenSingletSF.png}}
% \put(0.11,0.28){\scriptsize (d)}
% \end{picture}
\includegraphics[width=0.98\linewidth]{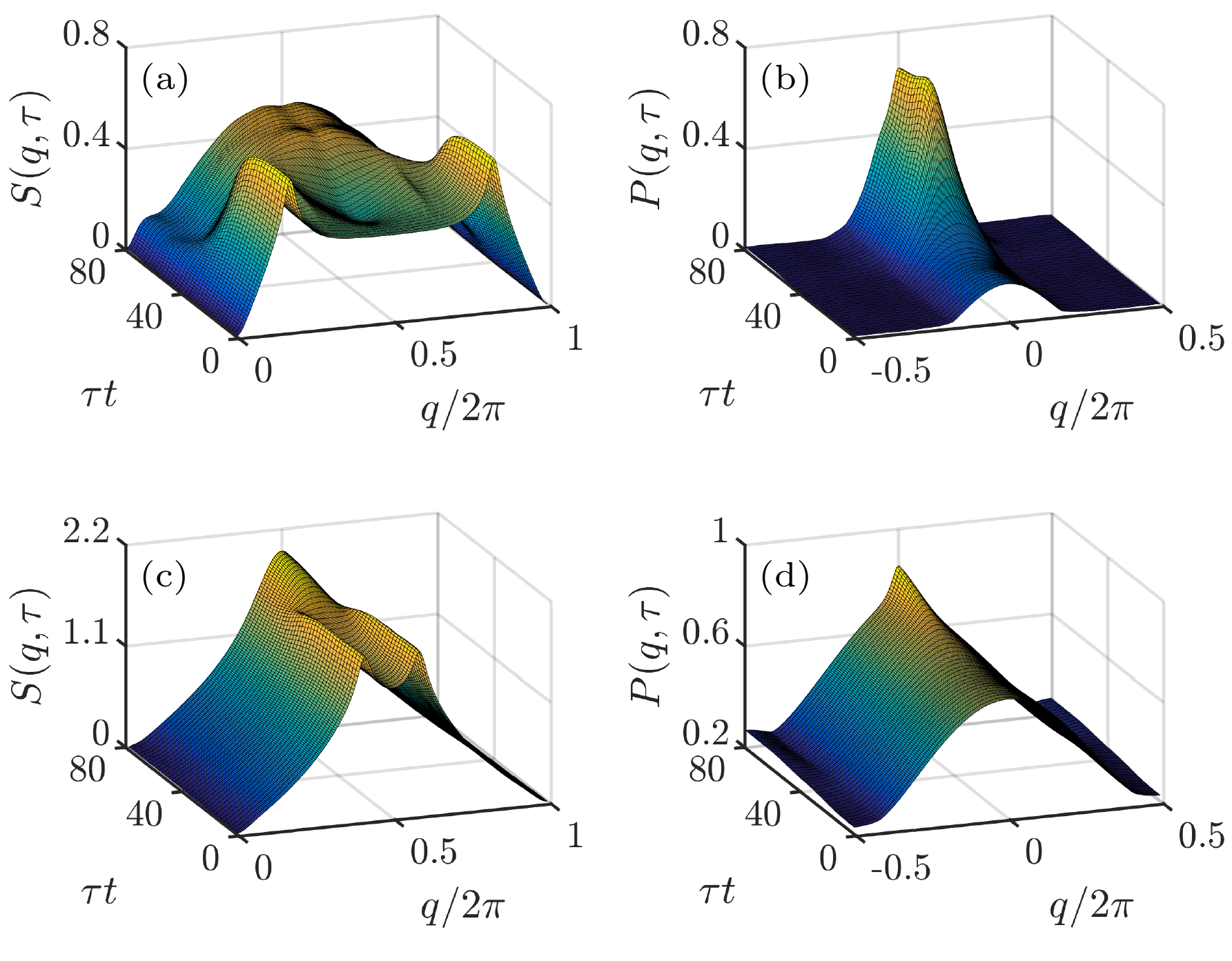}
\caption{An infinite Hubbard chain with the same system parameters as in \fir{fig3}, except for the chemical potential which is $\mu = -1.0t$ in (a) and (b), and $\mu = 1.5t$ in (c) and (d). The respective spin structure factors are shown in (a) and (c), while the singlet structure factors are shown in (b) and (d).} \label{fig11}
\end{center}
\end{figure}

For clarity and its relevance to organic salts, in the main text we focused on quarter-filling. To confirm that driving enhanced singlet pairing is still present away from half-filling, we repeated the iTEBD calculations presented in \fir{fig3}. However, we instead begin the time evolution from ground states obtained with chemical potentials $\mu = -1.0t$ and $\mu = 1.5t$, resulting in fillings of $\bar{n} \approx 0.35$ and $\bar{n} \approx 0.82$ respectively. The results are shown in \fir{fig11}. We find that irrespective of the filling, we obtain qualitatively the same features as in \fir{fig3}. Namely, the sharp peaks in the spin structure factor $S(q)$ are suppressed in favour of a broad peak about $q=\pi$, and a significant enhancement in the singlet structure factor abour $q=0$ is again observed. However, as discussed in \secr{sec:phasediagram}, the one dimensional $t$--$J$ model with $\alpha = 1/2$, features suppressed superconducting correlations at dense fillings, reflected here in the more modest pairing enhancement at $P(q=0)$. 

\section{Perturbative corrections} \label{sec:app_pert}
The virtue of constructing a Floquet Hamiltonian $\hat{H}_{\rm F}$ is that standard time-independent perturbative methods are applicable to it. In particular, given a Hamiltonian $\hat{H} = \hat{H}_0 + \lambda\hat{H}_1$, composed of a bare contribution $\hat{H}_0$ and a perturbation $\hat{H}_1$ with coupling strength $\lambda$, we wish to determine an effective Hamiltonian $\hat{H}_{\rm eff}$ describing the energy eigenvalues of $\hat{H}$ in a degenerate subspace of eigenstates of $\hat{H}_0$ with energy $E$. To second order in $\lambda$ we obtain $\hat{H}_{\rm eff}$ via a standard projection approach \cite{Essler2005} as
\begin{equation}
\hat{H}_{\rm eff} = \mathbbm{P}\hat{H}\mathbbm{P} - \mathbbm{P}\hat{H}\mathbbm{Q} \frac{1}{\mathbbm{Q}\hat{H}\mathbbm{Q} - E} \mathbbm{Q}\hat{H}\mathbbm{P}, \label{projection}
\end{equation}
where $\mathbb{P}$ is the projector onto the degenerate subspace of $\hat{H}_0$ and $\mathbbm{Q}$ is its orthogonal complement. For $\hat{H} = \hat{H}_{\rm hub}/U$, where $\hat{H}_0 = \hat{H}_{\rm int}/U$, $\lambda\hat{H}_1 = \hat{H}_{\rm hop}/U$ and $\mathbbm{P}=\mathbbm{P}_0$ this method yields the $t$--$J$ model given in \eqr{eqn:tjjmodel}. 

We now apply the same approach to $\hat{H}_{\rm F}$ in the enlarged Hilbert space $\mathcal{H}\otimes\mathcal{T}$ using a projector $\mathbbm{P}=\mathbbm{P}_0\mathbbm{M}_0$, where $\mathbbm{M}_0$ is the projector onto the $m=0$ Floquet sector, i.e. the DC Fourier component. This approach has the virtue of dealing with the strong-coupling limit and driving on the same footing \cite{Bukov2016a}. Specifically, it allows us to determine an effective time-independent Hamiltonian describing the stroboscopic evolution that contains perturbative corrections due to couplings to doubly occupied states and their Floquet replicas. We find that this again yields $t$--$J$ model with ${\alpha} = 1/2$, and $\tilde{t} = \mathcal{J}_0(\nu)t$ as before, but a super-exchange coupling given by

\begin{equation}
\tilde{J} = \tilde{J}_{\rm kin} + \tilde{J}_{\rm light}, 
\end{equation}
where
\begin{equation}
\tilde{J}_{\rm kin}/\tilde{t} = \frac{4t \mathcal{J}_{0}(\nu)}{U},
\end{equation}
and
\begin{equation}
\tilde{J}_{\rm light}/\tilde{t} = \frac{4t}{U\mathcal{J}_0(\nu)}\sum_{m=1}^\infty \left[\frac{\mathcal{J}_m(\nu)^2}{1+m\Omega/U} + \frac{\mathcal{J}_m(\nu)^2}{1-m\Omega/U}\right]. \label{floquet_J}
\end{equation}
Note that this second order result for the effective $\tilde{J}$ for sub-lattice driving is identical to that obtained by applying a driving term describing an AC electric field across the lattice \cite{Mentink2014}. In the limit $U \gg \Omega$, $\nu < 1$ the expression for $\tilde{J}$ reduces to \eqr{simpleresult} found from the $t$--$J$ model Floquet analysis.

In the limit of $\Omega\rightarrow\infty$ the energy separation between the Floquet sectors far exceeds the energy scales $t$ and $U$ within them. Consequently the perturbative contributions to $\tilde{J}$ from $m\neq 0$ sectors can be safely ignored leaving a suppressed super-exchange given by $\tilde{J}/\tilde{t} = 4\tilde{t}/U$. Indeed \eqr{floquet_J} indicates that above gap driving $\Omega > U$ will lead generically to a reduction in $\tilde{J}$. This is because while $m<0$ and $m>0$ contributions have the same numerator, for $m<0$ the denominator is negative and smaller than for $m>0$. It therefore acts against the $m \geq 0$ contributions to reduce $\tilde{J}$ pushing the system further into the metallic $\tilde{J}/\tilde{t} \ll 1$ regime.

For in-gap frequencies $t < \Omega < U$ an interesting interplay between strong interactions and driving leads to different behaviour. In this case the leading $m<0$ contributions, where $\mathcal{J}_m(\nu)^2$ is non-negligible, lie above the retained subspace so their denominators are positive. These contributions represent super-exchange processes in which the gap $U$ is bridged virtually by borrowing $m\Omega$ energy from the driving and then returned. The net effect of this is to strengthen the $m\geq 0$ contributions resulting in $\tilde{J}$ slightly increasing beyond its equilibrium value. Since the hopping continues to be suppressed in-gap driving can therefore access the regime $\tilde{J}/\tilde{t} > 1$.

% Figure showing the comparison to parameters extracted from a trimer quasi-spectrum, figure showing the simplified perturbative transitions.

%\bibliographystyle{apsrev4-1}
\bibliography{1d_floquet_superconductor}

%merlin.mbs apsrev4-1.bst 2010-07-25 4.21a (PWD, AO, DPC) hacked
%Control: key (0)
%Control: author (0) dotless jnrlst
%Control: editor formatted (1) identically to author
%Control: production of article title (0) allowed
%Control: page (1) range
%Control: year (0) verbatim
%Control: production of eprint (0) enabled
\begin{thebibliography}{77}%
\makeatletter
\providecommand \@ifxundefined [1]{%
 \@ifx{#1\undefined}
}%
\providecommand \@ifnum [1]{%
 \ifnum #1\expandafter \@firstoftwo
 \else \expandafter \@secondoftwo
 \fi
}%
\providecommand \@ifx [1]{%
 \ifx #1\expandafter \@firstoftwo
 \else \expandafter \@secondoftwo
 \fi
}%
\providecommand \natexlab [1]{#1}%
\providecommand \enquote  [1]{``#1''}%
\providecommand \bibnamefont  [1]{#1}%
\providecommand \bibfnamefont [1]{#1}%
\providecommand \citenamefont [1]{#1}%
\providecommand \href@noop [0]{\@secondoftwo}%
\providecommand \href [0]{\begingroup \@sanitize@url \@href}%
\providecommand \@href[1]{\@@startlink{#1}\@@href}%
\providecommand \@@href[1]{\endgroup#1\@@endlink}%
\providecommand \@sanitize@url [0]{\catcode `\\12\catcode `\$12\catcode
  `\&12\catcode `\#12\catcode `\^12\catcode `\_12\catcode `\%12\relax}%
\providecommand \@@startlink[1]{}%
\providecommand \@@endlink[0]{}%
\providecommand \url  [0]{\begingroup\@sanitize@url \@url }%
\providecommand \@url [1]{\endgroup\@href {#1}{\urlprefix }}%
\providecommand \urlprefix  [0]{URL }%
\providecommand \Eprint [0]{\href }%
\providecommand \doibase [0]{http://dx.doi.org/}%
\providecommand \selectlanguage [0]{\@gobble}%
\providecommand \bibinfo  [0]{\@secondoftwo}%
\providecommand \bibfield  [0]{\@secondoftwo}%
\providecommand \translation [1]{[#1]}%
\providecommand \BibitemOpen [0]{}%
\providecommand \bibitemStop [0]{}%
\providecommand \bibitemNoStop [0]{.\EOS\space}%
\providecommand \EOS [0]{\spacefactor3000\relax}%
\providecommand \BibitemShut  [1]{\csname bibitem#1\endcsname}%
\let\auto@bib@innerbib\@empty
%</preamble>
\bibitem [{\citenamefont {Rini}\ \emph {et~al.}(2007)\citenamefont {Rini},
  \citenamefont {Tobey}, \citenamefont {Dean}, \citenamefont {Itatani},
  \citenamefont {Tomioka}, \citenamefont {Tokura}, \citenamefont {Schoenlein},\
  and\ \citenamefont {Cavalleri}}]{Rini2007}%
  \BibitemOpen
  \bibfield  {author} {\bibinfo {author} {\bibfnamefont {M.}~\bibnamefont
  {Rini}}, \bibinfo {author} {\bibfnamefont {R.}~\bibnamefont {Tobey}},
  \bibinfo {author} {\bibfnamefont {N.}~\bibnamefont {Dean}}, \bibinfo {author}
  {\bibfnamefont {J.}~\bibnamefont {Itatani}}, \bibinfo {author} {\bibfnamefont
  {Y.}~\bibnamefont {Tomioka}}, \bibinfo {author} {\bibfnamefont
  {Y.}~\bibnamefont {Tokura}}, \bibinfo {author} {\bibfnamefont {R.~W.}\
  \bibnamefont {Schoenlein}}, \ and\ \bibinfo {author} {\bibfnamefont
  {A.}~\bibnamefont {Cavalleri}},\ }\bibfield  {title} {\enquote {\bibinfo
  {title} {Control of the electronic phase of a manganite by mode-selective
  vibrational excitation},}\ }\href {http://dx.doi.org/10.1038/nature06119}
  {\bibfield  {journal} {\bibinfo  {journal} {Nature}\ }\textbf {\bibinfo
  {volume} {449}},\ \bibinfo {pages} {72--74} (\bibinfo {year}
  {2007})}\BibitemShut {NoStop}%
\bibitem [{\citenamefont {Fausti}\ \emph {et~al.}(2011)\citenamefont {Fausti},
  \citenamefont {Tobey}, \citenamefont {Dean}, \citenamefont {Kaiser},
  \citenamefont {Dienst}, \citenamefont {Hoffmann}, \citenamefont {Pyon},
  \citenamefont {Takayama}, \citenamefont {Takagi},\ and\ \citenamefont
  {Cavalleri}}]{Fausti2011}%
  \BibitemOpen
  \bibfield  {author} {\bibinfo {author} {\bibfnamefont {D.}~\bibnamefont
  {Fausti}}, \bibinfo {author} {\bibfnamefont {R.~I.}\ \bibnamefont {Tobey}},
  \bibinfo {author} {\bibfnamefont {N.}~\bibnamefont {Dean}}, \bibinfo {author}
  {\bibfnamefont {S.}~\bibnamefont {Kaiser}}, \bibinfo {author} {\bibfnamefont
  {A.}~\bibnamefont {Dienst}}, \bibinfo {author} {\bibfnamefont {M.~C.}\
  \bibnamefont {Hoffmann}}, \bibinfo {author} {\bibfnamefont {S.}~\bibnamefont
  {Pyon}}, \bibinfo {author} {\bibfnamefont {T.}~\bibnamefont {Takayama}},
  \bibinfo {author} {\bibfnamefont {H.}~\bibnamefont {Takagi}}, \ and\ \bibinfo
  {author} {\bibfnamefont {A.}~\bibnamefont {Cavalleri}},\ }\bibfield  {title}
  {\enquote {\bibinfo {title} {{Light-induced superconductivity in a
  stripe-ordered cuprate}},}\ }\href {\doibase 10.1126/science.1197294}
  {\bibfield  {journal} {\bibinfo  {journal} {Science}\ }\textbf {\bibinfo
  {volume} {331}},\ \bibinfo {pages} {189--191} (\bibinfo {year}
  {2011})}\BibitemShut {NoStop}%
\bibitem [{\citenamefont {Schmitt}\ \emph {et~al.}(2008)\citenamefont
  {Schmitt}, \citenamefont {Kirchmann}, \citenamefont {Bovensiepen},
  \citenamefont {Moore}, \citenamefont {Rettig}, \citenamefont {Krenz},
  \citenamefont {Chu}, \citenamefont {Ru}, \citenamefont {Perfetti},
  \citenamefont {Lu}, \citenamefont {Wolf}, \citenamefont {Fisher},\ and\
  \citenamefont {Shen}}]{Schmitt2008}%
  \BibitemOpen
  \bibfield  {author} {\bibinfo {author} {\bibfnamefont {F.}~\bibnamefont
  {Schmitt}}, \bibinfo {author} {\bibfnamefont {P.~S.}\ \bibnamefont
  {Kirchmann}}, \bibinfo {author} {\bibfnamefont {U.}~\bibnamefont
  {Bovensiepen}}, \bibinfo {author} {\bibfnamefont {R.~G.}\ \bibnamefont
  {Moore}}, \bibinfo {author} {\bibfnamefont {L.}~\bibnamefont {Rettig}},
  \bibinfo {author} {\bibfnamefont {M.}~\bibnamefont {Krenz}}, \bibinfo
  {author} {\bibfnamefont {J.-H.}\ \bibnamefont {Chu}}, \bibinfo {author}
  {\bibfnamefont {N.}~\bibnamefont {Ru}}, \bibinfo {author} {\bibfnamefont
  {L.}~\bibnamefont {Perfetti}}, \bibinfo {author} {\bibfnamefont {D.~H.}\
  \bibnamefont {Lu}}, \bibinfo {author} {\bibfnamefont {M.}~\bibnamefont
  {Wolf}}, \bibinfo {author} {\bibfnamefont {I.~R.}\ \bibnamefont {Fisher}}, \
  and\ \bibinfo {author} {\bibfnamefont {Z.-X.}\ \bibnamefont {Shen}},\
  }\bibfield  {title} {\enquote {\bibinfo {title} {{Transient electronic
  structure and melting of a charge density wave in TbTe$_{3}$}},}\ }\href
  {\doibase 10.1126/science.1160778} {\bibfield  {journal} {\bibinfo  {journal}
  {Science}\ }\textbf {\bibinfo {volume} {321}},\ \bibinfo {pages} {1649--1652}
  (\bibinfo {year} {2008})}\BibitemShut {NoStop}%
\bibitem [{\citenamefont {Miyano}\ \emph {et~al.}(1997)\citenamefont {Miyano},
  \citenamefont {Tanaka}, \citenamefont {Tomioka},\ and\ \citenamefont
  {Tokura}}]{Miyano1997}%
  \BibitemOpen
  \bibfield  {author} {\bibinfo {author} {\bibfnamefont {K.}~\bibnamefont
  {Miyano}}, \bibinfo {author} {\bibfnamefont {T.}~\bibnamefont {Tanaka}},
  \bibinfo {author} {\bibfnamefont {Y.}~\bibnamefont {Tomioka}}, \ and\
  \bibinfo {author} {\bibfnamefont {Y.}~\bibnamefont {Tokura}},\ }\bibfield
  {title} {\enquote {\bibinfo {title} {Photoinduced insulator-to-metal
  transition in a perovskite manganite},}\ }\href {\doibase
  10.1103/PhysRevLett.78.4257} {\bibfield  {journal} {\bibinfo  {journal}
  {Phys. Rev. Lett.}\ }\textbf {\bibinfo {volume} {78}},\ \bibinfo {pages}
  {4257--4260} (\bibinfo {year} {1997})}\BibitemShut {NoStop}%
\bibitem [{\citenamefont {Cavalleri}\ \emph {et~al.}(2001)\citenamefont
  {Cavalleri}, \citenamefont {T\'oth}, \citenamefont {Siders}, \citenamefont
  {Squier}, \citenamefont {R\'aksi}, \citenamefont {Forget},\ and\
  \citenamefont {Kieffer}}]{Cavalleri2001}%
  \BibitemOpen
  \bibfield  {author} {\bibinfo {author} {\bibfnamefont {A.}~\bibnamefont
  {Cavalleri}}, \bibinfo {author} {\bibfnamefont {Cs.}\ \bibnamefont {T\'oth}},
  \bibinfo {author} {\bibfnamefont {C.~W.}\ \bibnamefont {Siders}}, \bibinfo
  {author} {\bibfnamefont {J.~A.}\ \bibnamefont {Squier}}, \bibinfo {author}
  {\bibfnamefont {F.}~\bibnamefont {R\'aksi}}, \bibinfo {author} {\bibfnamefont
  {P.}~\bibnamefont {Forget}}, \ and\ \bibinfo {author} {\bibfnamefont {J.~C.}\
  \bibnamefont {Kieffer}},\ }\bibfield  {title} {\enquote {\bibinfo {title}
  {{Femtosecond structural dynamics in ${\mathrm{VO}}_{2}$ during an ultrafast
  solid-solid phase transition}},}\ }\href {\doibase
  10.1103/PhysRevLett.87.237401} {\bibfield  {journal} {\bibinfo  {journal}
  {Phys. Rev. Lett.}\ }\textbf {\bibinfo {volume} {87}},\ \bibinfo {pages}
  {237401} (\bibinfo {year} {2001})}\BibitemShut {NoStop}%
\bibitem [{\citenamefont {Perfetti}\ \emph {et~al.}(2008)\citenamefont
  {Perfetti}, \citenamefont {Loukakos}, \citenamefont {Lisowski}, \citenamefont
  {Bovensiepen}, \citenamefont {Wolf}, \citenamefont {Berger}, \citenamefont
  {Biermann},\ and\ \citenamefont {Georges}}]{Perfetti2008}%
  \BibitemOpen
  \bibfield  {author} {\bibinfo {author} {\bibfnamefont {L.}~\bibnamefont
  {Perfetti}}, \bibinfo {author} {\bibfnamefont {P.~A.}\ \bibnamefont
  {Loukakos}}, \bibinfo {author} {\bibfnamefont {M.}~\bibnamefont {Lisowski}},
  \bibinfo {author} {\bibfnamefont {U.}~\bibnamefont {Bovensiepen}}, \bibinfo
  {author} {\bibfnamefont {M.}~\bibnamefont {Wolf}}, \bibinfo {author}
  {\bibfnamefont {H.}~\bibnamefont {Berger}}, \bibinfo {author} {\bibfnamefont
  {S.}~\bibnamefont {Biermann}}, \ and\ \bibinfo {author} {\bibfnamefont
  {A.}~\bibnamefont {Georges}},\ }\bibfield  {title} {\enquote {\bibinfo
  {title} {{Femtosecond dynamics of electronic states in the Mott insulator
  $1$T-TaS$_{2}$ by time resolved photoelectron spectroscopy}},}\ }\href
  {http://stacks.iop.org/1367-2630/10/i=5/a=053019} {\bibfield  {journal}
  {\bibinfo  {journal} {New Journal of Physics}\ }\textbf {\bibinfo {volume}
  {10}},\ \bibinfo {pages} {053019} (\bibinfo {year} {2008})}\BibitemShut
  {NoStop}%
\bibitem [{\citenamefont {Beaurepaire}\ \emph {et~al.}(1996)\citenamefont
  {Beaurepaire}, \citenamefont {Merle}, \citenamefont {Daunois},\ and\
  \citenamefont {Bigot}}]{Beaurepaire1996}%
  \BibitemOpen
  \bibfield  {author} {\bibinfo {author} {\bibfnamefont {E.}~\bibnamefont
  {Beaurepaire}}, \bibinfo {author} {\bibfnamefont {J.-C.}\ \bibnamefont
  {Merle}}, \bibinfo {author} {\bibfnamefont {A.}~\bibnamefont {Daunois}}, \
  and\ \bibinfo {author} {\bibfnamefont {J.-Y.}\ \bibnamefont {Bigot}},\
  }\bibfield  {title} {\enquote {\bibinfo {title} {Ultrafast spin dynamics in
  ferromagnetic nickel},}\ }\href {\doibase 10.1103/PhysRevLett.76.4250}
  {\bibfield  {journal} {\bibinfo  {journal} {Phys. Rev. Lett.}\ }\textbf
  {\bibinfo {volume} {76}},\ \bibinfo {pages} {4250--4253} (\bibinfo {year}
  {1996})}\BibitemShut {NoStop}%
\bibitem [{\citenamefont {Stanciu}\ \emph {et~al.}(2007)\citenamefont
  {Stanciu}, \citenamefont {Hansteen}, \citenamefont {Kimel}, \citenamefont
  {Kirilyuk}, \citenamefont {Tsukamoto}, \citenamefont {Itoh},\ and\
  \citenamefont {Rasing}}]{Stanciu2007}%
  \BibitemOpen
  \bibfield  {author} {\bibinfo {author} {\bibfnamefont {C.~D.}\ \bibnamefont
  {Stanciu}}, \bibinfo {author} {\bibfnamefont {F.}~\bibnamefont {Hansteen}},
  \bibinfo {author} {\bibfnamefont {A.~V.}\ \bibnamefont {Kimel}}, \bibinfo
  {author} {\bibfnamefont {A.}~\bibnamefont {Kirilyuk}}, \bibinfo {author}
  {\bibfnamefont {A.}~\bibnamefont {Tsukamoto}}, \bibinfo {author}
  {\bibfnamefont {A.}~\bibnamefont {Itoh}}, \ and\ \bibinfo {author}
  {\bibfnamefont {Th.}\ \bibnamefont {Rasing}},\ }\bibfield  {title} {\enquote
  {\bibinfo {title} {{All-optical magnetic recording with circularly polarized
  light}},}\ }\href {\doibase 10.1103/PhysRevLett.99.047601} {\bibfield
  {journal} {\bibinfo  {journal} {Phys. Rev. Lett.}\ }\textbf {\bibinfo
  {volume} {99}},\ \bibinfo {pages} {047601} (\bibinfo {year}
  {2007})}\BibitemShut {NoStop}%
\bibitem [{\citenamefont {Ehrke}\ \emph {et~al.}(2011)\citenamefont {Ehrke},
  \citenamefont {Tobey}, \citenamefont {Wall}, \citenamefont {Cavill},
  \citenamefont {F\"orst}, \citenamefont {Khanna}, \citenamefont {Garl},
  \citenamefont {Stojanovic}, \citenamefont {Prabhakaran}, \citenamefont
  {Boothroyd}, \citenamefont {Gensch}, \citenamefont {Mirone}, \citenamefont
  {Reutler}, \citenamefont {Revcolevschi}, \citenamefont {Dhesi},\ and\
  \citenamefont {Cavalleri}}]{Ehrke2011}%
  \BibitemOpen
  \bibfield  {author} {\bibinfo {author} {\bibfnamefont {H.}~\bibnamefont
  {Ehrke}}, \bibinfo {author} {\bibfnamefont {R.~I.}\ \bibnamefont {Tobey}},
  \bibinfo {author} {\bibfnamefont {S.}~\bibnamefont {Wall}}, \bibinfo {author}
  {\bibfnamefont {S.~A.}\ \bibnamefont {Cavill}}, \bibinfo {author}
  {\bibfnamefont {M.}~\bibnamefont {F\"orst}}, \bibinfo {author} {\bibfnamefont
  {V.}~\bibnamefont {Khanna}}, \bibinfo {author} {\bibfnamefont {Th.}\
  \bibnamefont {Garl}}, \bibinfo {author} {\bibfnamefont {N.}~\bibnamefont
  {Stojanovic}}, \bibinfo {author} {\bibfnamefont {D.}~\bibnamefont
  {Prabhakaran}}, \bibinfo {author} {\bibfnamefont {A.~T.}\ \bibnamefont
  {Boothroyd}}, \bibinfo {author} {\bibfnamefont {M.}~\bibnamefont {Gensch}},
  \bibinfo {author} {\bibfnamefont {A.}~\bibnamefont {Mirone}}, \bibinfo
  {author} {\bibfnamefont {P.}~\bibnamefont {Reutler}}, \bibinfo {author}
  {\bibfnamefont {A.}~\bibnamefont {Revcolevschi}}, \bibinfo {author}
  {\bibfnamefont {S.~S.}\ \bibnamefont {Dhesi}}, \ and\ \bibinfo {author}
  {\bibfnamefont {A.}~\bibnamefont {Cavalleri}},\ }\bibfield  {title} {\enquote
  {\bibinfo {title} {{Photoinduced melting of antiferromagnetic order in
  ${\mathrm{La}}_{0.5}{\mathrm{Sr}}_{1.5}{\mathrm{MnO}}_{4}$ measured using
  ultrafast resonant soft x-ray diffraction}},}\ }\href {\doibase
  10.1103/PhysRevLett.106.217401} {\bibfield  {journal} {\bibinfo  {journal}
  {Phys. Rev. Lett.}\ }\textbf {\bibinfo {volume} {106}},\ \bibinfo {pages}
  {217401} (\bibinfo {year} {2011})}\BibitemShut {NoStop}%
\bibitem [{\citenamefont {F\"orst}\ \emph {et~al.}(2011)\citenamefont
  {F\"orst}, \citenamefont {Tobey}, \citenamefont {Wall}, \citenamefont
  {Bromberger}, \citenamefont {Khanna}, \citenamefont {Cavalieri},
  \citenamefont {Chuang}, \citenamefont {Lee}, \citenamefont {Moore},
  \citenamefont {Schlotter}, \citenamefont {Turner}, \citenamefont {Krupin},
  \citenamefont {Trigo}, \citenamefont {Zheng}, \citenamefont {Mitchell},
  \citenamefont {Dhesi}, \citenamefont {Hill},\ and\ \citenamefont
  {Cavalleri}}]{Forst2011a}%
  \BibitemOpen
  \bibfield  {author} {\bibinfo {author} {\bibfnamefont {M.}~\bibnamefont
  {F\"orst}}, \bibinfo {author} {\bibfnamefont {R.~I.}\ \bibnamefont {Tobey}},
  \bibinfo {author} {\bibfnamefont {S.}~\bibnamefont {Wall}}, \bibinfo {author}
  {\bibfnamefont {H.}~\bibnamefont {Bromberger}}, \bibinfo {author}
  {\bibfnamefont {V.}~\bibnamefont {Khanna}}, \bibinfo {author} {\bibfnamefont
  {A.~L.}\ \bibnamefont {Cavalieri}}, \bibinfo {author} {\bibfnamefont {Y.-D.}\
  \bibnamefont {Chuang}}, \bibinfo {author} {\bibfnamefont {W.~S.}\
  \bibnamefont {Lee}}, \bibinfo {author} {\bibfnamefont {R.}~\bibnamefont
  {Moore}}, \bibinfo {author} {\bibfnamefont {W.~F.}\ \bibnamefont
  {Schlotter}}, \bibinfo {author} {\bibfnamefont {J.~J.}\ \bibnamefont
  {Turner}}, \bibinfo {author} {\bibfnamefont {O.}~\bibnamefont {Krupin}},
  \bibinfo {author} {\bibfnamefont {M.}~\bibnamefont {Trigo}}, \bibinfo
  {author} {\bibfnamefont {H.}~\bibnamefont {Zheng}}, \bibinfo {author}
  {\bibfnamefont {J.~F.}\ \bibnamefont {Mitchell}}, \bibinfo {author}
  {\bibfnamefont {S.~S.}\ \bibnamefont {Dhesi}}, \bibinfo {author}
  {\bibfnamefont {J.~P.}\ \bibnamefont {Hill}}, \ and\ \bibinfo {author}
  {\bibfnamefont {A.}~\bibnamefont {Cavalleri}},\ }\bibfield  {title} {\enquote
  {\bibinfo {title} {{Driving magnetic order in a manganite by ultrafast
  lattice excitation}},}\ }\href {\doibase 10.1103/PhysRevB.84.241104}
  {\bibfield  {journal} {\bibinfo  {journal} {Phys. Rev. B}\ }\textbf {\bibinfo
  {volume} {84}},\ \bibinfo {pages} {241104} (\bibinfo {year}
  {2011})}\BibitemShut {NoStop}%
\bibitem [{\citenamefont {Mitrano}\ \emph {et~al.}(2016)\citenamefont
  {Mitrano}, \citenamefont {Cantaluppi}, \citenamefont {Nicoletti},
  \citenamefont {Kaiser}, \citenamefont {Perucchi}, \citenamefont {Lupi},
  \citenamefont {Di~Pietro}, \citenamefont {Pontiroli}, \citenamefont
  {Ricc{\`o}}, \citenamefont {Clark}, \citenamefont {Jaksch},\ and\
  \citenamefont {Cavalleri}}]{Mitrano2016}%
  \BibitemOpen
  \bibfield  {author} {\bibinfo {author} {\bibfnamefont {M.}~\bibnamefont
  {Mitrano}}, \bibinfo {author} {\bibfnamefont {A.}~\bibnamefont {Cantaluppi}},
  \bibinfo {author} {\bibfnamefont {D.}~\bibnamefont {Nicoletti}}, \bibinfo
  {author} {\bibfnamefont {S.}~\bibnamefont {Kaiser}}, \bibinfo {author}
  {\bibfnamefont {A.}~\bibnamefont {Perucchi}}, \bibinfo {author}
  {\bibfnamefont {S.}~\bibnamefont {Lupi}}, \bibinfo {author} {\bibfnamefont
  {P.}~\bibnamefont {Di~Pietro}}, \bibinfo {author} {\bibfnamefont
  {D.}~\bibnamefont {Pontiroli}}, \bibinfo {author} {\bibfnamefont
  {M.}~\bibnamefont {Ricc{\`o}}}, \bibinfo {author} {\bibfnamefont {S.~R.}\
  \bibnamefont {Clark}}, \bibinfo {author} {\bibfnamefont {D.}~\bibnamefont
  {Jaksch}}, \ and\ \bibinfo {author} {\bibfnamefont {A.}~\bibnamefont
  {Cavalleri}},\ }\bibfield  {title} {\enquote {\bibinfo {title} {{Possible
  light-induced superconductivity in K$_{3}$C$_{60}$ at high temperature}},}\
  }\href {http://dx.doi.org/10.1038/nature16522} {\bibfield  {journal}
  {\bibinfo  {journal} {Nature}\ }\textbf {\bibinfo {volume} {530}},\ \bibinfo
  {pages} {461--464} (\bibinfo {year} {2016})}\BibitemShut {NoStop}%
\bibitem [{\citenamefont {Denny}\ \emph {et~al.}(2015)\citenamefont {Denny},
  \citenamefont {Clark}, \citenamefont {Laplace}, \citenamefont {Cavalleri},\
  and\ \citenamefont {Jaksch}}]{Denny2015}%
  \BibitemOpen
  \bibfield  {author} {\bibinfo {author} {\bibfnamefont {S.~J.}\ \bibnamefont
  {Denny}}, \bibinfo {author} {\bibfnamefont {S.~R.}\ \bibnamefont {Clark}},
  \bibinfo {author} {\bibfnamefont {Y.}~\bibnamefont {Laplace}}, \bibinfo
  {author} {\bibfnamefont {A.}~\bibnamefont {Cavalleri}}, \ and\ \bibinfo
  {author} {\bibfnamefont {D.}~\bibnamefont {Jaksch}},\ }\bibfield  {title}
  {\enquote {\bibinfo {title} {{Proposed parametric cooling of bilayer cuprate
  superconductors by terahertz excitation}},}\ }\href {\doibase
  10.1103/PhysRevLett.114.137001} {\bibfield  {journal} {\bibinfo  {journal}
  {Phys. Rev. Lett.}\ }\textbf {\bibinfo {volume} {114}},\ \bibinfo {pages}
  {137001} (\bibinfo {year} {2015})}\BibitemShut {NoStop}%
\bibitem [{\citenamefont {Hu}\ \emph {et~al.}(2014)\citenamefont {Hu},
  \citenamefont {Kaiser}, \citenamefont {Nicoletti}, \citenamefont {Hunt},
  \citenamefont {Gierz}, \citenamefont {Hoffmann}, \citenamefont {Le~Tacon},
  \citenamefont {Loew}, \citenamefont {Keimer},\ and\ \citenamefont
  {Cavalleri}}]{Hu2014}%
  \BibitemOpen
  \bibfield  {author} {\bibinfo {author} {\bibfnamefont {W.}~\bibnamefont
  {Hu}}, \bibinfo {author} {\bibfnamefont {S.}~\bibnamefont {Kaiser}}, \bibinfo
  {author} {\bibfnamefont {D.}~\bibnamefont {Nicoletti}}, \bibinfo {author}
  {\bibfnamefont {C.~R.}\ \bibnamefont {Hunt}}, \bibinfo {author}
  {\bibfnamefont {I.}~\bibnamefont {Gierz}}, \bibinfo {author} {\bibfnamefont
  {M.~C.}\ \bibnamefont {Hoffmann}}, \bibinfo {author} {\bibfnamefont
  {M.}~\bibnamefont {Le~Tacon}}, \bibinfo {author} {\bibfnamefont
  {T.}~\bibnamefont {Loew}}, \bibinfo {author} {\bibfnamefont {B.}~\bibnamefont
  {Keimer}}, \ and\ \bibinfo {author} {\bibfnamefont {A.}~\bibnamefont
  {Cavalleri}},\ }\bibfield  {title} {\enquote {\bibinfo {title} {{Optically
  enhanced coherent transport in YBa$_{2}$Cu$_{3}$O$_{6.5}$ by ultrafast
  redistribution of interlayer coupling}},}\ }\href
  {http://dx.doi.org/10.1038/nmat3963} {\bibfield  {journal} {\bibinfo
  {journal} {Nat Mater}\ }\textbf {\bibinfo {volume} {13}},\ \bibinfo {pages}
  {705--711} (\bibinfo {year} {2014})}\BibitemShut {NoStop}%
\bibitem [{\citenamefont {Kaiser}\ \emph
  {et~al.}(2014{\natexlab{a}})\citenamefont {Kaiser}, \citenamefont {Hunt},
  \citenamefont {Nicoletti}, \citenamefont {Hu}, \citenamefont {Gierz},
  \citenamefont {Liu}, \citenamefont {Le~Tacon}, \citenamefont {Loew},
  \citenamefont {Haug}, \citenamefont {Keimer},\ and\ \citenamefont
  {Cavalleri}}]{Kaiser2014a}%
  \BibitemOpen
  \bibfield  {author} {\bibinfo {author} {\bibfnamefont {S.}~\bibnamefont
  {Kaiser}}, \bibinfo {author} {\bibfnamefont {C.~R.}\ \bibnamefont {Hunt}},
  \bibinfo {author} {\bibfnamefont {D.}~\bibnamefont {Nicoletti}}, \bibinfo
  {author} {\bibfnamefont {W.}~\bibnamefont {Hu}}, \bibinfo {author}
  {\bibfnamefont {I.}~\bibnamefont {Gierz}}, \bibinfo {author} {\bibfnamefont
  {H.~Y.}\ \bibnamefont {Liu}}, \bibinfo {author} {\bibfnamefont
  {M.}~\bibnamefont {Le~Tacon}}, \bibinfo {author} {\bibfnamefont
  {T.}~\bibnamefont {Loew}}, \bibinfo {author} {\bibfnamefont {D.}~\bibnamefont
  {Haug}}, \bibinfo {author} {\bibfnamefont {B.}~\bibnamefont {Keimer}}, \ and\
  \bibinfo {author} {\bibfnamefont {A.}~\bibnamefont {Cavalleri}},\ }\bibfield
  {title} {\enquote {\bibinfo {title} {{Optically induced coherent transport
  far above ${T}_{c}$ in underdoped
  ${\mathrm{YBa}}_{2}{\mathrm{Cu}}_{3}{\mathrm{O}}_{6+\ensuremath{\delta}}$}},}\
  }\href {\doibase 10.1103/PhysRevB.89.184516} {\bibfield  {journal} {\bibinfo
  {journal} {Phys. Rev. B}\ }\textbf {\bibinfo {volume} {89}},\ \bibinfo
  {pages} {184516} (\bibinfo {year} {2014}{\natexlab{a}})}\BibitemShut
  {NoStop}%
\bibitem [{\citenamefont {Mankowsky}\ \emph {et~al.}(2014)\citenamefont
  {Mankowsky}, \citenamefont {Subedi}, \citenamefont {F\"{o}rst}, \citenamefont
  {Mariager}, \citenamefont {Chollet}, \citenamefont {Lemke}, \citenamefont
  {Robinson}, \citenamefont {Glownia}, \citenamefont {Minitti}, \citenamefont
  {Frano}, \citenamefont {Fechner}, \citenamefont {Spaldin}, \citenamefont
  {Loew}, \citenamefont {Keimer}, \citenamefont {Georges},\ and\ \citenamefont
  {Cavalleri}}]{Mankowsky2014}%
  \BibitemOpen
  \bibfield  {author} {\bibinfo {author} {\bibfnamefont {R.}~\bibnamefont
  {Mankowsky}}, \bibinfo {author} {\bibfnamefont {A.}~\bibnamefont {Subedi}},
  \bibinfo {author} {\bibfnamefont {M.}~\bibnamefont {F\"{o}rst}}, \bibinfo
  {author} {\bibfnamefont {S.~O.}\ \bibnamefont {Mariager}}, \bibinfo {author}
  {\bibfnamefont {M.}~\bibnamefont {Chollet}}, \bibinfo {author} {\bibfnamefont
  {H.~T.}\ \bibnamefont {Lemke}}, \bibinfo {author} {\bibfnamefont {J.~S.}\
  \bibnamefont {Robinson}}, \bibinfo {author} {\bibfnamefont {J.~M.}\
  \bibnamefont {Glownia}}, \bibinfo {author} {\bibfnamefont {M.~P.}\
  \bibnamefont {Minitti}}, \bibinfo {author} {\bibfnamefont {A.}~\bibnamefont
  {Frano}}, \bibinfo {author} {\bibfnamefont {M.}~\bibnamefont {Fechner}},
  \bibinfo {author} {\bibfnamefont {N.~A.}\ \bibnamefont {Spaldin}}, \bibinfo
  {author} {\bibfnamefont {T.}~\bibnamefont {Loew}}, \bibinfo {author}
  {\bibfnamefont {B.}~\bibnamefont {Keimer}}, \bibinfo {author} {\bibfnamefont
  {A.}~\bibnamefont {Georges}}, \ and\ \bibinfo {author} {\bibfnamefont
  {A.}~\bibnamefont {Cavalleri}},\ }\bibfield  {title} {\enquote {\bibinfo
  {title} {{Nonlinear lattice dynamics as a basis for enhanced
  superconductivity in YBa$_{2}$Cu$_{3}$O$_{6.5}$}},}\ }\href {\doibase
  10.1038/nature13875} {\bibfield  {journal} {\bibinfo  {journal} {Nature}\
  }\textbf {\bibinfo {volume} {516}},\ \bibinfo {pages} {71--73} (\bibinfo
  {year} {2014})}\BibitemShut {NoStop}%
\bibitem [{\citenamefont {Sentef}\ \emph {et~al.}(2015)\citenamefont {Sentef},
  \citenamefont {Kemper}, \citenamefont {Georges},\ and\ \citenamefont
  {Kollath}}]{Sentef2015}%
  \BibitemOpen
  \bibfield  {author} {\bibinfo {author} {\bibfnamefont {M.A.}\ \bibnamefont
  {Sentef}}, \bibinfo {author} {\bibfnamefont {A.F.}\ \bibnamefont {Kemper}},
  \bibinfo {author} {\bibfnamefont {A.}~\bibnamefont {Georges}}, \ and\
  \bibinfo {author} {\bibfnamefont {C.}~\bibnamefont {Kollath}},\ }\bibfield
  {title} {\enquote {\bibinfo {title} {{Theory of light-enhanced
  phonon-mediated superconductivity}},}\ }\href
  {http://arxiv.org/abs/1505.07575} {\  (\bibinfo {year} {2015})},\ \Eprint
  {http://arxiv.org/abs/1505.07575} {arXiv:1505.07575} \BibitemShut {NoStop}%
\bibitem [{\citenamefont {Knap}\ \emph {et~al.}(2015)\citenamefont {Knap},
  \citenamefont {Badadi}, \citenamefont {Refael}, \citenamefont {Martin},\ and\
  \citenamefont {Demler}}]{Knap2015}%
  \BibitemOpen
  \bibfield  {author} {\bibinfo {author} {\bibfnamefont {M.}~\bibnamefont
  {Knap}}, \bibinfo {author} {\bibfnamefont {M.}~\bibnamefont {Badadi}},
  \bibinfo {author} {\bibfnamefont {G.}~\bibnamefont {Refael}}, \bibinfo
  {author} {\bibfnamefont {I.}~\bibnamefont {Martin}}, \ and\ \bibinfo {author}
  {\bibfnamefont {E.}~\bibnamefont {Demler}},\ }\bibfield  {title} {\enquote
  {\bibinfo {title} {{Dynamical Cooper pairing in non-equilibrium
  electron-phonon systems}},}\ }\href {http://arxiv.org/abs/1511.07874} {\
  (\bibinfo {year} {2015})},\ \Eprint {http://arxiv.org/abs/1511.07874}
  {arXiv:1511.07874} \BibitemShut {NoStop}%
\bibitem [{\citenamefont {Subedi}\ \emph {et~al.}(2014)\citenamefont {Subedi},
  \citenamefont {Cavalleri},\ and\ \citenamefont {Georges}}]{Subedi2014}%
  \BibitemOpen
  \bibfield  {author} {\bibinfo {author} {\bibfnamefont {A.}~\bibnamefont
  {Subedi}}, \bibinfo {author} {\bibfnamefont {A.}~\bibnamefont {Cavalleri}}, \
  and\ \bibinfo {author} {\bibfnamefont {A.}~\bibnamefont {Georges}},\
  }\bibfield  {title} {\enquote {\bibinfo {title} {{Theory of nonlinear
  phononics for coherent light control of solids}},}\ }\href {\doibase
  10.1103/PhysRevB.89.220301} {\bibfield  {journal} {\bibinfo  {journal} {Phys.
  Rev. B}\ }\textbf {\bibinfo {volume} {89}},\ \bibinfo {pages} {220301}
  (\bibinfo {year} {2014})}\BibitemShut {NoStop}%
\bibitem [{\citenamefont {Shirley}(1965)}]{Shirley1965}%
  \BibitemOpen
  \bibfield  {author} {\bibinfo {author} {\bibfnamefont {J.~H.}\ \bibnamefont
  {Shirley}},\ }\bibfield  {title} {\enquote {\bibinfo {title} {Solution of the
  schr\"odinger equation with a hamiltonian periodic in time},}\ }\href
  {\doibase 10.1103/PhysRev.138.B979} {\bibfield  {journal} {\bibinfo
  {journal} {Phys. Rev.}\ }\textbf {\bibinfo {volume} {138}},\ \bibinfo {pages}
  {B979--B987} (\bibinfo {year} {1965})}\BibitemShut {NoStop}%
\bibitem [{\citenamefont {Dunlap}\ and\ \citenamefont
  {Kenkre}(1986)}]{Dunlap1986}%
  \BibitemOpen
  \bibfield  {author} {\bibinfo {author} {\bibfnamefont {D.~H.}\ \bibnamefont
  {Dunlap}}\ and\ \bibinfo {author} {\bibfnamefont {V.~M.}\ \bibnamefont
  {Kenkre}},\ }\bibfield  {title} {\enquote {\bibinfo {title} {Dynamic
  localization of a charged particle moving under the influence of an electric
  field},}\ }\href {\doibase 10.1103/PhysRevB.34.3625} {\bibfield  {journal}
  {\bibinfo  {journal} {Phys. Rev. B}\ }\textbf {\bibinfo {volume} {34}},\
  \bibinfo {pages} {3625--3633} (\bibinfo {year} {1986})}\BibitemShut {NoStop}%
\bibitem [{\citenamefont {Buchleitner}\ \emph {et~al.}(2002)\citenamefont
  {Buchleitner}, \citenamefont {Delande},\ and\ \citenamefont
  {Zakrzewski}}]{Buchleitner2002}%
  \BibitemOpen
  \bibfield  {author} {\bibinfo {author} {\bibfnamefont {A.}~\bibnamefont
  {Buchleitner}}, \bibinfo {author} {\bibfnamefont {D.}~\bibnamefont
  {Delande}}, \ and\ \bibinfo {author} {\bibfnamefont {J.}~\bibnamefont
  {Zakrzewski}},\ }\bibfield  {title} {\enquote {\bibinfo {title}
  {Non-dispersive wave packets in periodically driven quantum systems},}\
  }\href {\doibase http://dx.doi.org/10.1016/S0370-1573(02)00270-3} {\bibfield
  {journal} {\bibinfo  {journal} {Physics Reports}\ }\textbf {\bibinfo {volume}
  {368}},\ \bibinfo {pages} {409 -- 547} (\bibinfo {year} {2002})}\BibitemShut
  {NoStop}%
\bibitem [{\citenamefont {Bukov}\ \emph {et~al.}(2015)\citenamefont {Bukov},
  \citenamefont {D'Alessio},\ and\ \citenamefont {Polkovnikov}}]{Bukov2015}%
  \BibitemOpen
  \bibfield  {author} {\bibinfo {author} {\bibfnamefont {M.}~\bibnamefont
  {Bukov}}, \bibinfo {author} {\bibfnamefont {L.}~\bibnamefont {D'Alessio}}, \
  and\ \bibinfo {author} {\bibfnamefont {A.}~\bibnamefont {Polkovnikov}},\
  }\bibfield  {title} {\enquote {\bibinfo {title} {{Universal high-frequency
  behavior of periodically driven systems: from dynamical stabilization to
  Floquet engineering}},}\ }\href
  {http://www.tandfonline.com/doi/full/10.1080/00018732.2015.1055918#.VdsWfrxViko}
  {\bibfield  {journal} {\bibinfo  {journal} {Adv. Phys.}\ }\textbf {\bibinfo
  {volume} {64}},\ \bibinfo {pages} {139--226} (\bibinfo {year}
  {2015})}\BibitemShut {NoStop}%
\bibitem [{\citenamefont {Mentink}\ \emph {et~al.}(2015)\citenamefont
  {Mentink}, \citenamefont {Balzer},\ and\ \citenamefont
  {Eckstein}}]{Mentink2014}%
  \BibitemOpen
  \bibfield  {author} {\bibinfo {author} {\bibfnamefont {J.}~\bibnamefont
  {Mentink}}, \bibinfo {author} {\bibfnamefont {K.}~\bibnamefont {Balzer}}, \
  and\ \bibinfo {author} {\bibfnamefont {M.}~\bibnamefont {Eckstein}},\
  }\bibfield  {title} {\enquote {\bibinfo {title} {{Ultrafast and reversible
  control of the exchange interaction in Mott insulators}},}\ }\href {\doibase
  10.1038/ncomms7708} {\bibfield  {journal} {\bibinfo  {journal} {Nat.
  Commun.}\ }\textbf {\bibinfo {volume} {6}},\ \bibinfo {pages} {1--8}
  (\bibinfo {year} {2015})}\BibitemShut {NoStop}%
\bibitem [{\citenamefont {Creffield}\ and\ \citenamefont
  {Platero}(2002)}]{Creffield2002}%
  \BibitemOpen
  \bibfield  {author} {\bibinfo {author} {\bibfnamefont {C.~E.}\ \bibnamefont
  {Creffield}}\ and\ \bibinfo {author} {\bibfnamefont {G.}~\bibnamefont
  {Platero}},\ }\bibfield  {title} {\enquote {\bibinfo {title} {{AC-driven
  localization in a two-electron quantum dot molecule}},}\ }\href {\doibase
  10.1103/PhysRevB.65.113304} {\bibfield  {journal} {\bibinfo  {journal} {Phys.
  Rev. B}\ }\textbf {\bibinfo {volume} {65}},\ \bibinfo {pages} {113304}
  (\bibinfo {year} {2002})}\BibitemShut {NoStop}%
\bibitem [{\citenamefont {Eckardt}\ \emph {et~al.}(2005)\citenamefont
  {Eckardt}, \citenamefont {Weiss},\ and\ \citenamefont
  {Holthaus}}]{Eckardt2005}%
  \BibitemOpen
  \bibfield  {author} {\bibinfo {author} {\bibfnamefont {A.}~\bibnamefont
  {Eckardt}}, \bibinfo {author} {\bibfnamefont {C.}~\bibnamefont {Weiss}}, \
  and\ \bibinfo {author} {\bibfnamefont {M.}~\bibnamefont {Holthaus}},\
  }\bibfield  {title} {\enquote {\bibinfo {title} {{Superfluid-insulator
  transition in a periodically driven optical lattice}},}\ }\href {\doibase
  10.1103/PhysRevLett.95.260404} {\bibfield  {journal} {\bibinfo  {journal}
  {Phys. Rev. Lett.}\ }\textbf {\bibinfo {volume} {95}},\ \bibinfo {pages}
  {260404} (\bibinfo {year} {2005})}\BibitemShut {NoStop}%
\bibitem [{\citenamefont {Kaiser}\ \emph
  {et~al.}(2014{\natexlab{b}})\citenamefont {Kaiser}, \citenamefont {Clark},
  \citenamefont {Nicoletti}, \citenamefont {Cotugno}, \citenamefont {Tobey},
  \citenamefont {Dean}, \citenamefont {Lupi}, \citenamefont {Okamoto},
  \citenamefont {Hasegawa}, \citenamefont {Jaksch},\ and\ \citenamefont
  {Cavalleri}}]{Kaiser2014b}%
  \BibitemOpen
  \bibfield  {author} {\bibinfo {author} {\bibfnamefont {S.}~\bibnamefont
  {Kaiser}}, \bibinfo {author} {\bibfnamefont {S.~R.}\ \bibnamefont {Clark}},
  \bibinfo {author} {\bibfnamefont {D.}~\bibnamefont {Nicoletti}}, \bibinfo
  {author} {\bibfnamefont {G.}~\bibnamefont {Cotugno}}, \bibinfo {author}
  {\bibfnamefont {R.~I.}\ \bibnamefont {Tobey}}, \bibinfo {author}
  {\bibfnamefont {N.}~\bibnamefont {Dean}}, \bibinfo {author} {\bibfnamefont
  {S.}~\bibnamefont {Lupi}}, \bibinfo {author} {\bibfnamefont {H.}~\bibnamefont
  {Okamoto}}, \bibinfo {author} {\bibfnamefont {T.}~\bibnamefont {Hasegawa}},
  \bibinfo {author} {\bibfnamefont {D.}~\bibnamefont {Jaksch}}, \ and\ \bibinfo
  {author} {\bibfnamefont {A.}~\bibnamefont {Cavalleri}},\ }\bibfield  {title}
  {\enquote {\bibinfo {title} {{Optical properties of a vibrationally modulated
  solid state Mott insulator.}}}\ }\href {\doibase 10.1038/srep03823}
  {\bibfield  {journal} {\bibinfo  {journal} {Sci. Rep.}\ }\textbf {\bibinfo
  {volume} {4}},\ \bibinfo {pages} {3823} (\bibinfo {year}
  {2014}{\natexlab{b}})}\BibitemShut {NoStop}%
\bibitem [{\citenamefont {Singla}\ \emph {et~al.}(2015)\citenamefont {Singla},
  \citenamefont {Cotugno}, \citenamefont {Kaiser}, \citenamefont {F\"orst},
  \citenamefont {Mitrano}, \citenamefont {Liu}, \citenamefont {Cartella},
  \citenamefont {Manzoni}, \citenamefont {Okamoto}, \citenamefont {Hasegawa},
  \citenamefont {Clark}, \citenamefont {Jaksch},\ and\ \citenamefont
  {Cavalleri}}]{Singla2015}%
  \BibitemOpen
  \bibfield  {author} {\bibinfo {author} {\bibfnamefont {R.}~\bibnamefont
  {Singla}}, \bibinfo {author} {\bibfnamefont {G.}~\bibnamefont {Cotugno}},
  \bibinfo {author} {\bibfnamefont {S.}~\bibnamefont {Kaiser}}, \bibinfo
  {author} {\bibfnamefont {M.}~\bibnamefont {F\"orst}}, \bibinfo {author}
  {\bibfnamefont {M.}~\bibnamefont {Mitrano}}, \bibinfo {author} {\bibfnamefont
  {H.~Y.}\ \bibnamefont {Liu}}, \bibinfo {author} {\bibfnamefont
  {A.}~\bibnamefont {Cartella}}, \bibinfo {author} {\bibfnamefont
  {C.}~\bibnamefont {Manzoni}}, \bibinfo {author} {\bibfnamefont
  {H.}~\bibnamefont {Okamoto}}, \bibinfo {author} {\bibfnamefont
  {T.}~\bibnamefont {Hasegawa}}, \bibinfo {author} {\bibfnamefont {S.~R.}\
  \bibnamefont {Clark}}, \bibinfo {author} {\bibfnamefont {D.}~\bibnamefont
  {Jaksch}}, \ and\ \bibinfo {author} {\bibfnamefont {A.}~\bibnamefont
  {Cavalleri}},\ }\bibfield  {title} {\enquote {\bibinfo {title}
  {{THz-frequency modulation of the Hubbard $U$ in an organic Mott
  insulator}},}\ }\href {\doibase 10.1103/PhysRevLett.115.187401} {\bibfield
  {journal} {\bibinfo  {journal} {Phys. Rev. Lett.}\ }\textbf {\bibinfo
  {volume} {115}},\ \bibinfo {pages} {187401} (\bibinfo {year}
  {2015})}\BibitemShut {NoStop}%
\bibitem [{Note1()}]{Note1}%
  \BibitemOpen
  \bibinfo {note} {For the main effects described here, a two-site periodicity
  is not essential, and may be realised with higher spatial
  periodicities.}\BibitemShut {Stop}%
\bibitem [{\citenamefont {Mori}\ \emph {et~al.}(1998)\citenamefont {Mori},
  \citenamefont {Tanaka},\ and\ \citenamefont {Mori}}]{Mori1998}%
  \BibitemOpen
  \bibfield  {author} {\bibinfo {author} {\bibfnamefont {H.}~\bibnamefont
  {Mori}}, \bibinfo {author} {\bibfnamefont {S.}~\bibnamefont {Tanaka}}, \ and\
  \bibinfo {author} {\bibfnamefont {T.}~\bibnamefont {Mori}},\ }\bibfield
  {title} {\enquote {\bibinfo {title} {{Systematic study of the electronic
  state in $\theta$-type BEDT-TTF organic conductors by changing the electronic
  correlation}},}\ }\href {\doibase 10.1103/PhysRevB.57.12023} {\bibfield
  {journal} {\bibinfo  {journal} {Phys. Rev. B}\ }\textbf {\bibinfo {volume}
  {57}},\ \bibinfo {pages} {12023--12029} (\bibinfo {year} {1998})}\BibitemShut
  {NoStop}%
\bibitem [{\citenamefont {Eckardt}\ \emph {et~al.}(2009)\citenamefont
  {Eckardt}, \citenamefont {Holthaus}, \citenamefont {Lignier}, \citenamefont
  {Zenesini}, \citenamefont {Ciampini}, \citenamefont {Morsch},\ and\
  \citenamefont {Arimondo}}]{Eckardt2009}%
  \BibitemOpen
  \bibfield  {author} {\bibinfo {author} {\bibfnamefont {A.}~\bibnamefont
  {Eckardt}}, \bibinfo {author} {\bibfnamefont {M.}~\bibnamefont {Holthaus}},
  \bibinfo {author} {\bibfnamefont {H.}~\bibnamefont {Lignier}}, \bibinfo
  {author} {\bibfnamefont {A.}~\bibnamefont {Zenesini}}, \bibinfo {author}
  {\bibfnamefont {D.}~\bibnamefont {Ciampini}}, \bibinfo {author}
  {\bibfnamefont {O.}~\bibnamefont {Morsch}}, \ and\ \bibinfo {author}
  {\bibfnamefont {E.}~\bibnamefont {Arimondo}},\ }\bibfield  {title} {\enquote
  {\bibinfo {title} {{Exploring dynamic localization with a Bose-Einstein
  condensate}},}\ }\href {\doibase 10.1103/PhysRevA.79.013611} {\bibfield
  {journal} {\bibinfo  {journal} {Phys. Rev. A}\ }\textbf {\bibinfo {volume}
  {79}},\ \bibinfo {pages} {013611} (\bibinfo {year} {2009})}\BibitemShut
  {NoStop}%
\bibitem [{\citenamefont {Jotzu}\ \emph {et~al.}(2014)\citenamefont {Jotzu},
  \citenamefont {Messer}, \citenamefont {Desbuquois}, \citenamefont {Lebrat},
  \citenamefont {Uehlinger}, \citenamefont {Greif},\ and\ \citenamefont
  {Esslinger}}]{Jotzu2014}%
  \BibitemOpen
  \bibfield  {author} {\bibinfo {author} {\bibfnamefont {G.}~\bibnamefont
  {Jotzu}}, \bibinfo {author} {\bibfnamefont {M.}~\bibnamefont {Messer}},
  \bibinfo {author} {\bibfnamefont {R.}~\bibnamefont {Desbuquois}}, \bibinfo
  {author} {\bibfnamefont {M.}~\bibnamefont {Lebrat}}, \bibinfo {author}
  {\bibfnamefont {T.}~\bibnamefont {Uehlinger}}, \bibinfo {author}
  {\bibfnamefont {D.}~\bibnamefont {Greif}}, \ and\ \bibinfo {author}
  {\bibfnamefont {T.}~\bibnamefont {Esslinger}},\ }\bibfield  {title} {\enquote
  {\bibinfo {title} {{Experimental realization of the topological Haldane model
  with ultracold fermions}},}\ }\href {http://dx.doi.org/10.1038/nature13915}
  {\bibfield  {journal} {\bibinfo  {journal} {Nature}\ }\textbf {\bibinfo
  {volume} {515}},\ \bibinfo {pages} {237--240} (\bibinfo {year}
  {2014})}\BibitemShut {NoStop}%
\bibitem [{\citenamefont {Bloch}\ \emph {et~al.}(2008)\citenamefont {Bloch},
  \citenamefont {Dalibard},\ and\ \citenamefont {Zwerger}}]{Bloch2008}%
  \BibitemOpen
  \bibfield  {author} {\bibinfo {author} {\bibfnamefont {I.}~\bibnamefont
  {Bloch}}, \bibinfo {author} {\bibfnamefont {J.}~\bibnamefont {Dalibard}}, \
  and\ \bibinfo {author} {\bibfnamefont {W.}~\bibnamefont {Zwerger}},\
  }\bibfield  {title} {\enquote {\bibinfo {title} {{Many-body physics with
  ultracold gases}},}\ }\href {\doibase 10.1103/RevModPhys.80.885} {\bibfield
  {journal} {\bibinfo  {journal} {Rev. Mod. Phys.}\ }\textbf {\bibinfo {volume}
  {80}},\ \bibinfo {pages} {885--964} (\bibinfo {year} {2008})}\BibitemShut
  {NoStop}%
\bibitem [{\citenamefont {Lewenstein}\ \emph {et~al.}(2012)\citenamefont
  {Lewenstein}, \citenamefont {Sanpera},\ and\ \citenamefont
  {Ahufinger}}]{Lewenstein2012}%
  \BibitemOpen
  \bibfield  {author} {\bibinfo {author} {\bibfnamefont {M.}~\bibnamefont
  {Lewenstein}}, \bibinfo {author} {\bibfnamefont {A.}~\bibnamefont {Sanpera}},
  \ and\ \bibinfo {author} {\bibfnamefont {V.}~\bibnamefont {Ahufinger}},\
  }\href@noop {} {\emph {\bibinfo {title} {{Ultracold atoms in optical
  lattices: Simulating quantum many-body systems}}}}\ (\bibinfo  {publisher}
  {Oxford University Press},\ \bibinfo {year} {2012})\BibitemShut {NoStop}%
\bibitem [{\citenamefont {Struck}\ \emph {et~al.}(2012)\citenamefont {Struck},
  \citenamefont {\"{O}lschl\"{a}ger}, \citenamefont {Weinberg}, \citenamefont
  {Hauke}, \citenamefont {Simonet}, \citenamefont {Eckardt}, \citenamefont
  {Lewenstein}, \citenamefont {Sengstock},\ and\ \citenamefont
  {Windpassinger}}]{Struck2012}%
  \BibitemOpen
  \bibfield  {author} {\bibinfo {author} {\bibfnamefont {J.}~\bibnamefont
  {Struck}}, \bibinfo {author} {\bibfnamefont {C.}~\bibnamefont
  {\"{O}lschl\"{a}ger}}, \bibinfo {author} {\bibfnamefont {M.}~\bibnamefont
  {Weinberg}}, \bibinfo {author} {\bibfnamefont {P.}~\bibnamefont {Hauke}},
  \bibinfo {author} {\bibfnamefont {J.}~\bibnamefont {Simonet}}, \bibinfo
  {author} {\bibfnamefont {A.}~\bibnamefont {Eckardt}}, \bibinfo {author}
  {\bibfnamefont {M.}~\bibnamefont {Lewenstein}}, \bibinfo {author}
  {\bibfnamefont {K.}~\bibnamefont {Sengstock}}, \ and\ \bibinfo {author}
  {\bibfnamefont {P.}~\bibnamefont {Windpassinger}},\ }\bibfield  {title}
  {\enquote {\bibinfo {title} {{Tunable gauge potential for neutral and
  spinless particles in driven optical lattices}},}\ }\href {\doibase
  10.1103/PhysRevLett.108.225304} {\bibfield  {journal} {\bibinfo  {journal}
  {Phys. Rev. Lett.}\ }\textbf {\bibinfo {volume} {108}},\ \bibinfo {pages}
  {225304} (\bibinfo {year} {2012})}\BibitemShut {NoStop}%
\bibitem [{\citenamefont {{Desbuquois}}\ \emph {et~al.}(2017)\citenamefont
  {{Desbuquois}}, \citenamefont {{Messer}}, \citenamefont {{G{\"o}rg}},
  \citenamefont {{Sandholzer}}, \citenamefont {{Jotzu}},\ and\ \citenamefont
  {{Esslinger}}}]{Desbuquois2017}%
  \BibitemOpen
  \bibfield  {author} {\bibinfo {author} {\bibfnamefont {R.}~\bibnamefont
  {{Desbuquois}}}, \bibinfo {author} {\bibfnamefont {M.}~\bibnamefont
  {{Messer}}}, \bibinfo {author} {\bibfnamefont {F.}~\bibnamefont
  {{G{\"o}rg}}}, \bibinfo {author} {\bibfnamefont {K.}~\bibnamefont
  {{Sandholzer}}}, \bibinfo {author} {\bibfnamefont {G.}~\bibnamefont
  {{Jotzu}}}, \ and\ \bibinfo {author} {\bibfnamefont {T.}~\bibnamefont
  {{Esslinger}}},\ }\bibfield  {title} {\enquote {\bibinfo {title}
  {{Controlling the Floquet state population and observing micromotion in a
  periodically driven two-body quantum system}},}\ }\href@noop {} {\  (\bibinfo
  {year} {2017})},\ \Eprint {http://arxiv.org/abs/1703.07767}
  {arXiv:1703.07767} \BibitemShut {NoStop}%
\bibitem [{\citenamefont {Bilitewski}\ and\ \citenamefont
  {Cooper}(2016)}]{Bilitewski2016}%
  \BibitemOpen
  \bibfield  {author} {\bibinfo {author} {\bibfnamefont {T.}~\bibnamefont
  {Bilitewski}}\ and\ \bibinfo {author} {\bibfnamefont {N.~R.}\ \bibnamefont
  {Cooper}},\ }\bibfield  {title} {\enquote {\bibinfo {title} {Synthetic
  dimensions in the strong-coupling limit: Supersolids and pair superfluids},}\
  }\href {\doibase 10.1103/PhysRevA.94.023630} {\bibfield  {journal} {\bibinfo
  {journal} {Phys. Rev. A}\ }\textbf {\bibinfo {volume} {94}},\ \bibinfo
  {pages} {023630} (\bibinfo {year} {2016})}\BibitemShut {NoStop}%
\bibitem [{\citenamefont {Lignier}\ \emph {et~al.}(2007)\citenamefont
  {Lignier}, \citenamefont {Sias}, \citenamefont {Ciampini}, \citenamefont
  {Singh}, \citenamefont {Zenesini}, \citenamefont {Morsch},\ and\
  \citenamefont {Arimondo}}]{Lignier2007}%
  \BibitemOpen
  \bibfield  {author} {\bibinfo {author} {\bibfnamefont {H.}~\bibnamefont
  {Lignier}}, \bibinfo {author} {\bibfnamefont {C.}~\bibnamefont {Sias}},
  \bibinfo {author} {\bibfnamefont {D.}~\bibnamefont {Ciampini}}, \bibinfo
  {author} {\bibfnamefont {Y.}~\bibnamefont {Singh}}, \bibinfo {author}
  {\bibfnamefont {A.}~\bibnamefont {Zenesini}}, \bibinfo {author}
  {\bibfnamefont {O.}~\bibnamefont {Morsch}}, \ and\ \bibinfo {author}
  {\bibfnamefont {E.}~\bibnamefont {Arimondo}},\ }\bibfield  {title} {\enquote
  {\bibinfo {title} {Dynamical control of matter-wave tunneling in periodic
  potentials},}\ }\href {\doibase 10.1103/PhysRevLett.99.220403} {\bibfield
  {journal} {\bibinfo  {journal} {Phys. Rev. Lett.}\ }\textbf {\bibinfo
  {volume} {99}},\ \bibinfo {pages} {220403} (\bibinfo {year}
  {2007})}\BibitemShut {NoStop}%
\bibitem [{\citenamefont {Schollw\"{o}ck}(2011)}]{Schollwock2011}%
  \BibitemOpen
  \bibfield  {author} {\bibinfo {author} {\bibfnamefont {U.}~\bibnamefont
  {Schollw\"{o}ck}},\ }\bibfield  {title} {\enquote {\bibinfo {title} {{The
  density-matrix renormalization group in the age of matrix product states}},}\
  }\href {\doibase 10.1016/j.aop.2010.09.012} {\bibfield  {journal} {\bibinfo
  {journal} {Ann. Phys. (N. Y).}\ }\textbf {\bibinfo {volume} {326}},\ \bibinfo
  {pages} {96--192} (\bibinfo {year} {2011})}\BibitemShut {NoStop}%
\bibitem [{\citenamefont {Verstraete}\ \emph {et~al.}(2008)\citenamefont
  {Verstraete}, \citenamefont {Murg},\ and\ \citenamefont
  {Cirac}}]{Verstraete2008}%
  \BibitemOpen
  \bibfield  {author} {\bibinfo {author} {\bibfnamefont {F.}~\bibnamefont
  {Verstraete}}, \bibinfo {author} {\bibfnamefont {V.}~\bibnamefont {Murg}}, \
  and\ \bibinfo {author} {\bibfnamefont {J.I.}\ \bibnamefont {Cirac}},\
  }\bibfield  {title} {\enquote {\bibinfo {title} {{Matrix product states,
  projected entangled pair states, and variational renormalization group
  methods for quantum spin systems}},}\ }\href {\doibase
  10.1080/14789940801912366} {\bibfield  {journal} {\bibinfo  {journal}
  {Advances in Physics}\ }\textbf {\bibinfo {volume} {57}},\ \bibinfo {pages}
  {143--224} (\bibinfo {year} {2008})}\BibitemShut {NoStop}%
\bibitem [{\citenamefont {Vidal}(2003)}]{Vidal2003}%
  \BibitemOpen
  \bibfield  {author} {\bibinfo {author} {\bibfnamefont {G.}~\bibnamefont
  {Vidal}},\ }\bibfield  {title} {\enquote {\bibinfo {title} {{Efficient
  classical simulation of slightly entangled quantum computations}},}\ }\href
  {\doibase 10.1103/PhysRevLett.91.147902} {\bibfield  {journal} {\bibinfo
  {journal} {Phys. Rev. Lett.}\ }\textbf {\bibinfo {volume} {91}},\ \bibinfo
  {pages} {147902} (\bibinfo {year} {2003})}\BibitemShut {NoStop}%
\bibitem [{\citenamefont {Vidal}(2007)}]{Vidal2007}%
  \BibitemOpen
  \bibfield  {author} {\bibinfo {author} {\bibfnamefont {G.}~\bibnamefont
  {Vidal}},\ }\bibfield  {title} {\enquote {\bibinfo {title} {Classical
  simulation of infinite-size quantum lattice systems in one spatial
  dimension},}\ }\href {\doibase 10.1103/PhysRevLett.98.070201} {\bibfield
  {journal} {\bibinfo  {journal} {Phys. Rev. Lett.}\ }\textbf {\bibinfo
  {volume} {98}},\ \bibinfo {pages} {070201} (\bibinfo {year}
  {2007})}\BibitemShut {NoStop}%
\bibitem [{\citenamefont {{Al-Assam}}\ \emph {et~al.}(2016)\citenamefont
  {{Al-Assam}}, \citenamefont {{Clark}},\ and\ \citenamefont
  {{Jaksch}}}]{Alassam2016}%
  \BibitemOpen
  \bibfield  {author} {\bibinfo {author} {\bibfnamefont {S.}~\bibnamefont
  {{Al-Assam}}}, \bibinfo {author} {\bibfnamefont {S.~R.}\ \bibnamefont
  {{Clark}}}, \ and\ \bibinfo {author} {\bibfnamefont {D.}~\bibnamefont
  {{Jaksch}}},\ }\bibfield  {title} {\enquote {\bibinfo {title} {{Tensor
  Network Theory - Part 1: Overview of core library and tensor operations}},}\
  }\href@noop {} {\  (\bibinfo {year} {2016})},\ \Eprint
  {http://arxiv.org/abs/1610.02244} {arXiv:1610.02244} \BibitemShut {NoStop}%
\bibitem [{\citenamefont {Giarmarchi}(2003)}]{Giamarchi2003}%
  \BibitemOpen
  \bibfield  {author} {\bibinfo {author} {\bibfnamefont {T.}~\bibnamefont
  {Giarmarchi}},\ }\href@noop {} {\emph {\bibinfo {title} {{Quantum physics in
  one dimension}}}}\ (\bibinfo  {publisher} {Clarendon Press},\ \bibinfo {year}
  {2003})\BibitemShut {NoStop}%
\bibitem [{\citenamefont {Ejima}\ \emph {et~al.}(2005)\citenamefont {Ejima},
  \citenamefont {Gebhard},\ and\ \citenamefont {Nishimoto}}]{Ejima2005}%
  \BibitemOpen
  \bibfield  {author} {\bibinfo {author} {\bibfnamefont {S.}~\bibnamefont
  {Ejima}}, \bibinfo {author} {\bibfnamefont {F.}~\bibnamefont {Gebhard}}, \
  and\ \bibinfo {author} {\bibfnamefont {S.}~\bibnamefont {Nishimoto}},\
  }\bibfield  {title} {\enquote {\bibinfo {title} {{Tomonaga-Luttinger
  parameters for doped Mott insulators}},}\ }\href
  {http://stacks.iop.org/0295-5075/70/i=4/a=492} {\bibfield  {journal}
  {\bibinfo  {journal} {EPL (Europhysics Letters)}\ }\textbf {\bibinfo {volume}
  {70}},\ \bibinfo {pages} {492} (\bibinfo {year} {2005})}\BibitemShut
  {NoStop}%
\bibitem [{\citenamefont {Moreno}\ \emph {et~al.}(2011)\citenamefont {Moreno},
  \citenamefont {Muramatsu},\ and\ \citenamefont {Manmana}}]{Moreno2011}%
  \BibitemOpen
  \bibfield  {author} {\bibinfo {author} {\bibfnamefont {A.}~\bibnamefont
  {Moreno}}, \bibinfo {author} {\bibfnamefont {A.}~\bibnamefont {Muramatsu}}, \
  and\ \bibinfo {author} {\bibfnamefont {S.~R.}\ \bibnamefont {Manmana}},\
  }\bibfield  {title} {\enquote {\bibinfo {title} {{Ground-state phase diagram
  of the one-dimensional $t$--$J$ model}},}\ }\href {\doibase
  10.1103/PhysRevB.83.205113} {\bibfield  {journal} {\bibinfo  {journal} {Phys.
  Rev. B - Condens. Matter Mater. Phys.}\ }\textbf {\bibinfo {volume} {83}},\
  \bibinfo {pages} {1--13} (\bibinfo {year} {2011})},\ \Eprint
  {http://arxiv.org/abs/1012.4028} {1012.4028} \BibitemShut {NoStop}%
\bibitem [{\citenamefont {Chen}\ and\ \citenamefont {Lee}(1993)}]{Chen1993}%
  \BibitemOpen
  \bibfield  {author} {\bibinfo {author} {\bibfnamefont {Y.~C.}\ \bibnamefont
  {Chen}}\ and\ \bibinfo {author} {\bibfnamefont {T.~K.}\ \bibnamefont {Lee}},\
  }\bibfield  {title} {\enquote {\bibinfo {title} {{New phase in the
  one-dimensional $t$--$J$ model}},}\ }\href {\doibase
  10.1103/PhysRevB.47.11548} {\bibfield  {journal} {\bibinfo  {journal} {Phys.
  Rev. B}\ }\textbf {\bibinfo {volume} {47}},\ \bibinfo {pages} {11548--11551}
  (\bibinfo {year} {1993})}\BibitemShut {NoStop}%
\bibitem [{\citenamefont {Läuchli}\ and\ \citenamefont
  {Kollath}(2008)}]{Lauchli2008}%
  \BibitemOpen
  \bibfield  {author} {\bibinfo {author} {\bibfnamefont {A.~M.}\ \bibnamefont
  {Läuchli}}\ and\ \bibinfo {author} {\bibfnamefont {C.}~\bibnamefont
  {Kollath}},\ }\bibfield  {title} {\enquote {\bibinfo {title} {{Spreading of
  correlations and entanglement after a quench in the one-dimensional
  Bose–Hubbard model}},}\ }\href
  {http://stacks.iop.org/1742-5468/2008/i=05/a=P05018} {\bibfield  {journal}
  {\bibinfo  {journal} {Journal of Statistical Mechanics: Theory and
  Experiment}\ }\textbf {\bibinfo {volume} {2008}},\ \bibinfo {pages} {P05018}
  (\bibinfo {year} {2008})}\BibitemShut {NoStop}%
\bibitem [{\citenamefont {Verstraete}\ \emph {et~al.}(2004)\citenamefont
  {Verstraete}, \citenamefont {Garc\'{\i}a-Ripoll},\ and\ \citenamefont
  {Cirac}}]{Verstraete2004}%
  \BibitemOpen
  \bibfield  {author} {\bibinfo {author} {\bibfnamefont {F.}~\bibnamefont
  {Verstraete}}, \bibinfo {author} {\bibfnamefont {J.~J.}\ \bibnamefont
  {Garc\'{\i}a-Ripoll}}, \ and\ \bibinfo {author} {\bibfnamefont {J.~I.}\
  \bibnamefont {Cirac}},\ }\bibfield  {title} {\enquote {\bibinfo {title}
  {Matrix product density operators: Simulation of finite-temperature and
  dissipative systems},}\ }\href {\doibase 10.1103/PhysRevLett.93.207204}
  {\bibfield  {journal} {\bibinfo  {journal} {Phys. Rev. Lett.}\ }\textbf
  {\bibinfo {volume} {93}},\ \bibinfo {pages} {207204} (\bibinfo {year}
  {2004})}\BibitemShut {NoStop}%
\bibitem [{\citenamefont {Zwolak}\ and\ \citenamefont
  {Vidal}(2004)}]{Zwolak2004}%
  \BibitemOpen
  \bibfield  {author} {\bibinfo {author} {\bibfnamefont {M.}~\bibnamefont
  {Zwolak}}\ and\ \bibinfo {author} {\bibfnamefont {G.}~\bibnamefont {Vidal}},\
  }\bibfield  {title} {\enquote {\bibinfo {title} {{Mixed-state dynamics in
  one-dimensional quantum lattice systems: A time-dependent superoperator
  renormalization algorithm}},}\ }\href {\doibase
  10.1103/PhysRevLett.93.207205} {\bibfield  {journal} {\bibinfo  {journal}
  {Phys. Rev. Lett.}\ }\textbf {\bibinfo {volume} {93}},\ \bibinfo {pages}
  {207205} (\bibinfo {year} {2004})}\BibitemShut {NoStop}%
\bibitem [{\citenamefont {Blakie}\ and\ \citenamefont
  {Bezett}(2005)}]{Blakie2005}%
  \BibitemOpen
  \bibfield  {author} {\bibinfo {author} {\bibfnamefont {P.~B.}\ \bibnamefont
  {Blakie}}\ and\ \bibinfo {author} {\bibfnamefont {A.}~\bibnamefont
  {Bezett}},\ }\bibfield  {title} {\enquote {\bibinfo {title} {{Adiabatic
  cooling of fermions in an optical lattice}},}\ }\href {\doibase
  10.1103/PhysRevA.71.033616} {\bibfield  {journal} {\bibinfo  {journal} {Phys.
  Rev. A}\ }\textbf {\bibinfo {volume} {71}},\ \bibinfo {pages} {033616}
  (\bibinfo {year} {2005})}\BibitemShut {NoStop}%
\bibitem [{\citenamefont {Sotiriadis}\ \emph {et~al.}(2009)\citenamefont
  {Sotiriadis}, \citenamefont {Calabrese},\ and\ \citenamefont
  {Cardy}}]{Cardy2009}%
  \BibitemOpen
  \bibfield  {author} {\bibinfo {author} {\bibfnamefont {S.}~\bibnamefont
  {Sotiriadis}}, \bibinfo {author} {\bibfnamefont {P.}~\bibnamefont
  {Calabrese}}, \ and\ \bibinfo {author} {\bibfnamefont {J.}~\bibnamefont
  {Cardy}},\ }\bibfield  {title} {\enquote {\bibinfo {title} {{Quantum quench
  from a thermal initial state}},}\ }\href
  {http://stacks.iop.org/0295-5075/87/i=2/a=20002} {\bibfield  {journal}
  {\bibinfo  {journal} {EPL (Europhysics Letters)}\ }\textbf {\bibinfo {volume}
  {87}},\ \bibinfo {pages} {20002} (\bibinfo {year} {2009})}\BibitemShut
  {NoStop}%
\bibitem [{\citenamefont {Bukov}\ \emph
  {et~al.}(2016{\natexlab{a}})\citenamefont {Bukov}, \citenamefont
  {Kolodrubetz},\ and\ \citenamefont {Polkovnikov}}]{Bukov2016a}%
  \BibitemOpen
  \bibfield  {author} {\bibinfo {author} {\bibfnamefont {M.}~\bibnamefont
  {Bukov}}, \bibinfo {author} {\bibfnamefont {M.}~\bibnamefont {Kolodrubetz}},
  \ and\ \bibinfo {author} {\bibfnamefont {A.}~\bibnamefont {Polkovnikov}},\
  }\bibfield  {title} {\enquote {\bibinfo {title} {{Schrieffer-Wolff
  transformation for periodically driven systems: Strongly correlated systems
  with artificial gauge fields}},}\ }\href {\doibase
  10.1103/PhysRevLett.116.125301} {\bibfield  {journal} {\bibinfo  {journal}
  {Phys. Rev. Lett.}\ }\textbf {\bibinfo {volume} {116}},\ \bibinfo {pages}
  {125301} (\bibinfo {year} {2016}{\natexlab{a}})}\BibitemShut {NoStop}%
\bibitem [{\citenamefont {{H. L. Essler}}\ \emph {et~al.}(2005)\citenamefont
  {{H. L. Essler}}, \citenamefont {Frahm}, \citenamefont {Gohmann},
  \citenamefont {Klumper},\ and\ \citenamefont {{E. Korepin}}}]{Essler2005}%
  \BibitemOpen
  \bibfield  {author} {\bibinfo {author} {\bibfnamefont {F.}~\bibnamefont {{H.
  L. Essler}}}, \bibinfo {author} {\bibfnamefont {H.}~\bibnamefont {Frahm}},
  \bibinfo {author} {\bibfnamefont {F.}~\bibnamefont {Gohmann}}, \bibinfo
  {author} {\bibfnamefont {A.}~\bibnamefont {Klumper}}, \ and\ \bibinfo
  {author} {\bibfnamefont {V.}~\bibnamefont {{E. Korepin}}},\ }\href@noop {}
  {\emph {\bibinfo {title} {{The one-dimensional Hubbard model}}}}\ (\bibinfo
  {publisher} {Cambridge University Press},\ \bibinfo {year}
  {2005})\BibitemShut {NoStop}%
\bibitem [{Note2()}]{Note2}%
  \BibitemOpen
  \bibinfo {note} {We note that while the $\protect \mathaccentV
  {hat}05E{b}_{ij}$ and $\protect \mathaccentV {hat}05E{b}_{ij}^{\dagger }$
  operators commute on disjoint pairs of sites they do not obey bosonic
  commutation relations if the pairs overlap.}\BibitemShut {Stop}%
\bibitem [{\citenamefont {Ammon}\ \emph {et~al.}(1995)\citenamefont {Ammon},
  \citenamefont {Troyer},\ and\ \citenamefont {Tsunetsugu}}]{Ammon1995}%
  \BibitemOpen
  \bibfield  {author} {\bibinfo {author} {\bibfnamefont {B.}~\bibnamefont
  {Ammon}}, \bibinfo {author} {\bibfnamefont {M.}~\bibnamefont {Troyer}}, \
  and\ \bibinfo {author} {\bibfnamefont {H.}~\bibnamefont {Tsunetsugu}},\
  }\bibfield  {title} {\enquote {\bibinfo {title} {{Effect of the three-site
  hopping term on the $t$--$J$ model}},}\ }\href {\doibase
  10.1103/PhysRevB.52.629} {\bibfield  {journal} {\bibinfo  {journal} {Phys.
  Rev. B}\ }\textbf {\bibinfo {volume} {52}},\ \bibinfo {pages} {629--636}
  (\bibinfo {year} {1995})}\BibitemShut {NoStop}%
\bibitem [{\citenamefont {Spa\l{}ek}(1988)}]{Spalek1987}%
  \BibitemOpen
  \bibfield  {author} {\bibinfo {author} {\bibfnamefont {J.}~\bibnamefont
  {Spa\l{}ek}},\ }\bibfield  {title} {\enquote {\bibinfo {title} {{Effect of
  pair hopping and magnitude of intra-atomic interaction on exchange-mediated
  superconductivity}},}\ }\href {\doibase 10.1103/PhysRevB.37.533} {\bibfield
  {journal} {\bibinfo  {journal} {Phys. Rev. B}\ }\textbf {\bibinfo {volume}
  {37}},\ \bibinfo {pages} {533--536} (\bibinfo {year} {1988})}\BibitemShut
  {NoStop}%
\bibitem [{\citenamefont {Ogata}\ \emph {et~al.}(1991)\citenamefont {Ogata},
  \citenamefont {Luchini}, \citenamefont {Sorella},\ and\ \citenamefont
  {Assaad}}]{Ogata1992}%
  \BibitemOpen
  \bibfield  {author} {\bibinfo {author} {\bibfnamefont {Masao}\ \bibnamefont
  {Ogata}}, \bibinfo {author} {\bibfnamefont {M.~U.}\ \bibnamefont {Luchini}},
  \bibinfo {author} {\bibfnamefont {S.}~\bibnamefont {Sorella}}, \ and\
  \bibinfo {author} {\bibfnamefont {F.~F.}\ \bibnamefont {Assaad}},\ }\bibfield
   {title} {\enquote {\bibinfo {title} {{Phase diagram of the one-dimensional
  $t$--$J$ model}},}\ }\href {\doibase 10.1103/PhysRevLett.66.2388} {\bibfield
  {journal} {\bibinfo  {journal} {Phys. Rev. Lett.}\ }\textbf {\bibinfo
  {volume} {66}},\ \bibinfo {pages} {2388--2391} (\bibinfo {year}
  {1991})}\BibitemShut {NoStop}%
\bibitem [{\citenamefont {Deutsch}(1991)}]{Deutsch1991}%
  \BibitemOpen
  \bibfield  {author} {\bibinfo {author} {\bibfnamefont {J.~M.}\ \bibnamefont
  {Deutsch}},\ }\bibfield  {title} {\enquote {\bibinfo {title} {{Quantum
  statistical mechanics in a closed system}},}\ }\href {\doibase
  10.1103/PhysRevA.43.2046} {\bibfield  {journal} {\bibinfo  {journal} {Phys.
  Rev. A}\ }\textbf {\bibinfo {volume} {43}},\ \bibinfo {pages} {2046--2049}
  (\bibinfo {year} {1991})}\BibitemShut {NoStop}%
\bibitem [{\citenamefont {Srednicki}(1994)}]{Srednicki1994}%
  \BibitemOpen
  \bibfield  {author} {\bibinfo {author} {\bibfnamefont {M.}~\bibnamefont
  {Srednicki}},\ }\bibfield  {title} {\enquote {\bibinfo {title} {{Chaos and
  quantum thermalization}},}\ }\href {\doibase 10.1103/PhysRevE.50.888}
  {\bibfield  {journal} {\bibinfo  {journal} {Phys. Rev. E}\ }\textbf {\bibinfo
  {volume} {50}},\ \bibinfo {pages} {888--901} (\bibinfo {year}
  {1994})}\BibitemShut {NoStop}%
\bibitem [{\citenamefont {Rigol}\ \emph {et~al.}(2008)\citenamefont {Rigol},
  \citenamefont {Dunjko},\ and\ \citenamefont {Olshanii}}]{Rigol2008}%
  \BibitemOpen
  \bibfield  {author} {\bibinfo {author} {\bibfnamefont {M.}~\bibnamefont
  {Rigol}}, \bibinfo {author} {\bibfnamefont {V.}~\bibnamefont {Dunjko}}, \
  and\ \bibinfo {author} {\bibfnamefont {M.}~\bibnamefont {Olshanii}},\
  }\bibfield  {title} {\enquote {\bibinfo {title} {{Thermalization and its
  mechanism for generic isolated quantum systems}},}\ }\href
  {http://dx.doi.org/10.1038/nature06838} {\bibfield  {journal} {\bibinfo
  {journal} {Nature}\ }\textbf {\bibinfo {volume} {452}},\ \bibinfo {pages}
  {854--858} (\bibinfo {year} {2008})}\BibitemShut {NoStop}%
\bibitem [{\citenamefont {Lazarides}\ \emph {et~al.}(2014)\citenamefont
  {Lazarides}, \citenamefont {Das},\ and\ \citenamefont
  {Moessner}}]{Lazarides2014}%
  \BibitemOpen
  \bibfield  {author} {\bibinfo {author} {\bibfnamefont {A.}~\bibnamefont
  {Lazarides}}, \bibinfo {author} {\bibfnamefont {A.}~\bibnamefont {Das}}, \
  and\ \bibinfo {author} {\bibfnamefont {R.}~\bibnamefont {Moessner}},\
  }\bibfield  {title} {\enquote {\bibinfo {title} {{Equilibrium states of
  generic quantum systems subject to periodic driving}},}\ }\href {\doibase
  10.1103/PhysRevE.90.012110} {\bibfield  {journal} {\bibinfo  {journal} {Phys.
  Rev. E}\ }\textbf {\bibinfo {volume} {90}},\ \bibinfo {pages} {012110}
  (\bibinfo {year} {2014})}\BibitemShut {NoStop}%
\bibitem [{\citenamefont {D'Alessio}\ and\ \citenamefont
  {Rigol}(2014)}]{DAlessio2014}%
  \BibitemOpen
  \bibfield  {author} {\bibinfo {author} {\bibfnamefont {L.}~\bibnamefont
  {D'Alessio}}\ and\ \bibinfo {author} {\bibfnamefont {M.}~\bibnamefont
  {Rigol}},\ }\bibfield  {title} {\enquote {\bibinfo {title} {{Long-time
  behavior of isolated periodically driven interacting lattice systems}},}\
  }\href {\doibase 10.1103/PhysRevX.4.041048} {\bibfield  {journal} {\bibinfo
  {journal} {Phys. Rev. X}\ }\textbf {\bibinfo {volume} {4}},\ \bibinfo {pages}
  {041048} (\bibinfo {year} {2014})}\BibitemShut {NoStop}%
\bibitem [{\citenamefont {Ponte}\ \emph {et~al.}(2015)\citenamefont {Ponte},
  \citenamefont {Chandran}, \citenamefont {Papić},\ and\ \citenamefont
  {Abanin}}]{Ponte2015}%
  \BibitemOpen
  \bibfield  {author} {\bibinfo {author} {\bibfnamefont {P.}~\bibnamefont
  {Ponte}}, \bibinfo {author} {\bibfnamefont {A.}~\bibnamefont {Chandran}},
  \bibinfo {author} {\bibfnamefont {Z.}~\bibnamefont {Papić}}, \ and\ \bibinfo
  {author} {\bibfnamefont {D.~A.}\ \bibnamefont {Abanin}},\ }\bibfield  {title}
  {\enquote {\bibinfo {title} {{Periodically driven ergodic and many-body
  localized quantum systems}},}\ }\href {\doibase
  http://dx.doi.org/10.1016/j.aop.2014.11.008} {\bibfield  {journal} {\bibinfo
  {journal} {Annals of Physics}\ }\textbf {\bibinfo {volume} {353}},\ \bibinfo
  {pages} {196 -- 204} (\bibinfo {year} {2015})}\BibitemShut {NoStop}%
\bibitem [{\citenamefont {Maricq}(1982)}]{Maricq1982}%
  \BibitemOpen
  \bibfield  {author} {\bibinfo {author} {\bibfnamefont {M.~M.}\ \bibnamefont
  {Maricq}},\ }\bibfield  {title} {\enquote {\bibinfo {title} {{Application of
  average Hamiltonian theory to the NMR of solids}},}\ }\href {\doibase
  10.1103/PhysRevB.25.6622} {\bibfield  {journal} {\bibinfo  {journal} {Phys.
  Rev. B}\ }\textbf {\bibinfo {volume} {25}},\ \bibinfo {pages} {6622--6632}
  (\bibinfo {year} {1982})}\BibitemShut {NoStop}%
\bibitem [{\citenamefont {Berges}\ \emph {et~al.}(2004)\citenamefont {Berges},
  \citenamefont {Bors\'anyi},\ and\ \citenamefont {Wetterich}}]{Berges2004}%
  \BibitemOpen
  \bibfield  {author} {\bibinfo {author} {\bibfnamefont {J.}~\bibnamefont
  {Berges}}, \bibinfo {author} {\bibfnamefont {Sz.}\ \bibnamefont
  {Bors\'anyi}}, \ and\ \bibinfo {author} {\bibfnamefont {C.}~\bibnamefont
  {Wetterich}},\ }\bibfield  {title} {\enquote {\bibinfo {title}
  {Prethermalization},}\ }\href {\doibase 10.1103/PhysRevLett.93.142002}
  {\bibfield  {journal} {\bibinfo  {journal} {Phys. Rev. Lett.}\ }\textbf
  {\bibinfo {volume} {93}},\ \bibinfo {pages} {142002} (\bibinfo {year}
  {2004})}\BibitemShut {NoStop}%
\bibitem [{\citenamefont {Eckstein}\ \emph {et~al.}(2009)\citenamefont
  {Eckstein}, \citenamefont {Kollar},\ and\ \citenamefont
  {Werner}}]{Eckstein2009}%
  \BibitemOpen
  \bibfield  {author} {\bibinfo {author} {\bibfnamefont {M.}~\bibnamefont
  {Eckstein}}, \bibinfo {author} {\bibfnamefont {M.}~\bibnamefont {Kollar}}, \
  and\ \bibinfo {author} {\bibfnamefont {P.}~\bibnamefont {Werner}},\
  }\bibfield  {title} {\enquote {\bibinfo {title} {{Thermalization after an
  interaction quench in the Hubbard model}},}\ }\href {\doibase
  10.1103/PhysRevLett.103.056403} {\bibfield  {journal} {\bibinfo  {journal}
  {Phys. Rev. Lett.}\ }\textbf {\bibinfo {volume} {103}},\ \bibinfo {pages}
  {056403} (\bibinfo {year} {2009})}\BibitemShut {NoStop}%
\bibitem [{\citenamefont {Mathey}\ and\ \citenamefont
  {Polkovnikov}(2010)}]{Mathey2010}%
  \BibitemOpen
  \bibfield  {author} {\bibinfo {author} {\bibfnamefont {L.}~\bibnamefont
  {Mathey}}\ and\ \bibinfo {author} {\bibfnamefont {A.}~\bibnamefont
  {Polkovnikov}},\ }\bibfield  {title} {\enquote {\bibinfo {title} {{Light cone
  dynamics and reverse Kibble-Zurek mechanism in two-dimensional superfluids
  following a quantum quench}},}\ }\href {\doibase 10.1103/PhysRevA.81.033605}
  {\bibfield  {journal} {\bibinfo  {journal} {Phys. Rev. A}\ }\textbf {\bibinfo
  {volume} {81}},\ \bibinfo {pages} {033605} (\bibinfo {year}
  {2010})}\BibitemShut {NoStop}%
\bibitem [{\citenamefont {Bukov}\ \emph
  {et~al.}(2016{\natexlab{b}})\citenamefont {Bukov}, \citenamefont {Heyl},
  \citenamefont {Huse},\ and\ \citenamefont {Polkovnikov}}]{Bukov2016b}%
  \BibitemOpen
  \bibfield  {author} {\bibinfo {author} {\bibfnamefont {M.}~\bibnamefont
  {Bukov}}, \bibinfo {author} {\bibfnamefont {M.}~\bibnamefont {Heyl}},
  \bibinfo {author} {\bibfnamefont {D.~A.}\ \bibnamefont {Huse}}, \ and\
  \bibinfo {author} {\bibfnamefont {A.}~\bibnamefont {Polkovnikov}},\
  }\bibfield  {title} {\enquote {\bibinfo {title} {{Heating and many-body
  resonances in a periodically driven two-band system}},}\ }\href {\doibase
  10.1103/PhysRevB.93.155132} {\bibfield  {journal} {\bibinfo  {journal} {Phys.
  Rev. B}\ }\textbf {\bibinfo {volume} {93}},\ \bibinfo {pages} {155132}
  (\bibinfo {year} {2016}{\natexlab{b}})}\BibitemShut {NoStop}%
\bibitem [{\citenamefont {{Mendoza-Arenas}}\ \emph {et~al.}(2017)\citenamefont
  {{Mendoza-Arenas}}, \citenamefont {{Gomez-Ruiz}}, \citenamefont {{Eckstein}},
  \citenamefont {{Jaksch}},\ and\ \citenamefont {{Clark}}}]{Mendoza2017}%
  \BibitemOpen
  \bibfield  {author} {\bibinfo {author} {\bibfnamefont {J.~J.}\ \bibnamefont
  {{Mendoza-Arenas}}}, \bibinfo {author} {\bibfnamefont {F.~J.}\ \bibnamefont
  {{Gomez-Ruiz}}}, \bibinfo {author} {\bibfnamefont {M.}~\bibnamefont
  {{Eckstein}}}, \bibinfo {author} {\bibfnamefont {D.}~\bibnamefont
  {{Jaksch}}}, \ and\ \bibinfo {author} {\bibfnamefont {S.~R.}\ \bibnamefont
  {{Clark}}},\ }\bibfield  {title} {\enquote {\bibinfo {title} {{Ultra-fast
  control of magnetic relaxation in a periodically driven Hubbard model}},}\
  }\href@noop {} {\  (\bibinfo {year} {2017})},\ \Eprint
  {http://arxiv.org/abs/1701.04123} {arXiv:1701.04123} \BibitemShut {NoStop}%
\bibitem [{\citenamefont {Abanin}\ \emph
  {et~al.}(2015{\natexlab{a}})\citenamefont {Abanin}, \citenamefont
  {De~Roeck},\ and\ \citenamefont {Huveneers}}]{Abanin2015a}%
  \BibitemOpen
  \bibfield  {author} {\bibinfo {author} {\bibfnamefont {D.~A.}\ \bibnamefont
  {Abanin}}, \bibinfo {author} {\bibfnamefont {W.}~\bibnamefont {De~Roeck}}, \
  and\ \bibinfo {author} {\bibfnamefont {F.}~\bibnamefont {Huveneers}},\
  }\bibfield  {title} {\enquote {\bibinfo {title} {{Exponentially slow heating
  in periodically driven many-body systems}},}\ }\href {\doibase
  10.1103/PhysRevLett.115.256803} {\bibfield  {journal} {\bibinfo  {journal}
  {Phys. Rev. Lett.}\ }\textbf {\bibinfo {volume} {115}},\ \bibinfo {pages}
  {256803} (\bibinfo {year} {2015}{\natexlab{a}})}\BibitemShut {NoStop}%
\bibitem [{\citenamefont {Abanin}\ \emph
  {et~al.}(2015{\natexlab{b}})\citenamefont {Abanin}, \citenamefont
  {De~Roeck},\ and\ \citenamefont {Ho}}]{Abanin2015b}%
  \BibitemOpen
  \bibfield  {author} {\bibinfo {author} {\bibfnamefont {D.~A.}\ \bibnamefont
  {Abanin}}, \bibinfo {author} {\bibfnamefont {W.}~\bibnamefont {De~Roeck}}, \
  and\ \bibinfo {author} {\bibfnamefont {W.~W.}\ \bibnamefont {Ho}},\
  }\bibfield  {title} {\enquote {\bibinfo {title} {{Effective Hamiltonians,
  prethermalization and slow energy absorption in periodically driven many-body
  systems}},}\ }\href {http://arxiv.org/abs/1510.03405} {\  (\bibinfo {year}
  {2015}{\natexlab{b}})},\ \Eprint {http://arxiv.org/abs/1510.03405}
  {arXiv:1510.03405} \BibitemShut {NoStop}%
\bibitem [{\citenamefont {Kuwahara}\ \emph {et~al.}(2016)\citenamefont
  {Kuwahara}, \citenamefont {Mori},\ and\ \citenamefont
  {Saito}}]{Kuwahara2016}%
  \BibitemOpen
  \bibfield  {author} {\bibinfo {author} {\bibfnamefont {T.}~\bibnamefont
  {Kuwahara}}, \bibinfo {author} {\bibfnamefont {T.}~\bibnamefont {Mori}}, \
  and\ \bibinfo {author} {\bibfnamefont {K.}~\bibnamefont {Saito}},\ }\bibfield
   {title} {\enquote {\bibinfo {title} {{Floquet-Magnus theory and generic
  transient dynamics in periodically driven many-body quantum systems}},}\
  }\href {\doibase http://dx.doi.org/10.1016/j.aop.2016.01.012} {\bibfield
  {journal} {\bibinfo  {journal} {Annals of Physics}\ }\textbf {\bibinfo
  {volume} {367}},\ \bibinfo {pages} {96 -- 124} (\bibinfo {year}
  {2016})}\BibitemShut {NoStop}%
\bibitem [{\citenamefont {Mori}\ \emph {et~al.}(2016)\citenamefont {Mori},
  \citenamefont {Kuwahara},\ and\ \citenamefont {Saito}}]{Mori2016}%
  \BibitemOpen
  \bibfield  {author} {\bibinfo {author} {\bibfnamefont {T.}~\bibnamefont
  {Mori}}, \bibinfo {author} {\bibfnamefont {T.}~\bibnamefont {Kuwahara}}, \
  and\ \bibinfo {author} {\bibfnamefont {K.}~\bibnamefont {Saito}},\ }\bibfield
   {title} {\enquote {\bibinfo {title} {{Rigorous bound on energy absorption
  and generic relaxation in periodically driven quantum systems}},}\ }\href
  {\doibase 10.1103/PhysRevLett.116.120401} {\bibfield  {journal} {\bibinfo
  {journal} {Phys. Rev. Lett.}\ }\textbf {\bibinfo {volume} {116}},\ \bibinfo
  {pages} {120401} (\bibinfo {year} {2016})}\BibitemShut {NoStop}%
\bibitem [{\citenamefont {Sorella}\ \emph {et~al.}(2002)\citenamefont
  {Sorella}, \citenamefont {Martins}, \citenamefont {Becca}, \citenamefont
  {Gazza}, \citenamefont {Capriotti}, \citenamefont {Parola},\ and\
  \citenamefont {Dagotto}}]{Sorella2002}%
  \BibitemOpen
  \bibfield  {author} {\bibinfo {author} {\bibfnamefont {S.}~\bibnamefont
  {Sorella}}, \bibinfo {author} {\bibfnamefont {G.~B.}\ \bibnamefont
  {Martins}}, \bibinfo {author} {\bibfnamefont {F.}~\bibnamefont {Becca}},
  \bibinfo {author} {\bibfnamefont {C.}~\bibnamefont {Gazza}}, \bibinfo
  {author} {\bibfnamefont {L.}~\bibnamefont {Capriotti}}, \bibinfo {author}
  {\bibfnamefont {A.}~\bibnamefont {Parola}}, \ and\ \bibinfo {author}
  {\bibfnamefont {E.}~\bibnamefont {Dagotto}},\ }\bibfield  {title} {\enquote
  {\bibinfo {title} {{Superconductivity in the two-dimensional $t$--$J$
  model}},}\ }\href {\doibase 10.1103/PhysRevLett.88.117002} {\bibfield
  {journal} {\bibinfo  {journal} {Phys. Rev. Lett.}\ }\textbf {\bibinfo
  {volume} {88}},\ \bibinfo {pages} {117002} (\bibinfo {year}
  {2002})}\BibitemShut {NoStop}%
\bibitem [{\citenamefont {Langemeyer}\ and\ \citenamefont
  {Holthaus}(2014)}]{Langemeyer2014}%
  \BibitemOpen
  \bibfield  {author} {\bibinfo {author} {\bibfnamefont {M.}~\bibnamefont
  {Langemeyer}}\ and\ \bibinfo {author} {\bibfnamefont {M.}~\bibnamefont
  {Holthaus}},\ }\bibfield  {title} {\enquote {\bibinfo {title} {{Energy flow
  in periodic thermodynamics}},}\ }\href {\doibase 10.1103/PhysRevE.89.012101}
  {\bibfield  {journal} {\bibinfo  {journal} {Phys. Rev. E}\ }\textbf {\bibinfo
  {volume} {89}},\ \bibinfo {pages} {012101} (\bibinfo {year}
  {2014})}\BibitemShut {NoStop}%
\bibitem [{\citenamefont {Dehghani}\ \emph {et~al.}(2014)\citenamefont
  {Dehghani}, \citenamefont {Oka},\ and\ \citenamefont {Mitra}}]{Hossein2014}%
  \BibitemOpen
  \bibfield  {author} {\bibinfo {author} {\bibfnamefont {H.}~\bibnamefont
  {Dehghani}}, \bibinfo {author} {\bibfnamefont {T.}~\bibnamefont {Oka}}, \
  and\ \bibinfo {author} {\bibfnamefont {A.}~\bibnamefont {Mitra}},\ }\bibfield
   {title} {\enquote {\bibinfo {title} {{Dissipative Floquet topological
  systems}},}\ }\href {\doibase 10.1103/PhysRevB.90.195429} {\bibfield
  {journal} {\bibinfo  {journal} {Phys. Rev. B}\ }\textbf {\bibinfo {volume}
  {90}},\ \bibinfo {pages} {195429} (\bibinfo {year} {2014})}\BibitemShut
  {NoStop}%
\bibitem [{\citenamefont {Daley}\ \emph {et~al.}(2005)\citenamefont {Daley},
  \citenamefont {Clark}, \citenamefont {Jaksch},\ and\ \citenamefont
  {Zoller}}]{Daley2005}%
  \BibitemOpen
  \bibfield  {author} {\bibinfo {author} {\bibfnamefont {A.~J.}\ \bibnamefont
  {Daley}}, \bibinfo {author} {\bibfnamefont {S.~R.}\ \bibnamefont {Clark}},
  \bibinfo {author} {\bibfnamefont {D.}~\bibnamefont {Jaksch}}, \ and\ \bibinfo
  {author} {\bibfnamefont {P.}~\bibnamefont {Zoller}},\ }\bibfield  {title}
  {\enquote {\bibinfo {title} {{Numerical analysis of coherent many-body
  currents in a single atom transistor}},}\ }\href {\doibase
  10.1103/PhysRevA.72.043618} {\bibfield  {journal} {\bibinfo  {journal} {Phys.
  Rev. A}\ }\textbf {\bibinfo {volume} {72}},\ \bibinfo {pages} {043618}
  (\bibinfo {year} {2005})}\BibitemShut {NoStop}%
\end{thebibliography}%

\end{document}